\documentclass[12pt,preprint,letter]{aastex}
\usepackage{amsmath,lscape}

\newcommand{\rs}{R$_{S}$}

\newcommand{\flt}{{\it \_flt~}}

\newcommand{\vmpe}{\langle v_{\phi} \rangle}
\newcommand{\vmre}{\langle v_{R} \rangle}
\newcommand{\vmze}{\langle v_{z} \rangle}
\newcommand{\vmle}{\langle v_{l} \rangle}
\newcommand{\vmbe}{\langle v_{b} \rangle}

\newcommand{\vcon}{$\langle v_{c} \rangle$}

\newcommand{\vmp}{$\langle v_\phi \rangle$~}
\newcommand{\vmr}{$\langle v_{R} \rangle$~}
\newcommand{\vmz}{$\langle v_{z} \rangle$~}
\newcommand{\vml}{$\langle v_{l} \rangle$~}
\newcommand{\vmb}{$\langle v_{b} \rangle$~}

\slugcomment{Accepted version}

\shorttitle{Stellar Proper Motions in the Bulge with HST}
\shortauthors{Clarkson et al.}

\begin{document}
\title{Stellar Proper Motions in the Galactic Bulge from deep HST ACS/WFC Photometry \footnote{Based on observations made with the NASA/ESA {\it Hubble
  Space Telescope}, obtained at the Space Telescope Science Institute,
which is operated by the Association of Universities for Research in
Astronomy, Inc., under NASA contract NAS 5-2655.}}

\author{Will Clarkson\altaffilmark{1}, Kailash Sahu\altaffilmark{1}, Jay Anderson\altaffilmark{1} T. Ed Smith\altaffilmark{1}, Thomas M. Brown\altaffilmark{1}, R. Michael Rich\altaffilmark{2}, Stefano Casertano\altaffilmark{1}, Howard E. Bond\altaffilmark{1}, Mario Livio\altaffilmark{1}, Dante Minniti\altaffilmark{3}, Nino Panagia\altaffilmark{1},
  Alvio Renzini\altaffilmark{4}, Jeff Valenti\altaffilmark{1}, \& Manuela Zoccali\altaffilmark{4}}

\affil{Space Telescope Science Institute, 3700 San Martin
  Drive, Baltimore, MD 21218, USA, email clarkson@stsci.edu}

\altaffiltext{1}{Space Telescope Science Institute, 3700 San Martin
  Drive, Baltimore, MD 21218, USA}
\altaffiltext{2}{UCLA, Los Angeles, California 90095-1562, USA}
\altaffiltext{3}{Universidad Catolica de Chile, Av, Vicuna Mackenna 4680, Santiago 22, Chile}
\altaffiltext{4}{INAF - Osservatorio di Padova, Vicolo dell'Osservatorio 5, 35122, Padova, Italy}

\begin{abstract}
We present stellar proper motions in the Galactic bulge from the
Sagittarius Window Eclipsing Extrasolar Search (SWEEPS) project using
ACS/WFC on HST. Proper motions are extracted for more than 180,000
objects, with $>$81,000 measured to accuracy better than 0.3 mas
yr$^{-1}$~in both coordinates. We report several results based on these
measurements: 1. Kinematic separation of bulge from disk allows a sample
of $>$15,000 bulge objects to be extracted based on $\ge 6\sigma$
detections of proper motion, with $<0.2\%$~contamination from the disk.
This includes the first detection of a candidate bulge Blue Straggler
population. 2. Armed with a photometric distance modulus on a star by
star basis, and using the large number of stars with high-quality
proper motion measurements to overcome intrinsic scatter, we dissect
the kinematic properties of the bulge as a function of distance along
the line of sight. This allows us to extract the stellar circular
speed curve from proper motions alone, which we compare with the
circular speed curve obtained from radial velocities. 3. We trace the
variation of the $\{l,b\}$~velocity ellipse as a function of
depth. 4. Finally, we use the density-weighted $\{l,b\}$~proper motion
ellipse produced from the tracer stars to assess the kinematic
membership of the sixteen transiting planet candidates discovered in
the Sagittarius Window; the kinematic distribution of the planet
candidates is consistent with that of the disk and bulge stellar
populations.
\end{abstract}

\keywords:{instrumentation: high angular resolution --- methods: data analysis --- techniques: photometric --- Galaxy: bulge --- Galaxy: disk --- Galaxy: kinematics and dynamics}

\section{Introduction}\label{s_intro}

The formation and evolution of merger-built bulges and secularly-grown
pseudobulges and bars is crucial to the evolution of spiral galaxies,
and indeed their formation history is used to test models for the
formation of structure in the Universe such as the $\Lambda$CDM
framework; 
(see Kormendy \& Kennicutt 2004 for a review). The inner region of our
own Milky Way shows evidence for a bulge
(e.g. Blitz \& Spergel 1991) and at least one Bar-like structure (e.g. Benjamin et al. 2005; L{\'o}pez-Corredoira
et~al. 2007). Furthermore, for our own bulge the stars are close
enough that detailed stellar data may be obtained on a star-by-star
basis, such as radial velocities and proper motions, which are simply
not yet available for most external galaxies.

These studies of our own bulge show it to be a distinct stellar
population from the disk of the Milky Way with a wide range of
abundances (Rich 1988)\footnote{We refer to the thin disk and the thick
  disk together simply as the ``disk,'' as both have similar
  kinematics (Gilmore, Wyse \& Kuijken 1989).}. High
$[\alpha/Fe]$~compared to the disk suggests rapid enrichment; Type-II
SNe produce the $\alpha$-elements and result from short-lived,
high-mass stars, while iron enrichment requires the rather slower
buildup of a significant number of Type I SNe (McWilliam \& Rich 1994;
Zoccali et~al. 2006; Lecureur et~al. 2007; Fulbright et al. 2007). A
recent spectroscopic comparison with halo objects shows the majority
of bulge stars to be $\alpha$-enhanced compared to the halo, thus the
majority bulge stellar population cannot have formed from the halo
(Fulbright et al. 2006; Fulbright et~al. 2007).

The detailed balance of populations of the bulge and its current
kinematics remain rather poorly constrained, and thus the picture of
its step-by-step formation and evolution is far from complete. Strong,
variable extinction caused by gas and dust in the intervening spiral
arms (e.g. Sumi 2004), contamination of the test population by stars
in the foreground disk, and the significant spatial depth of the bulge
along the line of sight, combine to make the separation of a
pure-bulge samply for further study a challenging task. While one can
identify a candidate main sequence, the main sequence turn-off
(hereafter MSTO) usually used as an age/metallicity diagnostic is
broadened by the age, metallicity and depth range of the bulge and
confused by the foreground disk population. Force-fitting the
Horizontal Branch in the color-magnitude diagram (hereafter CMD) and
comparison with globular cluster sequences suggests a majority
population $\sim$~10 Gyr in age, but with as much as 30\% of the stars
belonging to a young population (Holtzman et al. 1993; Ortolani et
al. 1995). Furthermore, a population of OH masers (Sevenster et
al. 1997) and a few hundred mass-losing AGB stars (van Loon et
al. 2003) {\it has} been detected in the inner Milky Way. These
objects are claimed to represent an intermediate-age ($\sim 1$-few
Gyr) population within the inner Milky Way. These objects populate the
circumnuclear molecular zone ($|l| < 1.5\degr$, $|b|<0.5\degr$) and are
thought to be tracers of a larger population that is difficult to
separate from the bulk stellar population in the color-magnitude
diagram (Lindqvist et al. 1992; Messineo et al. 2002; Habing et
al. 2006). In optical CMDs, young-intermediate age populations
overlap significantly with the main sequence of a young foreground
disk population; separation of the bulge from the disk population is
therefore critical if the detailed population distribution of the
bulge is to be obtained.

Feltzing \& Gilmore (2000) compare number counts along the CMD between
populous clusters at a range of galactic latitudes to estimate the
foreground disk contamination, producing a prediction of disk
contamination in two bulge fields and thus a differential estimate of
the bulge population itself. They find a bulge population almost
exclusively older than $\sim 10$~Gyr. This conclusion was reinforced
by a later study employing statistical subtraction to take off a
scaled contamination from a comparison field in the foreground disk;
globular cluster comparison suggested an old, metal-rich population to
fit the bulge well, but without a precise age distribution (Zoccali et
al. 2003). While a metallicity gradient with height $z$~above the
galactic midplane probably does exist (Zoccali et al. 2003), a {\it
  radial}~metallicity gradient has yet to be conclusively demonstrated
(Rich \& Origlia 2005; Minniti \& Zoccali 2008).

No evidence has yet been found for Blue Stragglers in the bulge. Blue
Stragglers are hydrogen-burning stars with apparent age younger than
that of the parent population, most likely the result of significant
mass deposition onto the star (e.g. Stryker 1993; Bailyn 1995). These
objects are found in clusters of all ages (e.g. Stryker 1993) and
would thus be expected to be present in the bulge. However they occupy
a region of the CMD that overlaps with the foreground disk so are
difficult to discriminate.

Stellar kinematics offer the cleanest method to separate bulge from
disk. The disk shows apparent streaming motion in front of the bulge
due to the motion of the Sun about the galactic center, while the
proper motion dispersion of the two populations should differ because
the two populations have widely different ages and likely differing
relaxation times (Binney \& Tremaine 1994). While statistical
subtraction tells us that some fraction of objects in a region of the
CMD may be bulge objects, kinematic constraints allow us to assign
likely bulge/disk membership to {\it individual objects}.

Beyond bulge-disk separation, stellar proper motions are valuable in
their own right as a probe of the present-day kinematics of the
bulge. Proper motions have suggested that the bulge stars show net rotation
in the same sense as the rest of the disk, though a smaller retrograde
population may also be required (Spaenhauer et al. 1992; Zhao et
al. 1994). Most previous studies converge on the model that the bulge
apparently rotates as a solid body, with circular velocity \vmp rising
linearly with distance from the galactic center until the stellar
population becomes dominated by disk stars (which show roughly
constant $\langle v_\phi\rangle$); broadly similar behaviour is
reproduced by bulge models including a rapidly rotating bar
(e.g. Sellwood 1981; Zhao 1996).
This consensus has recently been called into question by the radial
velocity study of Rich et al. (2007), which (with velocities
consistent with the PNe results of Beaulieu et al. 2000) suggests an
inflection point in the \vmp curve roughly $\sim$0.6 kpc from the
galactic center, outside which \vmp is constant at $\sim 50$km
s$^{-1}$. This suggests that solid body-like rotation may only persist
over the inner regions of the bulge/bar system. A change in the
general character of stellar orbits within the inner region of the
bulge may relate to an inner resonance (e.g. Sevenster 1999). Clearly,
an independent measure of the stellar \vmp curve is required. This is
best constructed from datasets for which systematics at different
galactocentric radii are similar, in particular the correction due to
solar motion about the galactic center; as we will demonstrate, proper
motions along a single sight-line provide just such a dataset provided
the observations are at sufficient depth to assemble large numbers of
objects in each distance bin to overcome intrinsic variation.

The bulge has now been the subject of a number of proper motion
studies. Early ground-based studies of 427 $K$~and
$M$-giants\footnote{The Spaenhauer et al. (1992) figure of 429 stars
  includes two repeated entries.} in Baade's Window
($l,b=1.02\degr,-3.93\degr$), using photographic plates separated by
nearly 33 years, showed a proper motion distribution with small
anisotropy, $\sigma_l/\sigma_b=1.15\pm0.06$~(Spaenhauer et
al. 1992). On the bulge minor axis, plate-scans over an area $25'
\times 25'$~in Plaut's Window ($l,b=0\degr,-8\degr$) across a 21-year
time interval were used to extract proper motions from 5088 objects
(at $14 < V < 18$), finding similar dispersion and anisotropy to the
Baade's Window sample (Mendez et al. 1996).

From the ground, five years of MACHO photometry allowed stars with
proper motions $\ga 18$~mas yr$^{-1}$~to be isolated, to search for
future microlensing events towards the bulge (50$\times0.5''\times
0.5''$~fields at $-1\degr < l < -10\degr$ and $-11\degr < b <
-1.5\degr$)~and Magellanic Clouds (Alcock et al. 2001). The OGLE-II
experiment yielded proper motions of some $5\times 10^6$~stars from
$49\times 0.24\degr \times 0.95\degr$~fields within ($-11\degr < l <
+11\degr$) and ($-6\degr < b < +3\degr$), with proper motions measured
to a precision of 0.8-3.5 mas yr$^{-1}$~(Eyer \& Wo\'zniak 2001; Sumi
et al. 2004), which was recently used as the basis for a study of the
trends in proper motion with location in the bulge (Rattenbury et
al. 2007a,b). Most recently, Plaut's Window has been the subject of a
second plate-based study of some 21,000 stars within a
$25'\times25'$~region, producing proper motions to $\sim 1$~mas
yr$^{-1}$~and using $\sim8700$~cross-matches with 2MASS to carefully
select bulge giants (Vieira et al. 2007).

Turning to space-based proper motion studies, a number of studies of
clusters that use proper motions to discriminate the cluster from the
field, produce bulge proper motions if the field includes a
significant number of bulge stars. Zoccali et al. (2001) reported the
the velocity dispersion of about $10^4$~bulge stars at
($l,b=5.25\degr,-3.02\degr$), as a by-product of their WFPC2 study of
the cluster NGC 6553, finding proper motion dispersion consistent with
previous ground-based studies of Baade's Window. Similar results were
obtained from the bulge stars in the field of a WFPC2 study of the
metal-rich cluster NGC 6528, at ($l,b=1.14\degr,-4.12\degr$; Feltzing
\& Johnson 2002). However, Kuijken \& Rich (2002) were the first to
dedicate HST observations to bulge proper motion studies, returning to
Baade's Window ($l,b=1.13\degr,-3.77\degr$) and the Sagittarius
low-extinction window ($l,b = 1.25\degr,-2.65\degr$) with WFPC2. This
isolated 3252 bulge stars in Baade's Window and 3867 in the
Sagittarius Window, concluding that the bulge can be best fit with a
10 Gyr-old open cluster-type sequence. No detection of $\{l,b\}$
covariance was reported, nor was any detection of blue straggler
candidates. An ACS/WFC and WFPC2 program is ongoing to reimage fields
for which observations are already present in the HST archive (Kuijken
2004; see Soto et al. 2007 for prelimiary results). Recently, ACS/HRC
was used to survey $35\times 35''\times 35''$~fields near Baade's
Window across a $5\degr\times2.5\degr$~region in the vicinity of
Baade's Window, in order to compare the bulk kinematic properties of
bulge fields as a function of location relative to the galactic center
(Koz{\l}owski et al. 2006). As pointed out by Vieira et al. (2007),
the results of the ACS/HRC study and the literature differ quite
strongly from those of Rattenbury et al. (2007b; particularly the
dispersions $\sigma_l$~and $\sigma_b$). This appears to be due to
differences in the selection of tracers used; the sample of Rattenbury
et al. (2007b) may be contaminated by evolved disk objects (Vieira et
al. 2007; see also Rattenbury et al. 2007c). Clearly, some care is
required in the selection of bulge tracer objects, particularly above
the MSTO.

We report on the use of ACS/WFC to extract precise proper motions from
a large number of stars in the Sagittarius Window towards the
bulge. The Sagittarius Window Eclipsing Extrasolar Planet Search
(SWEEPS project; Sahu et al. 2006) obtained an extremely well-sampled
photometric dataset with ACS/WFC of the Sagittarius Window towards the
bulge; this forms our first epoch. A repeat visit just over two years
later forms the second epoch, from which we extract proper motions. At
($l,b$)=(1.25$\degr$,-2.65$\degr$), the line of sight passes within
$\sim 300$~pc of the galactic center. This is far enough from the
center that the claimed population of young-intermediate age objects
traced by the OH masers and AGB stars, is likely to be an
insignificant contributor to the observed field; such objects are
likely confined to within $\sim100$~pc of the galactic mid-plane and
are found preferentially near the galactic center (Sevenster et
al. 1999; Frogel et al. 1999). Thus the stellar population of our
field of view consists of bulge, disk and halo objects.

Our first epoch has a total integration time $>$86ks each in
$F814W$~and $F606W$~and is the deepest ever observation of the
Galactic bulge. With $6\sigma$~detections of proper motions towards
the wings of the proper motion distributions where disk/bulge
separation is greatest, we push the disk contamination below the 0.3\%
level and extract the purest, largest sample of bulge stars yet
assembled.

This report is organized as follows: we provide the particulars of the
observations used in Section \ref{s_obs}. We report the procedures
used to produce precise position measurements and the resulting proper
motions in Section \ref{s_red}. Section \ref{s_pm} outlines the broad
features of the proper motion distribution of the population as a
whole. To correct for disk contamination in the sample of
kinematically selected bulge objects, we must first estimate the
fraction of disk and bulge objects in the observed sample; to do so
requires tracing the proper motion distribution as a function of
distance along the line of sight. We do so in Section \ref{s_dist};
before returning to the bulge/disk population distinction we use the
distance dependence of the proper motions to extract some kinematic
properties of the bulge in Section \ref{s_vdist}. We select a likely
bulge population in Section \ref{s_popn} and briefly examine its
properties. Finally, in Section \ref{s_cand} we assess the likely
kinematic membership of the sixteen SWEEPS planet candidates from Sahu
et al. (2006).

\section{Observations} \label{s_obs}

The target field, at (l,b) = (1.25$\degr$,-2.65$\degr$),
has been observed a total of five times with HST. The field was first
observed on 21 August 1994 with WFPC2, then again with WFPC2 on 08
August 2000. Proper motions based on these observations have been
reported by Kuijken \& Rich (2002). ACS/WFC first observed this and
four nearby regions on 9 June 2003 to allow optimal target selection
for the
SWEEPS planet-search. We focus on the final two epochs; the deep
SWEEPS observations of February 2004 and March 2006 - hereafter the
``first'' and ``second'' epochs in this text refer to the 2004 and
2006 epochs. 

Our field is the farthest from the minor axis in which deep proper
motion studies of the Galactic bulge have been performed. That said,
our line of sight passes within $\sim$200~pc of the center of the
Milky Way. With the Sun located $\sim$12-20~pc above the galactic
mid-plane (Joshi 2007), our line of sight reaches $\sim$0.5 kpc
beneath the Galactic mid-plane before intercepting the innermost
spiral arm (assuming the logarithmic spiral of e.g. Cordes 2004
accurately describes the Norma spiral arm on the far side of the
Galactic center). This is $\sim$1 disk scale-height beneath the
mid-disk (slightly more with the Milky Way disk-warp; e.g. Momany et
al. 2006), thus the disk contribution to the field on the far side of
the bulge should be rather lower than the near-side.

\subsection{ACS Epoch 1: Feb 2004}

ACS-WFC Epoch-1 observations took place between February 22-29 2004. A
total of 254 exposures in $F606W$~and 265 in $F814W$~were taken, each
with integration-time 339s. The target field was specifically chosen
to maximise the yield of potential host-stars in a single WFC pointing
and at the same time minimize the number of bright objects that would
black out regions of the chip through charge-bleeding (and thus reduce
the efficiency of the survey to planet-detection). Subpixel dithers
were set to well and redundantly sample intra-pixel sensitivity
variations. The integration-times chosen provide per-observation
photometric accuracy of about 0.04 mag at
$F814W$=23\footnote{Throughout this report, positive magnitudes refer
  to magnitudes in Vegamags, while negative values reflect the total
  counts recorded in an exposure, which we denote ``instrumental
  magnitudes,'' i.e. $M_{\rm{inst}}=-2.5\log(e-)$.}, with saturation
point just above the main-sequence turn-off at $F814W$=18.6. To
measure bright objects, three integrations each in $F814W$~and
$F606W$~were taken at 20s integration-time, providing unsaturated
measurements over the $18.5 \le F814W \le 13.8$~range, aiding
isochrone fits to the mixture of stellar populations
present. Information about these observations can be found in Sahu et
al. (2006), with further detail forthcoming (Sahu et al. 2008, in
prep).

An additional few observations were taken offset by 3$''$~in
detector-Y to cover the inter-chip gap; we do not use these bridging
observations in this work.

\subsection{ACS Epoch 2: March 2006}\label{ss_obs2}

Repeat observations of the SWEEPS field were taken with ACS/WFC on
March 09 2006 in $F814W$. Ten deep integrations were taken and two
shallow; a slightly longer visibility interval per HST orbit led to
349s integration times each for the deep observations. Subpixel
dithers were programmed in pairs, nominally at $\pm 0.25$~pix in
detector x,y from pixel-center, in reality $\pm (0.05-0.1)$~pix from
the programmed values. The two shallow observations were taken at 20s
integration times to provide proper motion estimates for bright
objects.

The two deepest ACS epoch images are quite well-aligned: not including
subpixel dithers, image centers from the two epochs are shifted with
respect to each other by (4.78,8.48) pixels (0.24$''$,0.424$''$) along
the detector, with mutual rotation $\la$8$''$, corresponding to a
pixel-offset $\sim$0.1 at the corners of the detector from rotation
alone.

\section{Analysis and Reduction} \label{s_red}

The Epoch-1 ACS dataset is among the deepest set of observations ever
taken in the optical with HST, with a strategy specifically set to
well sample the intra-pixel sensitivity variations with redundancy,
thus allowing an optimal combination of images into an oversampled
representation of the image-scene. The SWEEPS project achieved this
using an extension of the Gilliland techniques (Gilliland et al. 1999;
Gilliland et al. 2000), in which an image {\it model} of the scene is
produced for each filter. The flux at each pixel is represented by a
Legendre polynomial in the sub-pixel offsets $\Delta x$, $\Delta
y$~for each input frame (the polynomial coefficients and number of
terms to retain being determined by the counts in each pixel). For the
SWEEPS dataset this process was augmented through the fitting to each
image of a convolution kernel, that maps each input frame onto the
master-representation and thus accounts for focus-breathing. The
resulting continuous image-model can be evaluated at any sampling
desired as it tracks the estimate of flux within each WFC pixel at
each point. It is important to note that the input frames are never
resampled in the creation of this image-model, so photometric
precision is not lost. PSF-fitting photometry of this superimage
produced a master-list of magnitudes and positions for the SWEEPS
planet search (Sahu et al. 2006). The 339s exposures saturate just
above the bulge main sequence turn-off; catalogue magnitudes for
objects above the turn-off were produced using PSF-fitting photometry
of the 20s exposures. Objects saturated in the short exposures were
photometered summing over pixels into which charge has bled (Gilliland
2004). The result was a catalogue of 246,793 objects detected both in
$F814W$~and $F606W$.


\subsection{Precise position measurements with ACS/WFC}

The well-documented geometric distortion due to the optical layout of
ACS/WFC (e.g. Anderson \& King 2006) presents challenges when
attempting precise position measurements. Residuals that vary in a
complex way across the chips persist at the 0.05-0.1-pixel level after
the standard fourth-order polynomial distortion correction is applied
during the drizzle process; this alone makes the standard drizzled
image-stacks supplied by the pipeline problematic for our science
goals. We must use the raw images (or a combination thereof) to
preserve positional accuracy. After some experimentation (Appendix A)
it was found that position-scatter was minimized by using each image
within an epoch for a separate
estimate of the position and flux of each object. This produces
superior results to positions measured from a stack of images. This
approach rests on the existence of a highly-supersampled model for the
''effective PSF'' (the instrumental PSF as recorded by the detector,
hereafter ePSF; see Anderson \& King 2000 for discussion). For the
ACS/WFC, focus breathing is taken into account by adding a
perturbation-psf to the ePSF; this perturbation-psf is fit from each
image separately (Anderson \& King 2006).

\subsection{Multi-pass position-estimates}\label{s_sigma}

\begin{figure}
\plotone{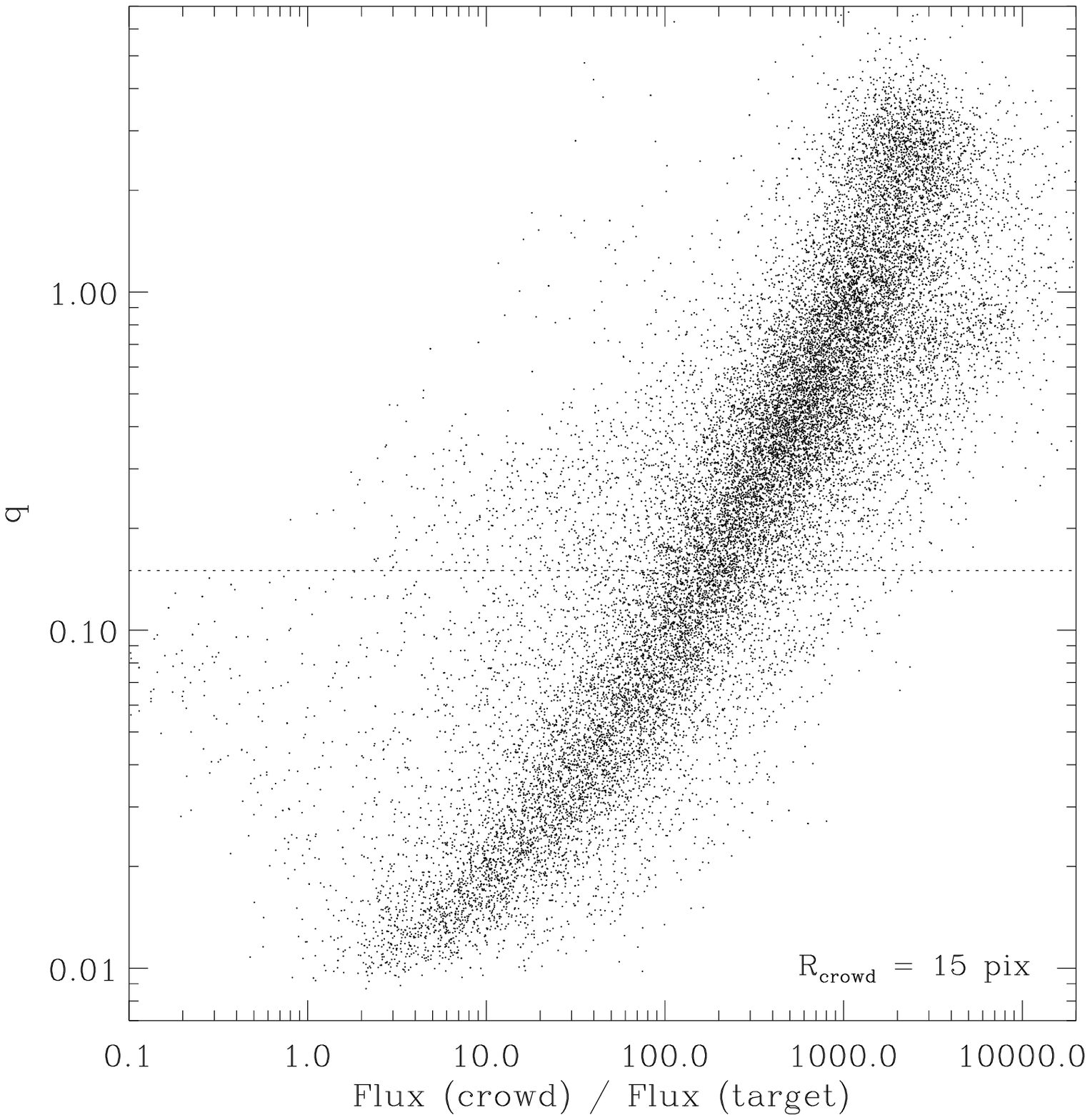}
\caption{The ratio $q$=total(ePSF-fitted flux - flux from aperture photometry)/total(ePSF-fitted flux) as a function of stellar crowding. From the master catalog of 246,793 objects we compute the ratio of the total measured flux from neighboring objects to that from the target itself. A random selection of fifteen percent of objects are plotted for clarity; we find the ratio $q$~correlates with the relative flux from crowding objects, so we adopt it as a crowding measure. Objects below the dotted line are retained when we construct a clean-bulge sample.}
\label{f_qual}
\end{figure}

With at least 246,793 objects in the frame (on average one every $\sim
8\times8$~pixels), the field is crowded but not pathologically so. To
measure positions on the frame, we thus used an improved version of
the Anderson \& King (2006) fitting routine {\it img2xym.F}, with a
3$\times$2 perturbation-psf grid rather than a single perturbation
PSF, which provides an improved measurement, and the capability to
subtract neighbors from each object before measuring its flux and
position, should similar estimates for the neighbors be available. To
minimize error, the ePSF was constructed from frames that have been
flatfielded and bias-subtracted but not resampled or corrected for
distortion. Thus measurements are performed in the raw coordinate
system of the detector (\flt space). The existence of extensive
globular cluster observations allowed distortion in the camera to be
constrained (see Anderson \& King 2003 \& 2006). This
distortion-correction is used by the Anderson \& King (2006) routines
produce output in both the raw \flt system and a transformed
coordinate set in which the application of the distortion-correction
removed higher-order distortions to the level of 0.005-0.01 pixel. We
will refer to this frame as the ``distortion-free'' frame, though it
should be remembered that residual distortion at the levels just
quoted may still be present. In addition to position and flux
measurements $\{x,y,m\}$, the Anderson \& King (2006) routines output
a quality factor $q$, which measures the difference between the
ePSF-fitted flux and the aperture flux (sum of pixel values) within a
5$\times$5-pixel region centered on the position of the star, scaled
by the ePSF-fitted flux. This ratio $q$~is seen to correlate with the
total flux from detected crowding objects in the master photometry
catalogue (Figure \ref{f_qual}), so we adopt it as a measure of
crowding due to neighboring objects (and thus including crowding
objects too close to have been isolated and measured in the
photometry).

We use multiple passes of stellar position and flux measurement to
take account of the tendency of neighbor-subtraction to build in a
dependency of the measurement of each star on those surrounding
it. Each pass requires a master list of mean estimates for each star
from the previous pass and a matchup list for each frame giving the
coordinates of a sample of well-measured, isolated objects in the
input frame and master-list, so that the transformation between the
master list and the frame itself can be properly taken into account
when subtracting neighbors.

The first pass with the Anderson \& King (2006) routines produces a
set of $\{x,y,m,q\}$~estimates for each image without neighbor
subtraction, and with nonuniform row ordering. We match these
measurements to the master catalogue of 246,793 objects from
photometry of the optimal superimage from Sahu et al (2006) to recast
each of the pass-1 estimates with the same row-ordering as the Sahu et
al (2006) catalogue. Note that we {\it never} use the position and
flux estimates from Sahu et al. (2006) again in this analysis due to
the shifting of flux that takes place when constructing the stack
(Appendix A); this catalog is used purely to enforce a uniform
row-ordering in the object catalogues for proper motion extraction. We
now construct a sigma-clipped median set of $\{x,y,m,q\}$~measurements
of each star from pass-1, accounting for residual trends in the manner
described below. This forms the input catalogue to the multi-pass
photometry.

In subsequent passes, this input catalogue is transformed back into
each individual \flt frame using the matchup lists just
produced. Neighbors are now subtracted from each object before
measurement; to avoid measurement instability and dependence on the
ordering in which objects are processed, the position and magnitude
measurements from the previous pass are used when subtracting
neighbors. The individual measurements are now used to produce a
sigma-clipped mean measurement for $\{x,y,m,q\}$, which updates the
input list for the next pass as well as the matchup lists. As a matter
of record, photometry using the modified Anderson \& King routines
with neighbor subtraction takes over an hour per frame on a 3GHz
Linux CPU with 1Gb RAM, so we split this step across multiple CPUs to
reduce the time necessary to process 265 images in the first
epoch. The collation, accounting for trends and refinement of the
input list to the next pass, requires the complete measurement list
for each star and can take up to eight hours per pass. This process is
repeated until convergence; there is little improvement between the
third and fourth pass of the photometry except at the faintest end of
the star-list, so we stop after four passes. Although there are only
ten deep images in the second epoch, the production of a clipped
master-list does improve the photometry for neighbor-subtraction so
we follow a similar process for the 2006 epoch.

\subsection{Reference frames and accounting for residual trends}

A reference image is chosen for each epoch with focus within a few
percent of the maximum focus sharpness and commanded orientation at
the middle of the dither pattern (for 2004 image j8q632crq was used,
for 2006 image j9ev01ntq was used). The Anderson \& King (2006)
estimates in \flt space are related to those in distortion-free space
by a well-constrained transformation, thus we are free to choose which
frame we use for fitting the transformation that maps each individual
set of measurements onto the master-set (essential for producing the
refined matchup lists); we work in distortion-corrected space so that
linear transformations can be used, reducing the transformation error.

\begin{figure}
\plotone{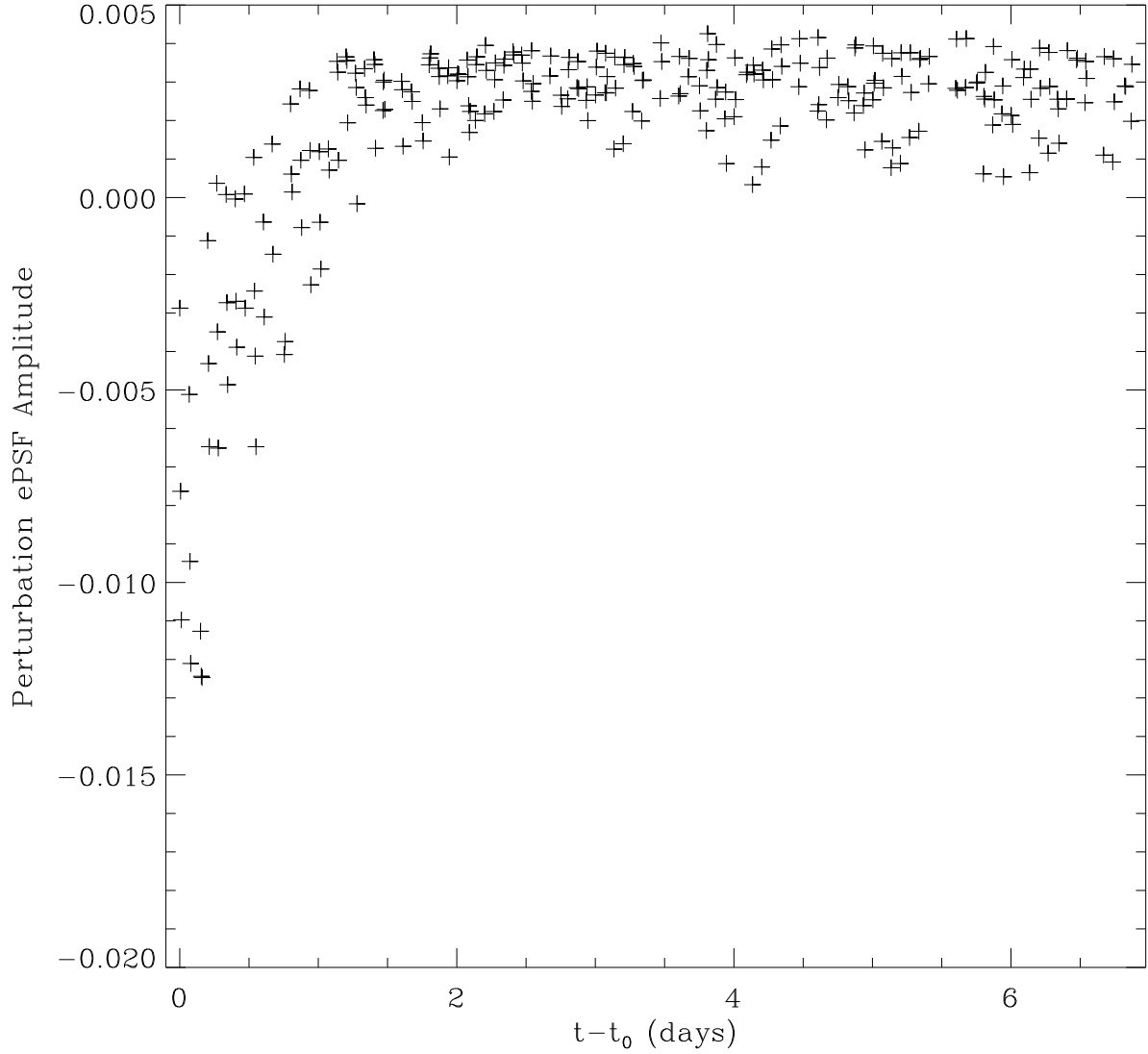}
\caption{Amplitude of the perturbation-PSF that must be added to the library ePSF to best represent the scene in each image. The history of this amplitude variation is a proxy for focus history. This measurement is from the first pass at position-measurement in which a single perturbation is fit to the entire frame; pass 2 onwards uses a spatially dependent perturbation PSF.}
\label{f_focus}
\end{figure}

The distortion correction transformation itself is claimed to be
accurate to the 0.005-0.01 pixel level (Anderson \& King 2006), this
tolerance introduces 0.01-0.5 mas uncertainty in position
measurement. Additionally, observatory-level systematics are present
that cause the position and flux to vary with time - this is clearly
seen in a plot of the amplitude of the perturbation-psf that must be
applied to the library psf to optimise the measurements in the first
pass (a proxy for focus; Figure \ref{f_focus}). These trends are not
an entirely smooth function of time and thus can add scatter of up to
$\sim$0.5mas to the position estimates. We account for and subtract
trends in $\Delta x$, $\Delta$y and $\Delta$m for each chip
independently, assessing the trends from the $\sim$4,000 best-measured
objects in the frame. Experimentation with polynomial surface-fitting
and a lookup-table approach suggested that for our data a polynomial
with cross-terms up to $x^3y^3$~provided the most robust subtraction
of trends. In cases where subtraction of the polynomial surface {\it
  increased} the scatter across the frame (an indication that residual
trends were not significant for the frame in question), the
subtraction was not used.

When positions and fluxes are measured with neighbor-subtraction, the
variance of measurement within an epoch is generally lower than from a
single-pass alone and the distribution is rather tighter. However
after the second pass (i.e. the first with neighbor-subtraction) a
small population appears that is fainter than instrumental magnitude
$\sim-13.2$~and with roughly constant position-rms $0.08$~pix or so
(Figure \ref{fig_rms}). These objects are apparently due to the
influence of neighboring objects, and were dealt with appropriately
(Appendix B). In addition, the effects of differential Charge Transfer
Efficiency (CTE) are noticeable both in position measurements taken at
differing integration times within the same epoch, and in magnitude
measurements taken at similar integration times in two epochs, due to
degradation of the detector in the low-Earth environment. However, any
differential CTE effect on our {\it proper motion} measurements is
predicted to be too small to measure, and indeed is not observed in
our positions (Appendix C).

\subsection{Positions to proper Motions} \label{sub_pm}

The result is two star lists, where mean positions in the 2004 epoch
are in the distortion-free frame of the most representative 2004 image
and the mean positions in the 2006 list are in the distortion-free
frame of the most representative 2006 image. The distortion-correction
of Anderson \& King (2006) was applied to both lists of coordinates
(and both sets of fluxes through the variation in pixel-area on the
sky). Residual imperfections in the distortion remaining after
subtracting trends will lead to distortion remaining between the two
position-lists.

The ACS/WFC distortion is now known to change monotonically with time,
as can be seen for example in the evolution of the skew terms in the
distortion solution used here. This change is rather small for most
purposes, roughly 0.3 pix in the 2002-2006 interval at the corners of
the image, and appears to be confined to the linear terms (Anderson
2007). However, evolution in distortion as traced by the first six
terms of the transformation are automatically accounted for in our
approach as we are fitting 6-term transformations between the epochs;
no evidence is seen for evolution in the higher terms of the
distortion correction.

The 2006 epoch is particularly prone to residual distortion due to the
comparatively low number of frames available to characterize and
remove the frame-by-frame residual trends. We therefore return to the
individual 2006-epoch position-lists, recomputing the proper motions
on an image-by-image basis. We use local transformations to best map
the frame near each target star onto the frame of the 2004
starlist. For each target star in each 2006 image, we use an AMOEBA
fit (Press et al. 1992) to find the 6-term transformation that maps
the positions of a set of nearby tracer stars from 2006 onto their
positions in the 2004 star-list (Figure \ref{fig_tracers}). This
transformation is then used to predict the position of the target star
in the 2004 epoch. The offset of this position from the true position
in 2004 is then the proper motion of the target star, estimated from
the individual 2006 star-list with reference to the 2004
master-list. The set of $3$-$10$~proper motion estimates (at least
three good measurements are needed) is then averaged together with
outlier removal to estimate the true proper motion of the target
star. Finally, the proper motions in image coordinates are transformed
into proper motions in galactic coordinates. This process is repeated
for every object in the master-photometry list, which is computationally 
accomplished most conveniently through an overnight run.

Care must be taken when selecting frame-to-frame reference stars in
this process. As we lack an extragalactic reference point, we assume
that all the stars in the frame are in motion and adopt as the zero
point the average motion of the bulge population as estimated from the
CMD. Of likely non-disk objects we include only those unsaturated
objects for which the combined x- and y-coordinate rms from the 2004
epoch is less than 0.007~pix, the crowding measure $q$~less than 0.025
and at least 200 measurements were taken in the 2004 epoch (so the
positions are well-characterized). As a precaution, the 2006
mean-position list produced above is used to remove tracer stars that
show discrepancy above 200mas (4~pix) between mean position in the two
epochs, indicative of unusually high proper motion, a problem with the
transformation between frames, or an object mismatch; this culling is
needed for a handful of objects. We do not include the target star
itself in the transformation (see Anderson et al.  2006), and are
careful to ensure that tracer stars only on the same chip as the
target star are used, to guard against any trends over time in the
distortion of the two chips relative to each other. These
considerations leave $\sim$8,000 reference objects over both chips. Of
these objects, the nearest 100 to the target star are selected to fit
the local transformation, with 5-pass sigma-clipping removing
typically 8-20 objects (Figure \ref{fig_tracers}). The area covered by
the reference stars is typically $500\times500$~pixels, or $\sim 1/8
\times 1/8$~the field of view.

This approach allows the proper motion error to be fully characterized
on a star by star basis. For each star, this error contains three
terms: the internal scatter $\epsilon_{2004}$~accompanying $N_{2004}$
repeated position measurements in the 2004 epoch, the rms variation in
position difference $\epsilon_{2}$~over the 2.04-yr interval between the
$N_{2006}$~set of position-pairs, and finally the relative
position-scatter that will be built up when $n_{tr}$~tracer stars are
allowed to move in random directions with intrinsic scatter
$\sigma_{pm}$~mas yr$^{-1}$~over a 2.04-year time interval. {\it Per
  co-ordinate}, then, the proper motion error is:
\begin{equation}
  \epsilon^2_{x} = \frac{1}{2.04^2}\left( \frac{\epsilon^2_{2004}}{N_{2004}} + \frac{\epsilon^2_{2}}{N_{2006}} + \frac{(2.04\sigma_{pm})^2}{n_{tr}-2} \right)
  \label{eq_e}
\end{equation}
\noindent For well-measured stars the first two terms are rather
small; in 2004 we find the brightest nonsaturated objects show
positional scatter $\epsilon_{2004}$~in the 1.5 {\it milli}pixel regime
(Figure \ref{fig_pmerror}); the intrinsic scatter $\sigma_{pm}$~over
the 2.04 year interval dominates the error estimate for these objects
However the advantage to this approach is that the additional
systematic introduced when two average position-lists are used under
the same distortion correction, has been fully accounted for. At the
faint end the measurement error dominates the error budget.

\subsection{Proper motions from saturated objects}

We will want to compare proper motions above the main sequence
turn-off with those below it. With 265+10 exposures total at 339s or
longer, we might expect better statistics when producing average
proper motions from the long exposures compared to the five available
short exposures. However, the integration time for the long exposures
was chosen so that saturation coincides roughly with the turn-off;
thus, position measurements and therefore proper motions may be
subject to greater error than the variance of position measurements
would indicate. Indeed, we note a CTE-like effect when comparing the
positions measured for highly saturated objects compared to the same
objects measured with much shorter integration times (Figure
\ref{ref_fig_satn}; see also Appendix C).

We thus extracted proper motions from the shorter exposures to compare
with the long exposures, to search for any saturation effects. The
process is similar to that used for the deeper exposures, however this
time too few integrations are available to make possible the
trend-fitting and removal procedures used for the deeper
exposures. Instead, proper motions are computed pairwise between each
of the six possible pairs of measurements between the three short
integrations in 2004 and the two in 2006. The results are then
robustly averaged with outlier removal to produce proper motion
estimates from the short integrations. The measured magnitude for each
star is the sigma-clipped mean of all five brightness measurements for
each star, to mitigate the effects of cosmic rays when subtracting
neighbors for object-measurement. The resulting proper motion
distribution is statistically indistinguishable to that from the
deeper integrations, as we show in Section \ref{s_pm}.

\subsection{Multi-epoch proper motions}\label{ss_multi}

To illustrate that we are indeed measuring proper motions, and not
spurious events, we construct multi-epoch proper motions, using all
five epochs at which this field has been observed (Section
\ref{s_obs}). Position-measurements from the WFPC2 observations and
the relatively poorly temporally-sampled 2003 ACS/WFC observations
have rather larger error than our 2004 and 2006 measurements, whilst
the transformation from WFPC2 to ACS/WFC systems is not yet fully
refined. To extract science from our proper motions we therefore
content ourselves with the 2004 and 2006 epochs. However, as can be
seen in Figure \ref{f_multi}, we clearly {\it are} measuring motion of
the stars across the field of view, with the distance traveled between
measurements proportional to the time interval between the
measurements (Figure \ref{f_multi}).

\section{Stellar Proper Motions} \label{s_pm}

The result of the procedures in the previous section is a set of
proper motions for 187,346 objects in our frame, 81,140 to better than
0.3 mas yr$^{-1}$~in {\it both} coordinates. From the shorter
observations, a list of
$>42,000$~objects is produced with proper motions at better than 1mas
yr$^{-1}$~accuracy. We aim to estimate proper motions in the
disk as separate from the bulge, so we first examine the proper
motions above the main sequence turn-off.

\subsection{Stellar proper motions above the MSTO}\label{ss_bright}

As noted by several previous authors, the region above the MSTO, where
the CMD splits relatively cleanly into disk-dominated and evolved
bulge-dominated populations, provides a convenient tracer 
which can be used to estimate
the kinematics of disk and bulge
populations. We will use this as a first estimate when we attempt to
divide the populations further by apparent distance below.

The longitudinal proper motions clearly separate the disk and bulge
populations, with the latitudinal proper motions also providing
separation but to a lesser extent. Fitting a single Gaussian to each
of the proper motion histograms of the two populations suggests
separations $(\Delta \mu_l, \Delta \mu_b)$=$(3.24 \pm 0.16 , 0.76 \pm
0.13)$~mas yr$^{-1}$~between the two populations. There is thus an
average tilt of order 13$\degr$~between the centroid proper motions of
the two populations (Figure \ref{f_short}). The proper motions in this
region from the short exposures are statistically identical to those
from the long exposures: $(\Delta \mu_l, \Delta \mu_b)$=$(3.21 \pm
0.15 , 0.81 \pm 0.13)$~mas yr$^{-1}$~between the two populations
(Figure \ref{f_short2}). However, the scatter in the resulting proper
distribution is lower, implying superior measurement. Because proper
motions above the MSTO show different systematics between the long and
short exposures, we cannot simply add the two sets of measurements in
quadrature. Instead, for objects that (i) are saturated in the long
exposures, and for which (ii) at least five proper motion measurements
are retained after sigma-clipping, we simply substitute proper motions
from long exposures with those from the short exposures. Because of
different selection biases for bulge objects above and below the MSTO
(which define the reference transformation between epochs), we are
careful to first take into account the difference between proper
motion zeropoints between long and short exposures; this difference
amounts to ($\Delta \mu_l,\Delta \mu_b$)=(-0.2,0.01) mas yr$^{-1}$.


\subsection{Stellar proper motions below the MSTO}

Kinematic distinction between bulge and disk is most vividly
illustrated by color-coding the CMD by the mean proper motions
$\mu_l,\mu_b$~and their dispersions $\sigma_{l},\sigma_b$. As noted
previously by Kuijken \& Rich (2002), $\mu_l$~and particularly
$\sigma_b$~show clear association with the
bulge population (Figure \ref{fig_hess}), the latter expected for a
dynamically older population. The conclusion that the bulge is mostly
an old stellar population, is thus independent of the details of any
isochrone fits. The longitudinal proper motion scatter is also a
tracer of the bulge stellar population, though less obviously than the
latitudinal scatter, as expected for a foreground disk with multiple
populations participating in streaming motion of various speeds around
the galactic plane. There is also a hint of a trend in $\mu_b$, however
from proper motions alone the significance of this trend is not clear.

The motion of the Sun with respect to the local standard of rest (LSR)
causes a trend in $\mu_l$~and $\mu_b$~that is strongest for foreground
disk objects in off-axis fields; indeed with multi-filter observations
to estimate distance and extinction directly, this can be used to
constrain stellar motions in the disk (Vieira et al. 2007). This trend
increases as $1/d$; any intrinsic disk-to-bulge proper motion trends
are superimposed on this trend. Because the Sun orbits the galactic
center slightly faster than the LSR, this trend will be in the
opposite sense to the $\mu_l$~trend we observe, in that the solar
reflex contribution to proper motions will be negative with respect to
our tracer bulge objects. For our field, the size of this correction
to bulge objects will be $\sim$0.1-0.2 mas yr$^{-1}$~from the near- to
far-side of the bulge. With two-filter photometry for our field, we do
not have distance estimates for most of our objects. We therefore
restrict detailed kinematic interpretation to those objects for which
distances can be estimated (see below); for these objects the solar
motion drops out as a constant velocity term. We note in passing that
the size of the solar motion trend is the same size as the apparent
trend in $\mu_b$~across the CMD (Figure \ref{fig_hess}).


\section{Photometric Parallax}\label{s_dist}

We need to assess the contamination from foreground disk stars in our
kinematically selected bulge sample; this requires us to estimate the
number of disk and bulge stars in our field. This exercise in turn
requires that we determine the kinematics as a function of distance
along the line of sight, so we now turn to dissection of proper motion
as a function of distance.

We estimate distances to each star by computing its distance modulus
relative to the isochrone of the mean-bulge population (from Sahu et
al. 2006). This isochrone was fit by dividing the CMD into strips of
equal color, taking the median in each strip and finding the
combination of extinction, distance modulus and metallicity that best
fits the resulting ridgeline for a canonical bulge age of 10
Gyr. Isochrones from VandenBerg, Bergbusch, \& Dowler (2006) were
used, transformed to the ACS bandpasses as reported in Brown et
al. (2005). The mean-bulge isochrone has (m-M)$_0 = 14.3$, foreground
extinction E(B-V)=0.64 and [$\alpha$/Fe]=0.3.  Note that although Sahu
et al (2006) used the standard extinction law (as opposed to the
anomalous extinction of Sumi (2004 and references therein), this
distance modulus does agree with the {\it corrected} distance modulus
of red clump giants in the bulge (Sumi 2004), and corresponds to a
mean-bulge distance of 7.24 kpc. With the geometrically determined
sun-galactic center distance $R_s=7.62 \pm 0.2$~kpc (Eisenhauer et
al. 2005), this bulge distance is consistent with the distance of
intersection of the bulge major axis with the line of sight for bulge
orientation angle $\beta \ga 14\degr$.

The majority of the interstellar extinction takes place in the
foreground screen of the galactic spiral arms (Stanek 1996; Cordes \&
Lazio 2003). For objects within about 3.3 kpc of the galactic center
(i.e. interior to the Norma spiral arm), this approximation is
adequate for detailed constraints on the bulge (Cordes 2004; Wainscoat
et al. 1992). Farther out from the galactic center, the line of sight
passes through the spiral arms Carina-Sagittarius (at approximately
6.7 \& 6.4 kpc from the galactic center on the near and far side of
the galaxy respectively),
Crux-Scutum (at 4.8 \& 4.8 kpc) and Norma (3.3 \& 2.9 kpc). The
interstellar dust distribution of the galactic disk shows scale height
$\sim$0.14~kpc (Bienaym\'e, Robin, \& Crez\'e 1987), which means our
line of sight passes more than one dust scale-height from the galactic
mid-plane just before intercepting the Norma spiral arm on the
near-side. Thus distances inferred for objects closer than $\sim$4~kpc
will be somewhat uncertain.

Some care must be taken interpreting the resultant distance modulus,
as the magnitude scatter is not due to distance effects alone. This is
most clearly seen by comparing the distance estimates obtained from
the main-sequence to those of HB clump+RCG Bump objects (which we
denote together as red-clump giant, or RCG objects; the small
contamination from evolved disk objects referred to in Zoccali et
al. 2003 and Vieira et al. 2007 is not an important contamination for
this estimate). Figure \ref{f_distcomp} shows the result: while the
{\it maximum} distance spread (1$\sigma$) of RCG objects corresponds
to 0.17 mag, below the main sequence the spread is twice this
amount. Were the disk to be highly overrepresented in the main
sequence sample, this would bias the distance-modulus distribution
towards the bright end; this is not seen. The two measures should
therefore be identical. The fact that the scatter of the main sequence
distance measure is a factor two higher than that from the RCG measure
suggests that the distance estimate for the main sequence is
contaminated by other effects which would include metallicity
variations, the presence of stellar binaries, etc. as described in
Section \ref{ss_recov}. When using photometric distance estimates to
draw conclusions about bulge kinematics, we use the scaling between
the RCG sample and the sample below the MSTO to correct all apparent
photometric distance modulii to estimated true distance modulii.

\subsection{Can we recover bulge kinematics from our data?}\label{ss_recov}

To assess the impact of this pollution of our distance estimate on our
ability to recover bulge kinematics, we first attempt to reproduce the
distance discrepancy, then compare the kinematics we would recover
with those simulated.

The CMD of a trial bulge population was simulated using the mean-bulge
isochrone from Sahu et al. (2006). First a stellar population was
produced along the mean-bulge isochrone (a salpeter IMF was assumed
for simplicity; the results turn out to be rather insensitive to the
IMF form adopted). The population was then perturbed by a distance
modulus distribution matching that measured from the RCG objects
(Figure \ref{f_distcomp}, upper right panel). The population was then
further perturbed due to a binary population. Because equal-mass
binaries are rather unlikely above the MSTO, the binary effect was
only added to objects beneath the turn-off. A metallicity distribution
to the magnitudes was then added using the Rich et al. (2007)
spectroscopic results to estimate the distribution of metallicity and
the Vandenberg isochrones to translate this into a magnitude
perturbation (see Brown et al. 2005); beneath the turn-off this
perturbs the magnitudes and (to a lesser extent) colors; above the
turn-off the main effect is on color.

The binary fraction of the bulge is unknown, as is the distribution of
relative brightness $f(\Delta I)$~of binary components. We first used
the solar neighborhood as a proxy, using the tabulation from Hipparcos
of S{\"o}derhjelm (2007), although repeated experiments suggest the
final result is somewhat insensitive to both the binary fraction and
$f(\Delta I)$, providing equal-light binaries are not dominant, and
the binary fraction is of order $\sim 0.3-0.6$. The difference in
spreads in apparent distances between the clump and main sequence is
reliably reproduced (Figure \ref{f_dtrial}).

We now assess our ability to recover input kinematics. For each trial,
stellar velocities were simulated following a simple model that
qualitatively matches the radial velocity results of Rich et al
(2007); the bulge is assumed to participate in solid-body rotation up
to some cut-off radius $R_C=0.4R_s$~with a flat mean circular speed
curve $\langle v_c \rangle=50$~km s$^{-1}$~exterior to this
radius. Velocity dispersion of 75 km s$^{-1}$~was added to the
canonical curve, and the resulting stellar velocities translated into
proper motions. Finally, the proper motions are translated back into
''observed'' velocities, this time without any information as to the
metallicity or binarity of each star. For reference, the relations
between the mean inferred transverse velocities $\langle v_l \rangle$,
$\langle v_b \rangle$~to the intrinsic radial, azimuthal and vertical
components \vmr, \vmp and \vmz are, for our pointing,
\begin{eqnarray}
  \vmle & = & \vmpe\sin(\alpha) + \vmre\cos(\alpha) - \langle v_l \rangle_0 \nonumber \\
  \vmbe & = & 0.9989\vmze - 0.0462\vmpe\cos(\alpha) - 0.0462\vmre \sin(\alpha) - \langle v_b\rangle_0 \nonumber \\
\label{eq_pmvel}
\end{eqnarray}
\noindent where the line of sight to the star and the velocity vector
of the star in a circular orbit make angle $\alpha$~to each other. (The
projected star-galactic centre distance $R$~is fixed by the position
of the field and the Sun-galactic centre distance $R_S$, allowing easy
estimation of $\alpha$~through the relation $R\cos(\alpha)=R_s
\sin(l)$). Error in the distance inferred thus propogates through to
error in the inferred circular speed curve.

The circular speed curve extracted from the simulated proper motions
is then fit to the simple model for the trial, and the result compared
to the input model. This process is repeated for many trials; Figure
\ref{f_binsim} shows the resulting distribution of recovered
parameters over 300,000 trials. Aside from cases in which the fit
fails to find a solution (both $R_c$~and $\langle v_c \rangle$~zero),
the cutoff radius $R_c$~tends to be about a factor two higher than
that input, and with a high scatter ($0.2R_S$). The recovered
tangential circular speed $\langle v_C \rangle$~is systematically lower than that
input (50 km s$^{-1}$) by a factor $\sim 2$; the most frequently
recovered value being $25 \pm 6$km s$^{-1}$. Thus, the form and
approximate cutoff of the input circular speed curve {\it can} be
recovered, but with a constant circular speed $\langle v_c
\rangle$~and cutoff radius $R_c$~about half and twice that input,
respectively.

\section{Stellar Kinematics vs Distance from the Galactic Center}\label{s_vdist}

We group objects by distance modulus to estimate the space motion of
the average star in each bin as a function of distance $d$~(Figure
\ref{fig_obs}; we remind the reader that we have corrected apparent
distance modulii to estimates of the true distance modulii; Section
\ref{ss_recov}). Within 3kpc of the mean-bulge, we use bins of equal
($d^3_{n+1}-d^3_n$), so that the number of objects in each bin traces
the mean volume density of stars in that bin. We assume that the bulge
stars in our selected sample are members of the same population, so
that the absolute magnitude distribution is the same for each bin
across the bulge. We also assume that the kinematic properties of the
bulge do not vary strongly with absolute magnitude, so that our
tendency to preferentially detect intrinsically fainter objects when
they are closer, does not bias the kinematics we produce. At $20 <
F814W <23$, our tracer objects are well above the brightness at which
photometric completeness becomes a significant systematic
(e.g. Piotto et al. 2007). Note that the proper motion dispersion
decreases once we probe the far side of the bulge, an indication that
photometric errors are not dominating our kinematics on the far side
of the bulge. Furthermore, while the proper motion dispersion
decreases on the bulge far side, the corresponding {\it velocity}
dispersion does not.
Our field of view encompasses an ever-larger area on the sky with
distance, so the far distance-bins may sample a wider range of
velocities than those close-by.


\subsection{Circular speed curve}

Without the line-of-sight counterpart $\langle v_{r} \rangle $~for
this field, Equation (\ref{eq_pmvel}) represents two conditions in
three unknown velocities; to break the degeneracy we must either
observe $\langle v_r \rangle$~as a function of line of sight distance,
or adopt a relationship between the three components of streaming
velocity; such a relationship is best obtained through numerical
modeling in a realistic potential (see e.g. Zhao 1996). We adopt a
simple prescription for \vmp and set the other two components to
zero. The ongoing radial velocity survey of Rich et al. (2007) finds
no evidence for minor-axis rotation, so \vmz $\sim$0 is at this stage
reasonable. Observations are rather more ambiguous for \vmr within a
few degrees of the galactic rotation axis (Rich et al. 2007); we set
this term to zero to determine if it is required to fit the observed
velocities. We assume \vmp follows the broad pattern of the ISM
(e.g. Clemens 1985), rising monotonically with galactocentric radius
interior to some cut-off radius $R_c$, exterior to which \vmp is
roughly constant at \vcon~ km s$^{-1}$. In this respect our
prescription for \vmp is similar to that of a rapidly rotating bar,
producing an apparently solid body-type rotation curve (e.g. Zhao
1996). We search for the combination of $R_c$, \vcon~and $R_s$~that
best reproduces the observed \vml variation. The sensitivity of proper
motion to intrinsic circular velocity decreases when motion is largely
along the line of sight, as occurs near the meridional plane for our
line of sight. Additionally, the most dramatic change in the balance
of components occurs over a relatively narrow range in line of sight
distance $d$, so that our chosen binning scheme (which preserves
information on the intrinsic stellar density) results in relatively
few bins over a complicated velocity variation and formally precise
fitting is difficult. We thus impose the additional condition that the
\vmp trend on near and far-sides of the bulge be similar. Finally, we
restrict ourselves to $R_S$~within 1$\sigma$~of the value determined
geometrically by Eisenhauer et al.  (2005). This artificially simple
model does reproduce the observed variation of \vml with distance
reasonably well, adopting $\langle v_{c}\rangle=25$~km s$^{-1}$,
$R_c=0.3-0.4$~kpc and $R_s= 7.7$~kpc (Figure \ref{f_rotcurve}). In
agreement with Zhao et al. (1994) we find the mean streaming motion
must be prograde.

To our knowledge, this is the first measurement of the circular speed
curve $\vmpe$~of the inner Milky Way to be detemined purely from
proper motions, and is thus an independent check of the stellar
circular speed curve determined by other means. Our stellar rotation
curve from proper motions agrees qualitatively with that recovered
from a sample of radial velocities of $K$~and $M$~giants across a swath at
$b=-4\degr$~(with 90-110 stars per field; Figure 3 of Rich et al
2007). The circular speed inflection of Rich. et al (2007) takes place
$\sim$~0.6kpc from the galactic center, which for the $b=-4\degr$~
swath corresponds to radial coordinate parallel to the galactic plane
$R_c \sim$0.3kpc. Thus, our turning point at $R_c=0.3-0.4$~kpc is
entirely consistent with that of Rich et al.(2007), suggesting
cylindrical, non solid-body rotation.

While the apparent departure points from solid body-like rotation
agree with radial velocity and proper motion studies, the amplitude of
\vcon~towards which both circular speed measurements trend at high
galactocentric radii, differ from each other by about a factor two. We
found in Section \ref{ss_recov} that the metallicity, age and binarity
uncertainties tend to produce \vcon~roughly half that of the intrinsic
value. Therefore, our circular speed curve is fully consistent with
that of the Rich et al. (2007) radial velocity survey in both turning
point and velocity amplitude. Because of the high intrinsic velocity
scatter, we are not able to set fine limits on the degree of any
disparity, however it seems clear that highly eccentric average
stellar orbits are not consistent with our data.

We note in passing that the circular speed curve of the ISM does not
provide a check of this discrepancy, as it measures the circular speed
of a population with a definite, well-defined pattern speed of
$\sim$200-230 km s$^{-1}$~at the galactocentric radii of interest
(e.g. Burton \& Liszt 1992). This is in stark contrast to the stellar
circular speed curve, which measures the average of a population with
high intrinsic velocity dispersion and likely several sub-populations
(which may include a population on {\it retrograde} orbits; e.g. Zhao
et al 1994). The importance of measured mean velocity on the mixing of
populations is also evident in the behaviour of \vmp at nearby
distances, which rises to rather more modest values than the
$\sim$220km s$^{-1}$~pattern speed of the disk (Figure
\ref{f_rotcurve}).

\subsection{Minor axis rotation?}

When binned to increase signal, \vmb exhibits an apparent change in
sign about either the $\alpha=0\degr$~distance or the distance of
maximum density (Figure \ref{fig_obs}, upper right panel), with peak
to peak amplitude $\sim 0.2$~mas yr$^{-1}$roughly 10 times lower than
the switch in \vml. Such a trend is in line with that expected due to
Solar reflex motion (Vieira et al. 2007); discrimination of any {\it
  intrinsic} minor-axis rotation from the signature of the Sun's own
motion with respect to the LSR, is deferred to future work.


\subsection{Velocity ellipse} \label{sec_vell}

The $\{l,b\}$~velocity ellipse is of great interest to bulge
kinematics as (i) it may provide a further discriminant between bulge
stellar populations, and (ii) within a given population it carries
information about the structure of the potential in the {\it z}
direction (e.g. Kuijken 2004). Previous studies have been inconclusive
on this matter; Zhao et al. (1994) found no relative
orientation between the two metallicity populations, though rather few
objects were used with which to detect such an offset significantly;
the followup study of Soto et al. (2007) used 66 and 227 objects at
low, high metallicity respectively, and did not detect significant
discrepancy between the two populations.
However, the recovery of intrinsic velocity components from those
observed is a rather strong function of distance, so a
depth-integrated study such as theirs can smear out intrinsic
variation.

We constructed the velocity ellipse as a function of apparent distance
using the tracer population beneath the MSTO (Figures \ref{fig_ellip}
\& \ref{fig_ellip2}). There is a clear signature of changing
$\sigma^2_{lb}$~with line of sight distance (also Figure
\ref{fig_obs}). Interpreting this in terms of stellar velocity
components $\sigma^2_R,\sigma^2_{\phi},\sigma^2_z$~is not
trivial. Though there is a clear resemblance to the prograde family of
orbits in the Zhao et al. (1994) models, those models give the
projected velocities for the Baade's Window field, which lies closer
to the assumed bulge rotation axis. We will apply the Zhao et
al. (1994) models to our field to predict the
$\sigma^2_l,\sigma^2_b$~and $\sigma^2_{lb}$~in future work, however we
may gain some insight by considering the relationship between the
observed and intrinsic velocity dispersions. For our line of sight,
the observed variances will in general relate to the intrinsic
dispersions as
\begin{eqnarray}
 \sigma^2_{l} & = & \sigma^2_{\phi}\sin^2(\alpha) + \sigma^2_R \cos^2(\alpha) 
   + 2 \sigma^2_{R\phi} \sin(\alpha)\cos(\alpha) \nonumber \\
 \sigma^2_{b} & = & 0.9978 \sigma^2_z - 0.1282\left\{ \sigma^2_{\phi z}\cos(\alpha) + \sigma^2_{Rz}\sin(\alpha) \right\} \nonumber \\
    & &  + 2.13\times10^{-3}\left\{ \sigma^2_{\phi} \cos^2(\alpha) + \sigma^2_R \sin^2(\alpha) + 2\sigma^2_{R\phi} \cos(\alpha)\sin(\alpha) \right\} \nonumber \\
 \sigma^2_{lb} & = & 0.9989 \left\{ \sigma^2_{\phi z} \sin(\alpha) + \sigma^2_{Rz} \cos(\alpha) \right\} \nonumber \\
 & & - 0.0462\left\{ \sin(\alpha)\cos(\alpha) \left( \sigma^2_{R} + \sigma^2_{\phi} \right) + \sigma^2_{R\phi} \right\}
\end{eqnarray}
\noindent which relates six unknowns to three observables (for assumed
\rs). Proper motion observations of two fields together would - if the
fields can be assumed to feel similar potential $\Phi(R,\phi,z)$~-
enable this system to be solved; alternatively a similar
depth-sensitive set of observations including the line of sight
dispersions $\sigma^2_r,\sigma^2_{rl},\sigma^2_{rb}$~would allow this
system to be solved to measure the intrinsic dispersions in this
field. Because the line of sight passes close to the galactic center,
the projection angle $\alpha$~is a strong function of distance. For
$|d-R_s| \ga 0.4$~kpc, $\cos(\alpha)\simeq 0$~while $|d-R_s| \la$
0.1~kpc corresponds to $\sin(\alpha)\simeq 0$. At both extremes, we
expect $\sigma^2_b\simeq \sigma^2_{z}$, while at $d \simeq R_s$
\begin{eqnarray}
  \sigma^2_l & \simeq & \sigma^2_R \nonumber \\
  \sigma^2_{lb} & \simeq & \sigma^2_{Rz} \nonumber \\
\label{eq_disp}
\end{eqnarray}
\noindent while for $|d-R_s| \ga 0.4$~kpc, the subscript $R$~is
replaced by $_c$~in (\ref{eq_disp}) above. Between the two regimes
the velocity variation is complex; in the triaxial bulge potential we
may well have $\sigma^2_{R\phi}\sim \sigma^2_{R z} \sim \sigma^2_{\phi
  z}$~(see e.g. Sellwood 1981).

In a complementary manner, Koz{\l}owski et al. (2006) have also
detected a nonzero tilt in the velocity ellipsoid from their study of
300-500 stars in each of 35 ACS/HRC fields. They find the tilt to be
$\sim24\degr$~in the same sense as the ISM observations (Koz{\l}owski
et al. 2006; no error on this estimate is given); we find
$\theta=34\degr \pm 8\degr$, which is likely consistent with their
measurement errors. Curiously, their proper motion covariance is
roughly constant within measurement error across a range of different
sight lines, which might argue against projection effects. However
because the ACS/HRC was used in their survey and the exposures were
comparatively short, tracing the tilt of the velocity ellipse as a
function of distance along the line of sight is difficult from their
data because far fewer objects are traced per apparent distance bin
than with ACS/WFC. We remind the reader that a non-tilted stellar
population can easily exhibit a tilted velocity ellipsoid if stars are
being selected in a small region of space, even if the potential is
spherical (e.g. Binney \& Tremaine 1994). The best way to use the tilt
of the velocity ellipsoid is probably to select objects closest to the
meridional plane from multiple sight-lines; this requires datasets of
sufficient depth to be confident that enough objects in the meridional
plane can indeed be selected. The Baade's window datasets of Kuijken
\& Rich (2002), newly complemented with the ACS/WFC observations of
the ChaMPlane project (e.g. van den Berg et al. 2006), constitute a
good example of the type of dataset useful for such an approach (see
also the discussion in Kuijken 2004).

Existing models compare the shape of the depth-integrated velocity
ellipse at different lines of sight (e.g. Koz{\l}owski et al. 2006;
Rattenbury et al. 2007a,b,c; see also Vieira et al. 2007). It would be
of great interest to trace the tilt variation with depth of the
velocity ellipsoid as a function of metallicity; if there is indeed a
separate stellar population exhibiting a tilt with respect to the
majority of the bulge objects, perhaps accompanying the tilted gas
(Burton \& Liszt 1992), such a population may be of a newer generation
and thus differ in metallicity from the rest of the bulge stellar
population. The study of Soto et al. (2007) showed little significant
difference between the tilts of the $\{l,b\}$~ellipsoids near the
rotation axis of the bulge, as a function of metallicity. However, as
we have shown, the $\{l,b\}$~velocity ellipsoid shows significant tilt
variation with distance, so the depth-integrated velocity ellipsoid of
a small population of objects should be treated with some
caution. Unfortunately, metallicity information for a large number of
proper motion stars in our field is not yet available. WFC3 will be
the ideal instrument to constrain the metallicity of our tracer stars
through multi-filter imaging observations.
 
\section{Disk and Bulge populations}\label{s_popn}

Having constrained the proper motion ellipse for the disk and bulge
populations as a function of distance, we are in a position to
estimate the relative contributions of disk and bulge to our field,
and thus the degree of contamination of the disk to the clean-bulge
sample we will extract using the proper motions.

\subsection{Bulge fraction}\label{ss_ndisk}

The uncertain binary fraction in the bulge makes extrication of the
bulge population from the disk difficult from the CMD (see also
Kuijken \& Rich 2002), so we turn to the vector point diagram to
estimate the relative contributions of the two populations. Direct
evaluation of the stellar density as a function of distance breaks
down at apparent distances nearer to the Sun than the innermost spiral
arm because the distance-estimate breaks down from differential
reddening. Furthermore, direct fitting to the vector point diagram is
challenging due to the high overlap between the populations; in
experiments, simulated populations were very difficult to reproduce by
such a procedure.

We take a simpler approach: we first assume the superposition of
populations along the line of sight can be approximated with a pair of
2D gaussians, and rely on the symmetry of these distributions to
correct the bulge population for the contamination induced by the
disk. The number of points within the 0.7$\sigma$~ellipse of each
population is evaluated to avoid counting objects twice due to overlap
of regions in proper motion space. Because of the high overlap, the
estimated population within each region is corrected by the expected
contribution from the other population (Figure \ref{f_fbulge}); the
result is a ``raw'' disk-fraction of 11.5$\%$. The discrepancy between
this value and the intrinsic value is assessed through simulation;
using the distance-dependent proper motion ellipses constrained
previously, we simulate populations with input disk fraction and
assess the difference between the value returned by this process, and
the value input (Figure \ref{f_fbulge}); this leads to a true value
$14\% \pm 1\%$~for the disk fraction among our population with proper
motion measurements.

\subsection{A clean-bulge sample}\label{ss_purebulge}


To select a clean-bulge sample, we modify slightly the kinematic
selection criteria of Kuijken \& Rich (2002); we use a cut on
longitudinal proper motion $\mu_l$~and on proper motion
measurement-errors $\epsilon_l, \epsilon_b$, but discard cuts on
$\mu_b$~because bulge and disk show similar latitudinal motion
(Figures \ref{f_short} and \ref{fig_hess}). The proper motion error is
an estimate of the measurement scatter; this estimate does {\it not}
take into account any bias in the measurement due to nearby brighter
objects. For this reason, we further impose a cut on the crowding
measure $q$~(Section \ref{s_sigma} and Figure \ref{f_qual}); we choose
to keep objects with $q<0.15$.

To set the proper motion cut-off, we return to the population above
the MSTO to find the value of $\mu_l$~at which the disk contribution
is low but the cut not so severe that too few objects are retained to
be useful. We use the same selection regions as Figure \ref{f_short}
to produce nominal ``disk'' and ``bulge'' populations. The ``disk''
population is markedly asymmetric in $\mu_l$. Indeed, this population
is best fit with a two-component model, of which the one with high
positive centroid proper motion ($\mu_{l,0}=+4.17$~mas yr$^{-1}$), is
identified with the disk (hereafter the ``true-disk'' population). The
component with centroid proper motion consistent with zero (Figure
\ref{f_strag}, top), is indistinguishable from the single component
that is needed to fit the $\mu_l$~distribution of the bulge population
above the MSTO (Figure \ref{f_strag}, bottom), indicating some
bulge/disk overlap in the ``disk'' region of the CMD. The proper motion cut
$\mu_l=-2.0$~mas yr$^{-1}$~of Kuijken \& Rich (2002) lies
approximately 2.9$\sigma$~from the center of the true-disk
distribution, so only about 0.19\% of disk objects have $\mu_l <
-2.0$~mas yr$^{-1}$. We thus adopt $\mu_l < -2.0$~mas yr$^{-1}$~as our
proper motion cutoff. To ensure that proper motions in the clean-bulge
sample are measured to at least $6\sigma$~significance, we impose the
proper motion error cutoff $\epsilon_{l},\epsilon_b < 0.3$~mas
yr$^{-1}$. This leaves 15,323 objects kinematically associated with
the bulge (Figure \ref{f_bulge}). Because we have set our
$\mu_l$~cutoff with reference to disk objects above the MSTO, our
cutoff is not affected by the small contribution of evolved disk stars
to the bulge RGB in the CMD (Vieira et al. 2007).


We now examine the contamination due to non-bulge stars that would
pass these kinematic cuts. The disk makes up approximately 14\% of the
stars with measured proper motions, while 81,140 of the 187,346
objects with proper motion measurements show errors
$\epsilon_{l},\epsilon_b < 0.3$~mas yr$^{-1}$. The likely disk
contamination to our proper motion-selected sample is therefore
81,140$\times$0.14$\times$0.0019=22~objects. Thus perhaps 0.14\% of
the ``clean-bulge'' sample may in reality be disk objects. The
galactic halo also provides a small contaminant; integrating the
Bahcall \& Soneira (1984) models along the line of sight predicts
$\sim 23$~halo objects in the ACS/WFC field of view. The halo shows
proper motion dispersion $\sigma_{l} \simeq 22/d$~mas
yr$^{-1}$~(Binney \& Merrifield 1998, with distance $d$~in kpc), about
a proper motion zeropoint similar to the bulge. We thus expect perhaps
40\% of halo objects to pass our proper motion cuts, or $\sim
9$~objects. Thus, the handful of objects high above the main sequence
(visible in Figure \ref{f_bulge}) are probably halo stars, and can
later be easily excised from the sample by position in the CMD. So the
total contamination due to non-bulge stars in our clean-bulge sample,
is of order 22+9 = 31 of 15,323 objects.

Isochrone-fitting to the bulge population is complicated by several
factors. The extinction curve along the line of sight is not
well-constrained, may vary spatially over the ACS/WFC field of view,
and may be anomalous (Sumi et al. 2004), making dereddening
challenging. The binary fraction of the bulge is unknown, causing an
unknown systematic between single-star isochrones and the observed
population. In addition, very few alpha-enhanced isochrones exist for
metallicities greater than +0.5. We therefore defer detailed fitting
of the bulge isochrone for future work and content ourselves with a
visual comparison to Vandenberg isochrones that have been transformed
using the adopted bulge extinction and distance in the manner
described by
Brown et al. (2005).

To estimate the best-fitting bulge sequence, a median bulge population
was estimated by binning the clean-bulge population along the
isochrone (Figure \ref{f_bulge}). Because the uncertain binary
fraction is not significant above the MSTO, we use the bulge-only CMD
in this region to assess the spread in age and metallicity that
overlap the observed range of colors. The 10Gyr, alpha-enhanced
solar-metallicity isochrone of Sahu et al (2006) is replaced by a
slightly older population at 11 Gyr; this better reproduces the median
population. Isochrones corresponding approximately to $\pm 1\sigma$~in
both metallicity and age encompass most of the color variation along
the post-MSTO bulge population. A very young, very metal-poor
isochrone ([Fe/H]=-1.009, 5 Gyr) overlaps the outliers along the bulge
RGB; younger, more metal-poor isochrones are not consistent with the
clean-bulge population (Figure \ref{f_bulge}). The transformed
isochrones do not reproduce an apparently metal-rich, old population
visible at the red edge of the post-MSTO bulge sequence; this might be
due to regions with higher extinction than our mean adopted value of
E(B-V)=0.64.

Cohen et al. (2008) recently compiled metallicities of a selection of
main-sequence objects in the bulge that had been sufficiently
amplified by microlensing to permit abundance analysis with comparable
precision to the more traditionally-used Giants (see also Minniti et
al. 1998). The three objects in their sample show $[Fe/H]\sim+0.3$,
leading to the claim that a random sample of bulge giants should be
less metal-rich than a similar sample from the main-sequence, perhaps
due to highly metal-rich objects losing so much mass that they never
undergo the Helium flash and evolve along a different track to the
rest of the tracer giants (Cohen et al. 2008). We find no support for
this hypothesis; allowing for the uncertain binary frequency which
affects the main-sequence far more than the bulge RGB, the
metallicities of the giants and main-sequence samples are consistent
with each other. More quantitative exploitation of the clean-bulge
sample to constrain the star formation history of the bulge is beyond
the scope of this article, and will be reported elsewhere.

\subsection{Blue straggler candidates} \label{ss_strag}

Within a few magnitudes above saturation, a population is visible with
a disk-like location in the CMD (blueward of and brighter than the
bulge MSTO) but with bulge-like proper motions (Figure \ref{fig_hess};
particularly obvious in $\mu_l$, but also visible in $\sigma_b$). When
we apply the kinematic cuts to extract a likely clean-bulge sample, 72
objects remain in this region of the cleaned CMD (Figure
\ref{f_strag}). As the halo is a somewhat evolved population, its
objects are not expected to lie in this region of the CMD, and with an
estimated 22 disk contaminants within the entire clean-bulge sample,
at least 69\% of this population must consist of bulge objects.

This population may consist of very young, very metal-poor objects,
although how such objects would form is not clear, and indeed
isochrones younger than about 5Gyr, which would be required to
describe the brightest, bluest objects in this population, are not
consistent with the CMD below the MSTO (Figure \ref{f_bulge}). The
alternative is that a significant fraction of these objects may be
bulge Blue Stragglers; their location in the CMD overlaps with the
region such objects are expected to occupy (e.g. Sarajedini
1992). Further details of these objects will be reported in a separate
communication (Clarkson et al. 2008 ApJ Lett in prep).



\section{Kinematics of the SWEEPS Planet Candidates} \label{s_cand}

We are now in a position to examine the sixteen SWEEPS transit planet
candidates for membership of disk or bulge populations; this
distinction will inform the history of their formation and
evolution. We construct a mean bulge proper motion best-fit ellipse by
taking a population-weighted average of the best-fit proper motion
(not velocity) ellipses using the kinematic tracer objects of the
previous section. We produce a mean (foreground) disk proper motion
ellipse using stellar tracers in the nearest distance-bin.

When we overplot the best-fitting mean-bulge and mean-disk proper
motion contours, we find an apparent grouping of four objects within
the 1$\sigma$~contour of the disk, and all but two of the rest within
the $1\sigma$~bulge contour (Figure \ref{f_scand}). Furthermore, the
object SWEEPS-04, which lies well within the 1$\sigma$~ellipse of the
disk population, resides on the upper disk sequence (Sahu et
al. 2006), where disk stars are expected to dominate\footnote{assuming
  its host is not a Blue Straggler, which would seem unlikely.}
However all objects are also within the 2$\sigma$~ellipse of the bulge
population.

\subsection{SWEEPS candidates as disk/bulge objects}

We use the angular distribution of candidates in proper motion space
to assess kinematic membership of the SWEEPS candidates. In
$\{\mu_l,\mu_b\}$~space, let $\Theta_i$~be the counter-clockwise angle
between the major axis of the best-fit bulge ellipse and the line
joining the center of the best-fit bulge ellipse to the $i$-th
candidate. The cumulative distribution function (CDF) of $\Theta_i$~is
then used as an indicator of the angular distribution of the SWEEPS
candidates in proper motion space. Should a large number of candidates
reside in the disk, one would expect a sharp steepening in the CDF
near $\Theta_d$, the angle between the major axis of the bulge ellipse
and the center of the disk distribution (Figure
\ref{f_candkin}). Alternatively, if all sixteen candidates were bulge
objects, then the CDF would be a straight line; no angle $\Theta_i$
would be preferred. We compare the observed cumulative distribution
function (CDF) of the SWEEPS candidates to a large number of trial
artificial datasets, in which sixteen objects are generated under the
best-fit bulge and disk proper motion distributions. For each trial,
the two-sided Kolmogorov-Smirnov (KS) statistic is computed between
the trial and the observed distribution, yielding the associated
formal probability that the SWEEPS candidates and the trial dataset
are both realisations of the same probability distribution. This
process is repeated for $10^5$ trial datasets. This test is repeated
for differing sizes of disk contribution $N_d$~to the total population
(for $0 \le N_d \le 16$) and the formal probability that the SWEEPS
sample matches the distribution using each $N_d$~is recovered.

To maximize use of available information we have also applied the 2D
Kolmogorov-Smirnov test to the set of positions in ($\mu_l,\mu_b$)
space of all the candidates. We use the implementation in Numerical
Recipes (Press et al. 1992; see also Metchev \& Grindlay 2002). In two
dimensions the equivalent K-S statistic $D_2$~is a function of the
input distribution. We thus evaluate the significance of the maximum
$D_2$~at each disk fraction $N_d$~using Monte Carlo simulations. This
produces an equivalent significance curve as a function of $N_d$
(Figure \ref{f_candkin}).

\subsection{The bulge and disk planet fractions}


Although the most probable disk population $N_d$~differs slightly
between the two tests, both are consistent (at $1\sigma$) with a disk
population in the range ($1 \le N_d \le 8$). If the fraction of stars
hosting jovian planets with periods less than 4.2 days were identical
between disk and bulge, we would expect the planet candidates to
follow the same disk/bulge distribution as the stars in general. Our
kinematic analysis would then suggest 14\% of planet candidates - two
candidates - would reside in the disk. This is entirely consistent
with the actual distribution of candidate kinematics. However, the
sample of SWEEPS transit planet candidates is too small to draw
meaningful conclusions about the fraction of planets in the disk
versus that of the bulge. Because we cannot state that the fraction of
planet candidates in disk and bulge are inconsistent with each other,
we cannot make any claims about the consistency or otherwise of the
fraction of stars hosting planets between the disk and bulge. 

Any inference on the true planet-host fraction in the bulge then
reflects our ignorance of the true ratio of astrophysical false
positives to true planets; we restrict ourselves to upper and lower
bounds. Considering the likely population of stellar triples,
grazing-incidence stellar binaries and low-mass stellar companions,
Sahu et al. (2006) estimate the maximum rate of astrophysical
false-positives among the candidates, of 9/16, or 56\%. We remind the
reader that this is probably a conservative upper limit, as no such
false-positives were found in the similar 47 Tuc analysis (Gilliland
et al. 2000; see Sahu et al. 2006 for further discussion of this
issue). Thus the lower bound on the fraction of true planets among the
candidates, is 44\%; the upper bound 100\%, predicting 6-14 true
detected planets in the bulge. Taking the detection efficiency, period
distribution and the probability of transit due to random orbital
inclinations into account, the frequency of stars hosting Jovian
planets with periods shorter than 4.2 days was estimated to be 0.42\%
(Sahu et al. 2006). The 6-14 true detected planets then imply an extra
uncertainty of perhaps a factor of two, since the planet frequency
consistent with observations depends not only on the fraction of true
planets but also the actual period, radius, transit phasing, and host
brightness of each planet.

We ask if the sub-population of five planet-host candidates with
periods less than one day (the ``Ultra-Short Period Planets'', or
USPP; Sahu et al. 2006) themselves are preferentially located in the
disk or bulge. Here there is no obvious correlation between period and
membership - two USPP fall within the 1$\sigma$~ellipse of the
best-fit disk, three fall within the 1$\sigma$~ellipse of the best-fit
bulge, and all are within 2$\sigma$~of the best-fit bulge. Thus the
USPP do not show any preferred kinematic association compared to the
non-USPP candidates; the best that can be said is that the USPP as a
family are unlikely to {\it all} be disk objects.

\section{Summary and Conclusions} \label{s_conclude}

We have measured proper motions for $> 180,000$~objects within the
Sagittarius low-reddening window towards the bulge
($l,b=1.65\degr,-2.65\degr$), and used them to extract a clean
clean-bulge sample of 15,323 objects - a sample roughly a factor four
larger than that afforded by WFPC2 across a six-year interval (Kuijken
\& Rich 2002). This clean-bulge sample contains perhaps 31
contaminants from both disk and halo, making it the purest bulge
population ever isolated. Constructing a median stellar sequence from
this bulge sample, we find that an 11 Gyr isochrone best represents
the bulge population, with most of the variation along the bulge
subgiant branch falling within the range $[Fe/H]=0.0 \pm 0.4$~and age
$11 \pm 3$~Gyr. Use of this sample to inform bulge age studies, in
conjunction with extensive completeness tests, will be the subject of
future reports. Work along these lines is particularly exciting when
we consider the parallel NICMOS observations we have undertaken, for
which similar selection should be possible and which brings the
possibility of tracing the bulge initial mass function to the
neighborhood of the Hydrogen-burning limit.


The large number of stars with proper motion measurements allows
kinematic features to be resolved in the color-magnitude diagram.
We construct the $\{l,b\}$~velocity ellipse as a function of line of
sight distance, demonstrating that its properties are quite sensitive
to the distance to the objects under consideration. Finally, we use
its proper motion analogue to attempt to classify the SWEEPS planetary
candidates by kinematic membership with bulge or disk. The proper
motion distribution of the candidates is consistent with the division
of the stellar population between bulge and disk, but the candidate
population is too small to draw further conclusions. We find no
evidence that Ultra Short Period Planet (USPP) candidates are
preferentially associated with the disk; instead we can only claim
that the distribution is entirely inconsistent with {\it all} the USPP
originating in the disk.

Although the metallicities of the majority of our targets are unknown
(with the exception of the bright objects, largely above the MSTO, for
which VLT spectroscopy has been possible), the large number of objects
with accurate proper motions allows the distance-uncertainty due to
metallicity to be overcome by virtue of a sample containing $>$500
objects per bin. Armed with photometric distances and therefore mean
transverse velocities for our distance-bins, we produce an independent
determination of the stellar circular speed curve $\vmpe$. Although the
sampling of this rotation curve is relatively sparse, we clearly
detect a transition away from solid body-like rotation at
galactocentric radius $R_c =0.3-0.4$kpc, in line with evidence from
radial velocities. Within the uncertainties of the distance estimates
we have used, our limiting circular speed at large radius is
consistent with that from radial velocity measurements.

With position-estimate dispersions reaching the 2-3{\it milli-}pixel
level (Section \ref{s_pm}), we are near the limit of
position-measurement with current HST instrumentation. The intrinsic
velocity dispersion of the bulge and disk nevertheless makes kinematic
classification of object groups with few members difficult. As the
bulge is likely a superposition of multiple stellar populations
(e.g. {Zoccali} et al. 2003), the intrinsic dispersion problem might
be overcome if populations can be separated by metallicity on a
star-by-star basis (c.f. {Soto} et al. 2007), which will require
further, multifilter observations of this field.

\section{Acknowledgements}

Support to W.I.C. through proposals GO-9750 and GO-10466 was provided
by NASA through grants from the Space Telescope Institute, which is
operated by the Association of Universities for Research in Astronomy,
Inc. under NASA contract. MZ and DM are supported by FONDAP Center for
Astrophysics Nr. 15010003.

We thank Ron Gilliland for his generous time and effort for this
project, and for insightful advice into many aspects of this work, and
Luigi Bedin, Nathaniel Paust, Roeland van der Marel, Imants Platais
and Vera Kozhurina-Platais for thoughtful and stimulating
discussion. It is a pleasure to acknowledge discussion with Mario Soto
and Kathy Vieira. We thank the anonymous referee for a number of
suggestions that helped to clarify the manuscript.

\appendix
\section{Proper Motions from Deep Photometry}
\label{appen}

Two approaches have in the past been taken to obtain proper motions
from two different-depth epochs of observations of crowded stellar
fields (at least one of which is dithered), both with some
success. The key difference lies in the production of the master-list
of positions from the deeper epoch. The first approach is to stack the
deepest-epoch images and build the master-list from the image
stack. This takes advantage of the subpixel dithers to produce a
(usually) twice-oversampled superimage, from which deep positions can
be obtained; the PSF is well-sampled in the resulting superimage and
thus the object-centers are better constrained than the individual
input images. This approach was used quite successfully by Richer et
al. (2004a, 2004b) when separating members of M4 from the rest of the
Field. The second approach is to measure positions on each image
individually amd combine the measurements with sigma-clipping to
produce the master-list. Positions are measured in the raw frame of
each image (the \flt frame), using a highly-supersampled model for the
''effective PSF'' (the instrumental PSF as recorded by the detector,
or ePSF), with a perturbation scaled to the data to account for
focus-breathing. This usually allows the measurement of positions from
the individual images to higher accuracy than possible from an ePSF
constructed from the data itself (Anderson \& King 2006), because the
observations used to produce the library ePSF are rather better
sampled than most program observations. Our dataset is exceptionally
well-sampled, so whether this advantage should hold here was less
clear when embarking upon the reductions for this project. We thus
tried both approaches for this dataset to allow a side-by-side
comparison of the two methods and picked the winner for further
evaluation of proper motions.

\subsection{Method 1. master-list from stacked images}

Positions in the second-epoch are mapped onto positions in the
master-list built from the epoch-1 image stack. Mutual misalignment
between images in the two epochs means the epoch-2 images cannot be
directly evaluated onto the super-image for comparison. In this
approach, images in the less well-populated epochs are {\it not}
stacked together before position determination as there are too few
images in each epoch to optimise the stacking; instead, each
individual input image provides a separate estimate of the proper
motion for each object. Because frame-to-master offsets are likely to
be of a high order, image-regions of size 1600 $\times$1600
twice-oversampled pixels (so $40''\times 40''$) are used to simplify
the transformation required (including a 200-pix buffer at the frame
edges where transformations will be least well-fit); each region
contains $\sim$7000 unsaturated stars in the master-list, of which
$\sim$60\% are at $F814W < 25.5$.

The optimal transformation from the master-image to each individual
second-epoch image is determined from an AMOEBA fit\footnote{When
  fitting coordinate transforms we try both AMOEBA (Press et al. 1992)
  and Levenberg-Marquardt (using the Markwardt MPFITFUN IDL
  implementation, available at
  http:\/\/cow.physics.wisc.edu\/$\sim$craigm\/idl\/fitting.html)
  approaches and select the approach that fits the transformation most
  robustly.} (Press et al. 1992) to the positions of marker stars;
typically 250 bulge marker-stars are used as candidates to the
transformation, which is cut down to $\sim$150-200 after
sigma-clipping of contaminating outliers. The standalone f77
implementation of STSDAS BLOT is then used to transform the
master-image to the individual second-epoch image, so that the less
well-determined input image is never re-sampled. Positions are then
found in both the input and the transformed master-image, and compared
directly to produce proper motions. The result is a separate estimate
of the proper motion for each input image, for each postage-stamp
subregion. Cosmic-rays and other artefacts are weeded out at the stage
of proper motion estimate-collation by sigma-clipping (c.f. Sahu,
Anderson, \& King 2002).

\subsection{Method 2. master-list from collated measurements}

The alternative approach, producing the master-list by combining
measurements from each individual image within an epoch, is preferable
if the ePSF in \flt space is well-enough constrained that its errors
are smaller than those introduced due to the small spatial shifting of
flux in the production of the epoch-1 superimage. The observations
used for the ePSF of Anderson \& King (2006; see also Anderson \& King
2003) provided a 4$\times$ supersampled, spatially-dependent
super-sampled PSF model in \flt space, as well as a distortion
solution that transforms raw positions into positions on the sky. The
distortion is accurate to $\sim$1\% of a pixel (Anderson \& King
2006). Thus, for the epoch of the calibration observations, positions
and fluxes can be measured to high accuracy {\it and} transformed to a
distortion-free frame using this solution.

\subsection{Method comparison}

We compare the two methods of proper motion determination by the
proper-motion distributions each produces over the same region of the
image; in this case the $80''\times 80'' $~surrounding the exoplanet
candidate-host SWEEPS-13. The resulting proper motion distribution
along pixel-x coordinate shows width $\sim 0.2$~mas yr$^{-1}$~(0.008
ACS/WFC pix over two years) greater when determined using the epoch-1
stack than from the image-to-image estimates, with the latter showing
closer agreement to the previously-published WFPC2-based proper
motions determined by Kuijken \& Rich (2002). The stack-based approach
thus provides proper motion estimates roughly seven percent less
precise than the image-to-image approach (Figure \ref{fig:compar}).

The cause of this discrepancy is most likely the combination of
deep-epoch images into the image-model; the convolution of input
images onto the master-scene has caused flux to be slightly
re-arranged; while the absolute number of electrons associated with a
given object is preserved (so the image-model is still ideal for
photometry), the distribution within the region occupied by the star
has been altered (Figure \ref{f_system}). The magnitude of this
systematic is reduced somewhat by the blot process; while it can be
mitigated in principle by re-fitting the ePSF to the postage stamp for
each region in each image, there is no guarantee that any spatial
variation of this shifting can be swept into the ePSF-fit. For this
reason, all proper motions we report here are based on the
image-to-image approach.

\section{Objects Without a Unique Mean Position Estimate}
\label{AppenB}


For roughly 6\% of objects, the techniques of Section \ref{s_sigma}
fail to converge on a single mean position estimate, but produce a
bimodal estimate. The majority of these objects (86\%, or 5.2\% of the
total population in the field of view) show low separation
($<0.15$~pix) between the two peaks in the estimate, and upon visual
inspection usually consist of a cloud of points (the main peak) plus a
trail (within which the secondary peak resides; Figure \ref{f_bimod},
Top). For a significant fraction of this bimodal population, the line
joining the two peaks shows preferential alignment to the direction
towards the nearest neighbor that is at least as bright as the object
itself (Figure \ref{f_bimod2}, Right). Thus, we conclude that the
trailing in these position estimates is due to the influence of a
bright neighbor with slightly incorrect subtraction.

Objects with separation between peaks greater than 0.15~pix make up
the remaining 14\% of the bimodal objects (or 0.8\% of the total
population), and generally show more well-separated clusters of
position estimates. In contrast to the low-separation bimodal objects,
the reported separation between peaks is close to the apparent
separation between islands upon visual inspection (Figure
\ref{f_bimod}, Left Bottom). Also unlike the low-separation bimodal
objects, the position-angle between the peaks shows no relation to the
direction to the nearest bright neighbor (Figure \ref{f_bimod2},
Left). However, for this population, the nearest bright neighbor is
on average closer than for the population as a whole. 

\begin{figure}
\plotone{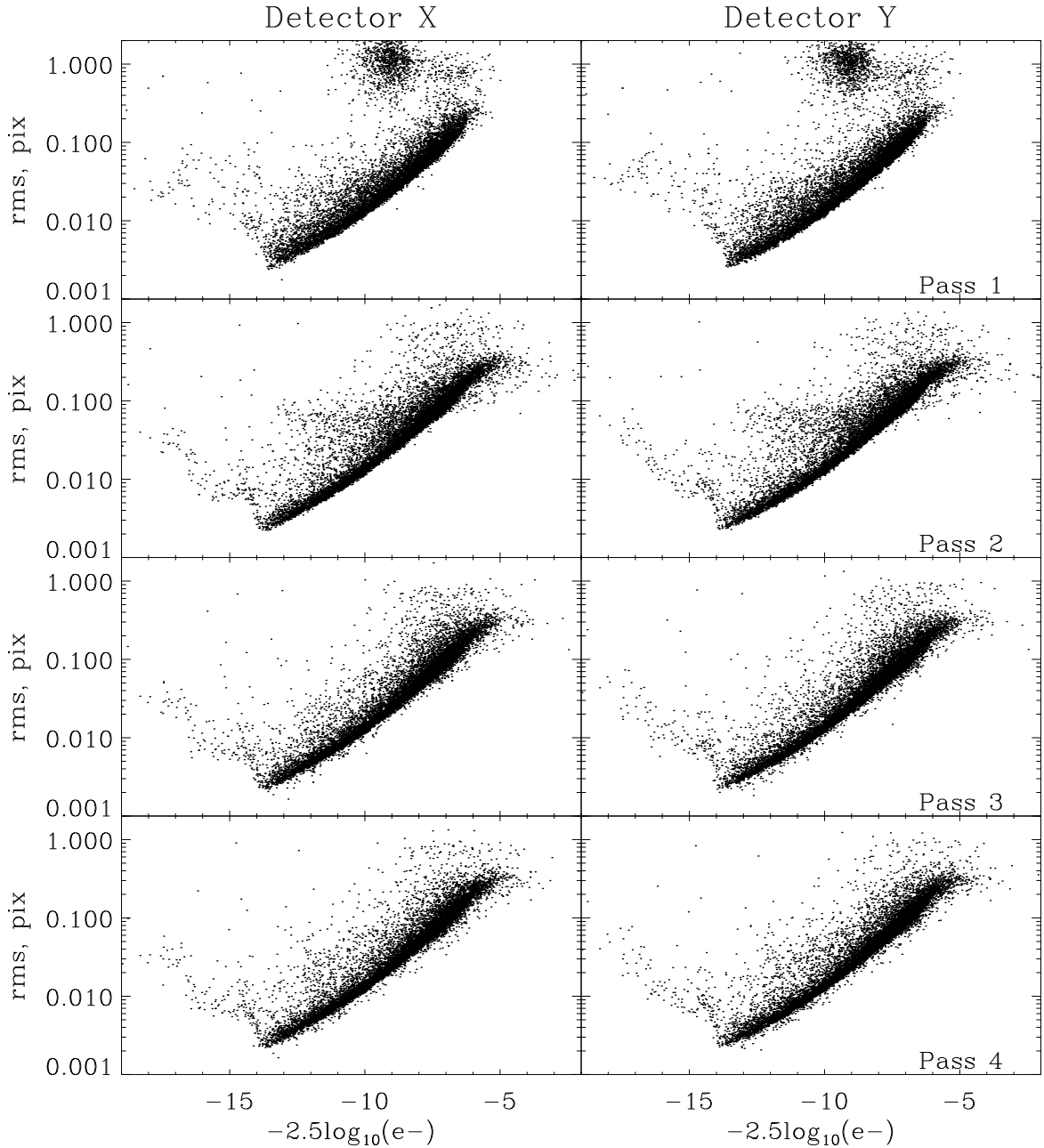}
\caption{Scatter in position-measurement as a function of object flux, stepping through the photometry passes. A single pass (top) shows a significant cloud of fainter objects with high coordinate dispersion. After a second pass using neighbor-subtraction (second row), a significant population is visible with $\sim 0.1$-pix rms. Each object in this population shows at least two clusters of position-estimates. When this is accounted for, these objects show rather less scatter in their position estimates (third row). 5\% of stars are plotted for clarity, and the same stars are plotted in all panels.}
\label{fig_rms}
\end{figure}

\begin{figure}
 \plotone{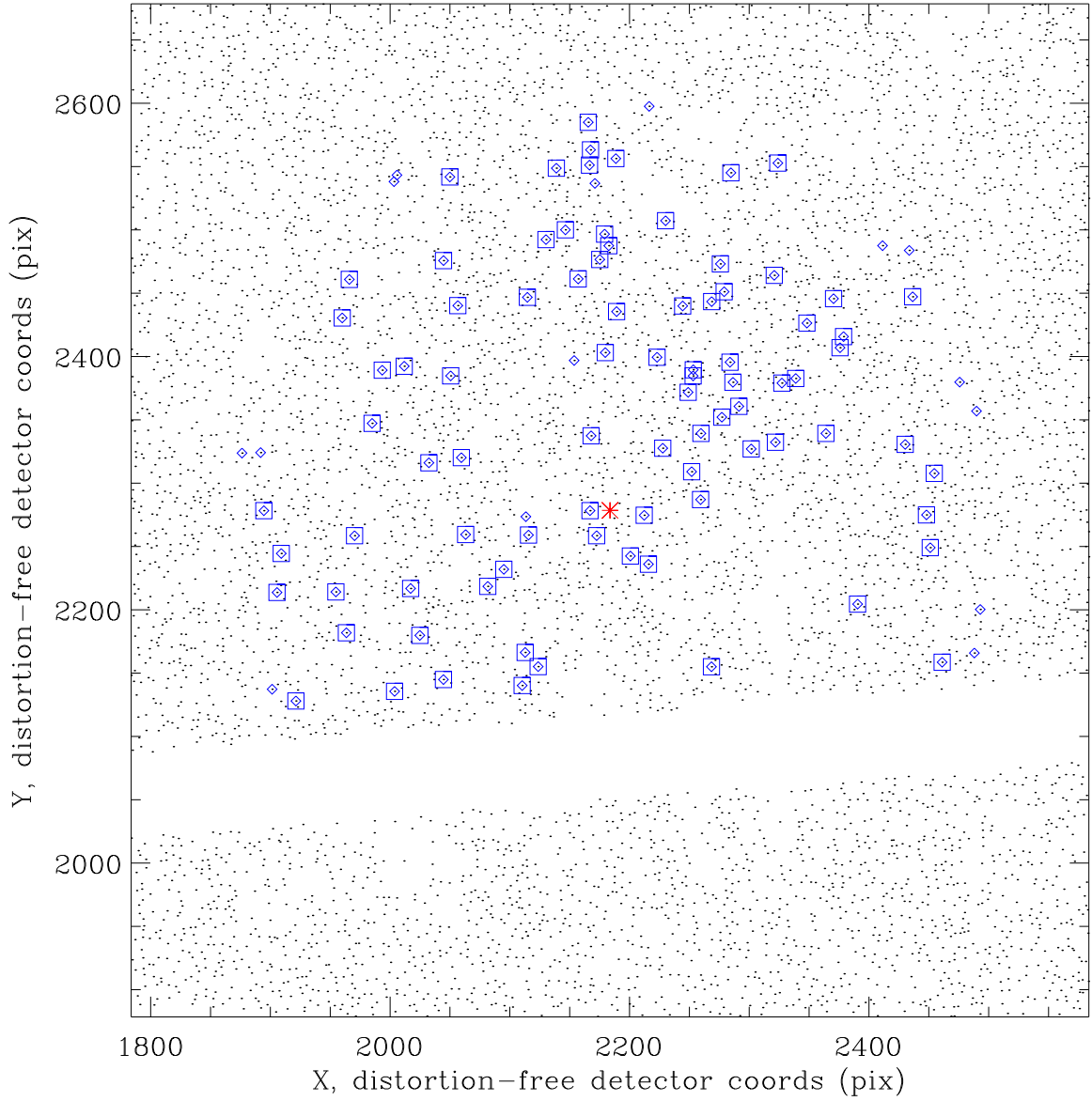}
 \caption{Example local transformation. The target star is denoted by an asterisk, with the nearest tracer objects denoted by diamonds. Boxes around the diamonds show the tracer stars surviving sigma-clipping to make the final transformation.}
\label{fig_tracers}
\end{figure}

\begin{figure}
  \plotone{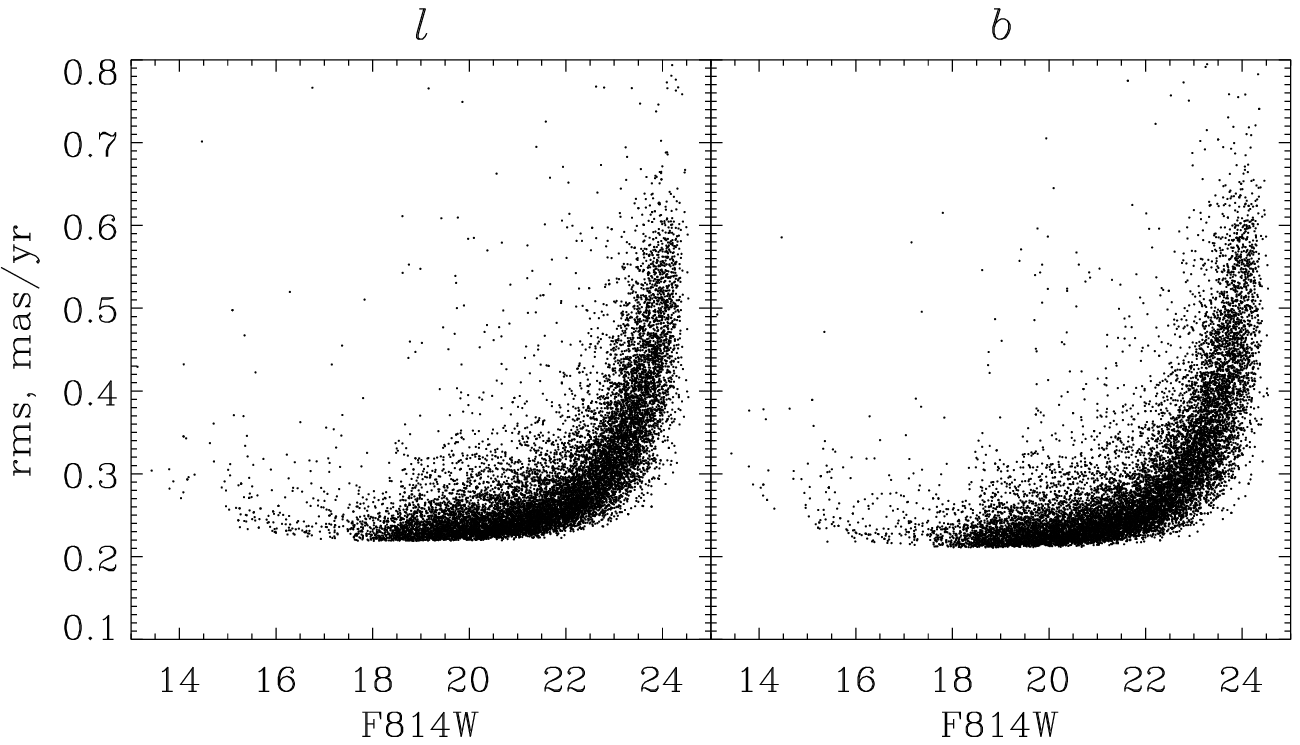}
  \plotone{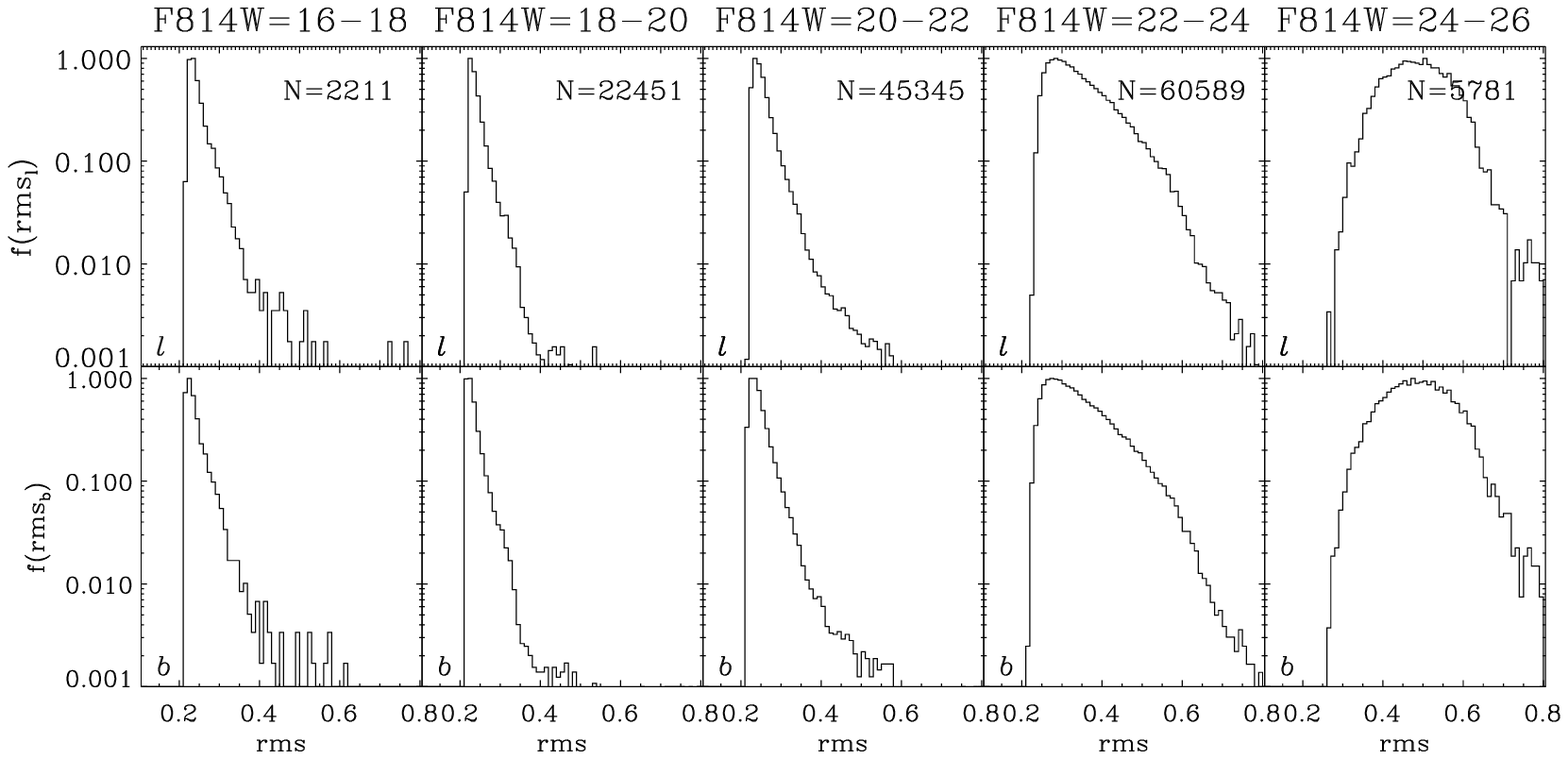}
  \caption{Proper Motion errors under the local-transformation approach (Section \ref{sub_pm}); 10\% of objects are plotted for clarity. {\it Top:} scatterplots of the error in $\mu_l$, $\mu_b$. {\it Bottom:} histograms of proper motion error selected by magnitude, each scaled to the histogram peak. The scatter in proper motion is given in mas yr$^{-1}$~in all panels; this establishes that internal proper motion scatter is $<0.5$mas yr$^{-1}$~for nearly all objects and $<$0.3 mas yr$^{-1}$~for a significant fraction (81,140/187,346).}
  \label{fig_pmerror}
\end{figure}

\begin{figure}
\plotone{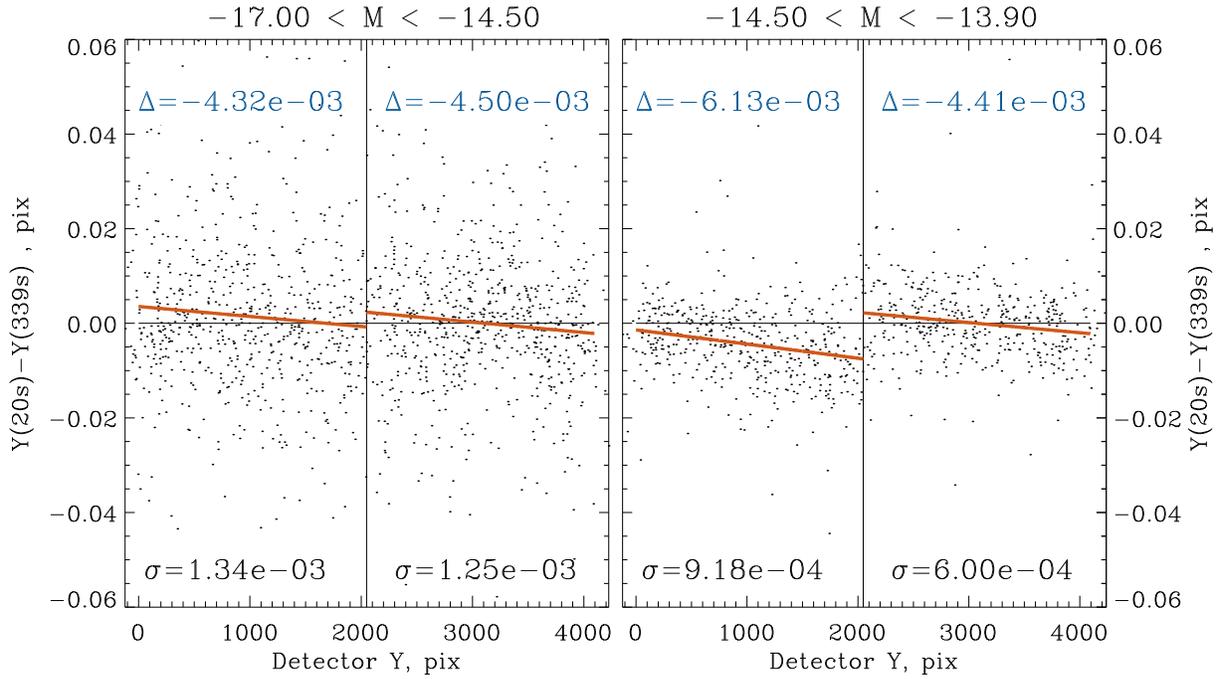}
\caption{Position discrepancy $\Delta Y$~between long and short integrations in 2004 of moderately saturated (right) and highly saturated (left) objects. Titles give the instrumental magnitude range in the 339s exposures, the dashed line the expected magnitude difference from the differences in exposure-time alone. Inset numbers give variation amplitude across each chip and the scatter $\sigma$~in the fit due to measurement errors.}
\label{ref_fig_satn}
\end{figure}

\begin{figure}
\plotone{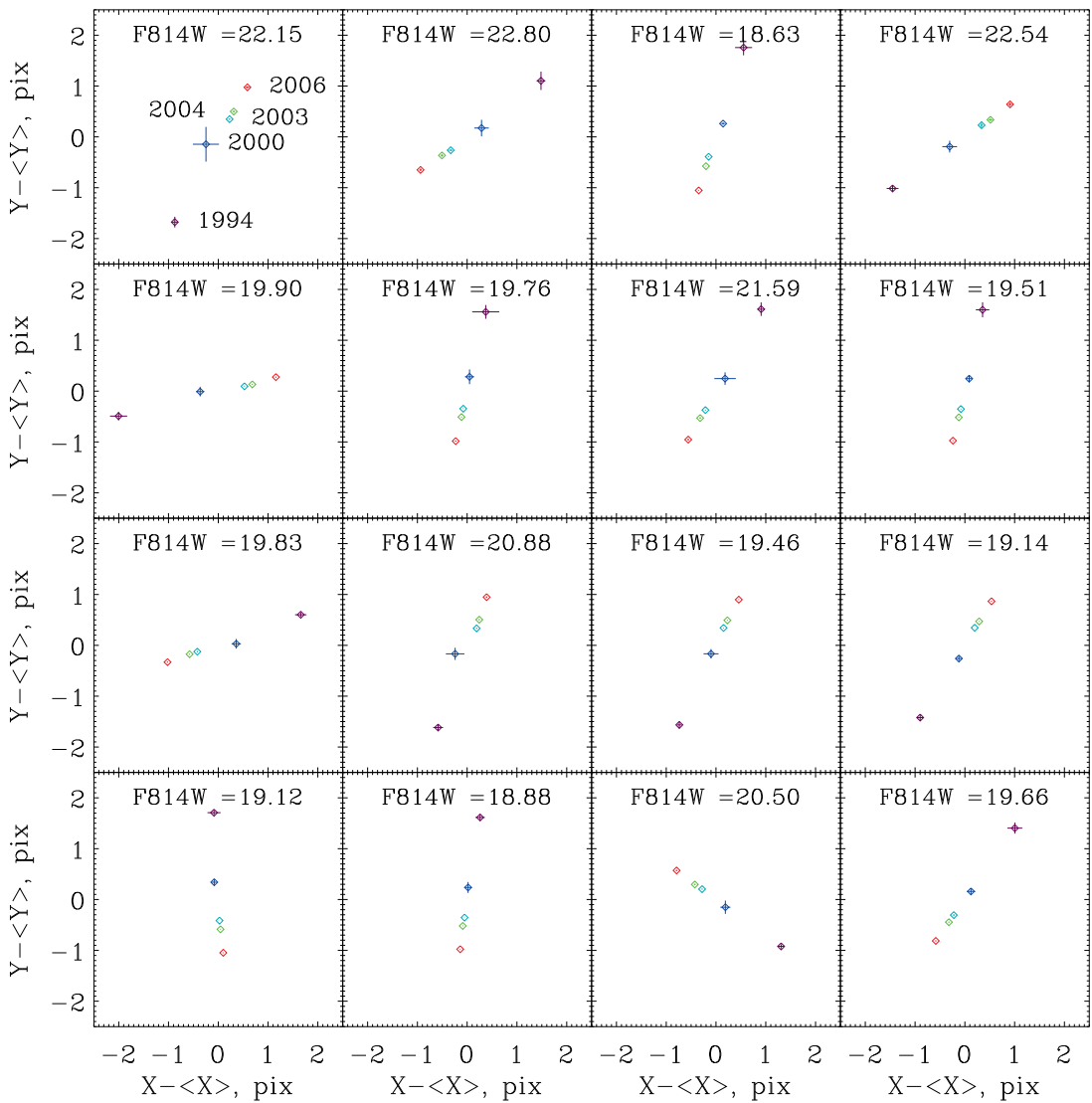}
\caption{Multi-epoch positions on-sky of a random selection of (for clarity) high-proper motion stars. Detector positions are given in ACS/WFC pixel coordinates (1pix=50mas), with the $F814W$~magnitude listed in the panel for each star. Epochs may be identified from the position-separation in most cases; the top-left panel explicitly shows the epochs for one example. Epochs 1994 and 2000 represent WFPC2 measurements, all the others are ACS/WFC. See Section \ref{ss_multi}.}
\label{f_multi}
\end{figure}

\begin{figure}
\plotone{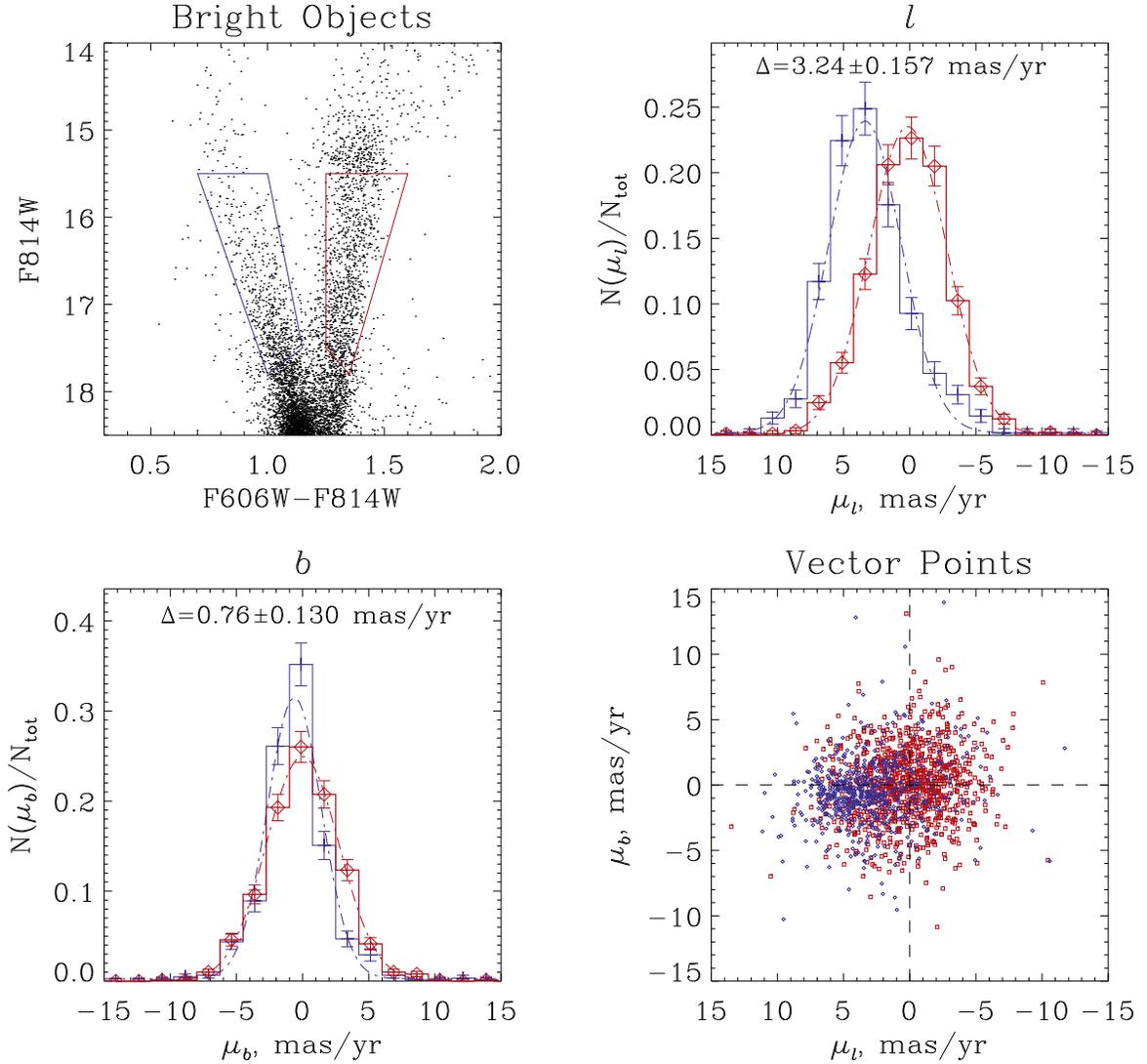}
\caption{Proper motions above the main sequence turn-off, estimated from the 339s and 349s exposures. {\it Top-Left:} CMD of the region above the turn-off, with the disk-dominated (left) and bulge-dominated (right) regions indicated. {Top-Right}: Longitudinal proper motion histogram (scaled by the number of objects in each population), with Poisson errorbars overplotted and peak separation $\Delta$~indicated. There is clear separation between the bulge-dominated (red) and disk-dominated (blue) populations. Continuous dashed lines show gaussian fits to each population. {\it Bottom Left:} latitudinal proper motion distribution. {\it  Bottom Right:} vector point diagram for objects above the main sequence turn-off (bulge red, disk blue).}
\label{f_short}
\end{figure}

\begin{figure}
\plotone{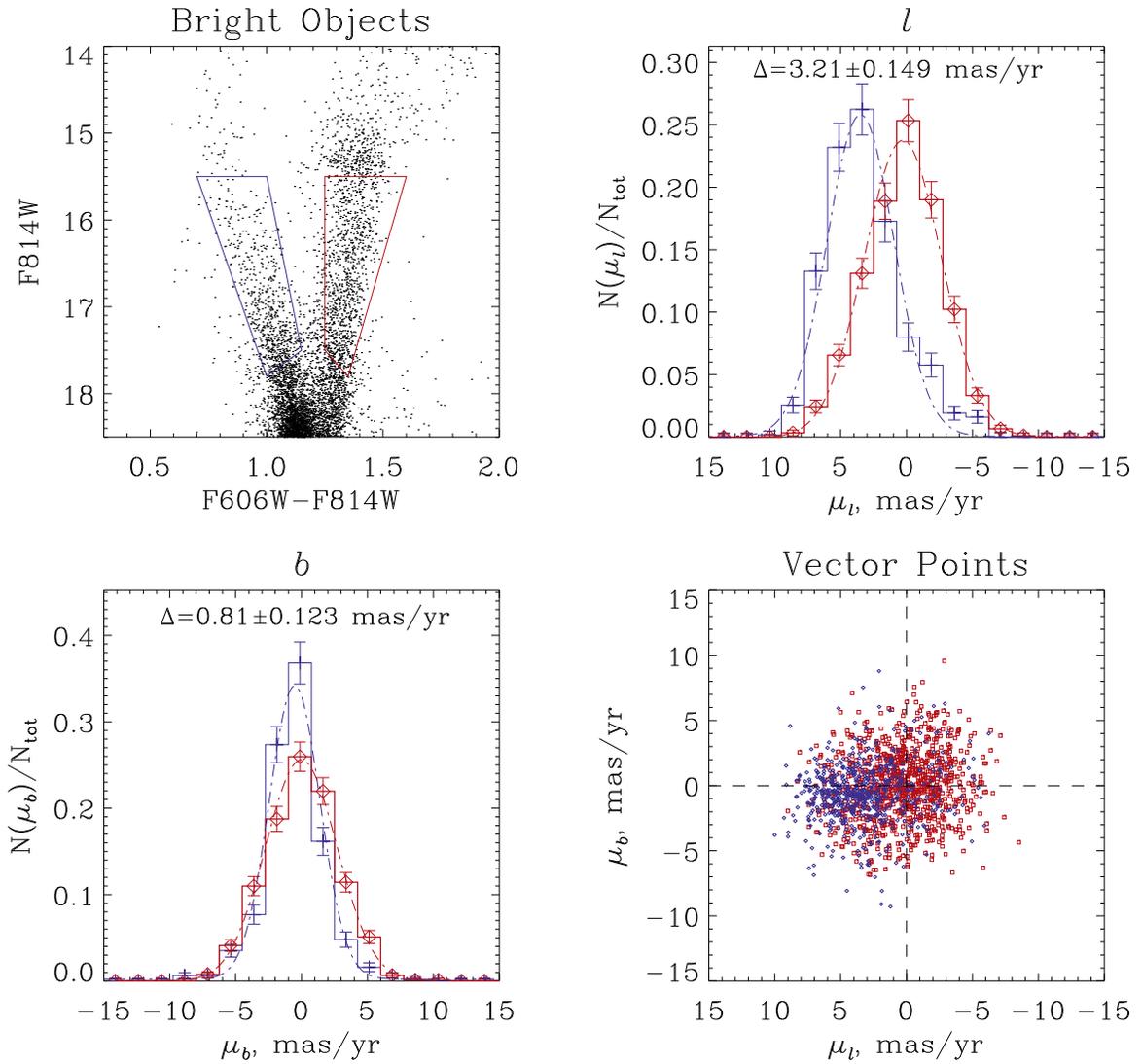}
\caption{As Figure \ref{f_short}, but from the 20s exposures in each epoch.}
\label{f_short2}
\end{figure}

\begin{figure}
\plotone{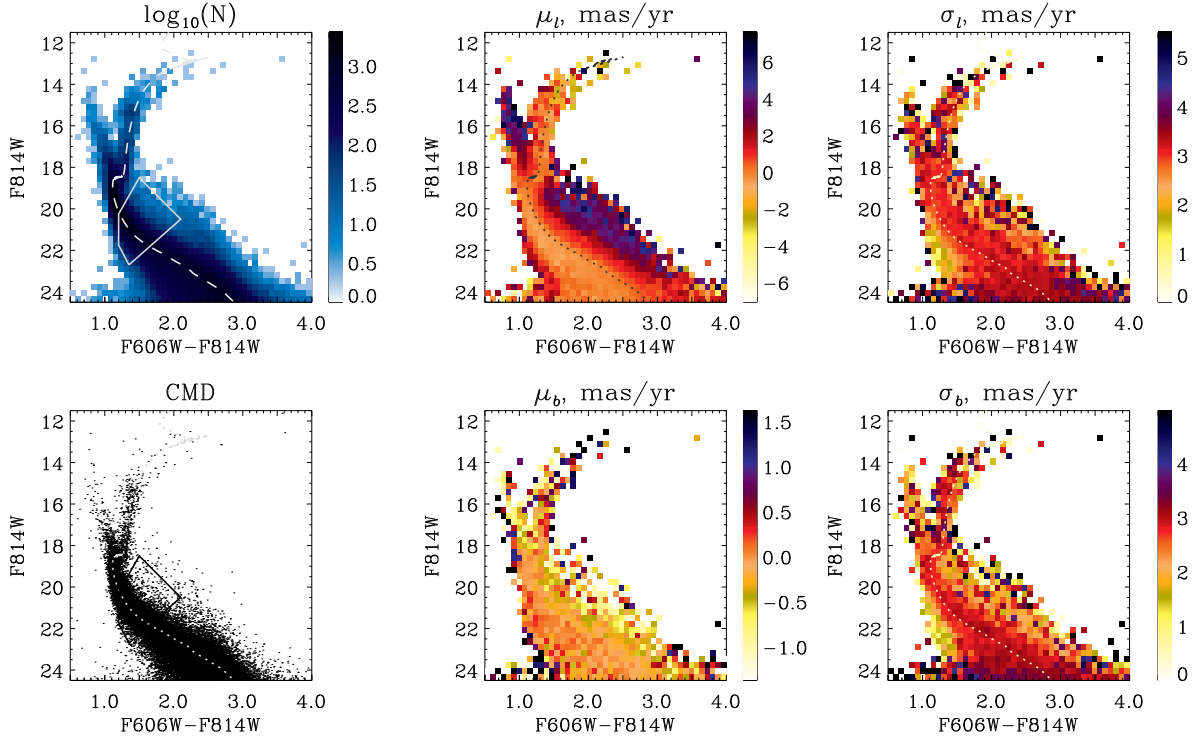}
\caption{{\it Top row:} 2D histograms of the entire population for which proper motions are available, color coded by star-counts (left), mean proper motion $\mu_l$~(middle) and latitudinal proper motion dispersion $\sigma_{l}$~(right), following the approach first presented in Kuijken \& Rich (2002). {\it Bottom row:} the unbinned CMD (left), histograms of $\mu_b$~(middle) and $\sigma_{b}$~(right). Dotted line: mean-bulge isochrone. The polygon in the leftmost pair of panels gives the region in the CMD corresponding to the kinematic tracer stars. In the unbinned CMD, 20\% of stars are plotted for clarity.} 
\label{fig_hess}
\end{figure}

\begin{figure}
\plotone{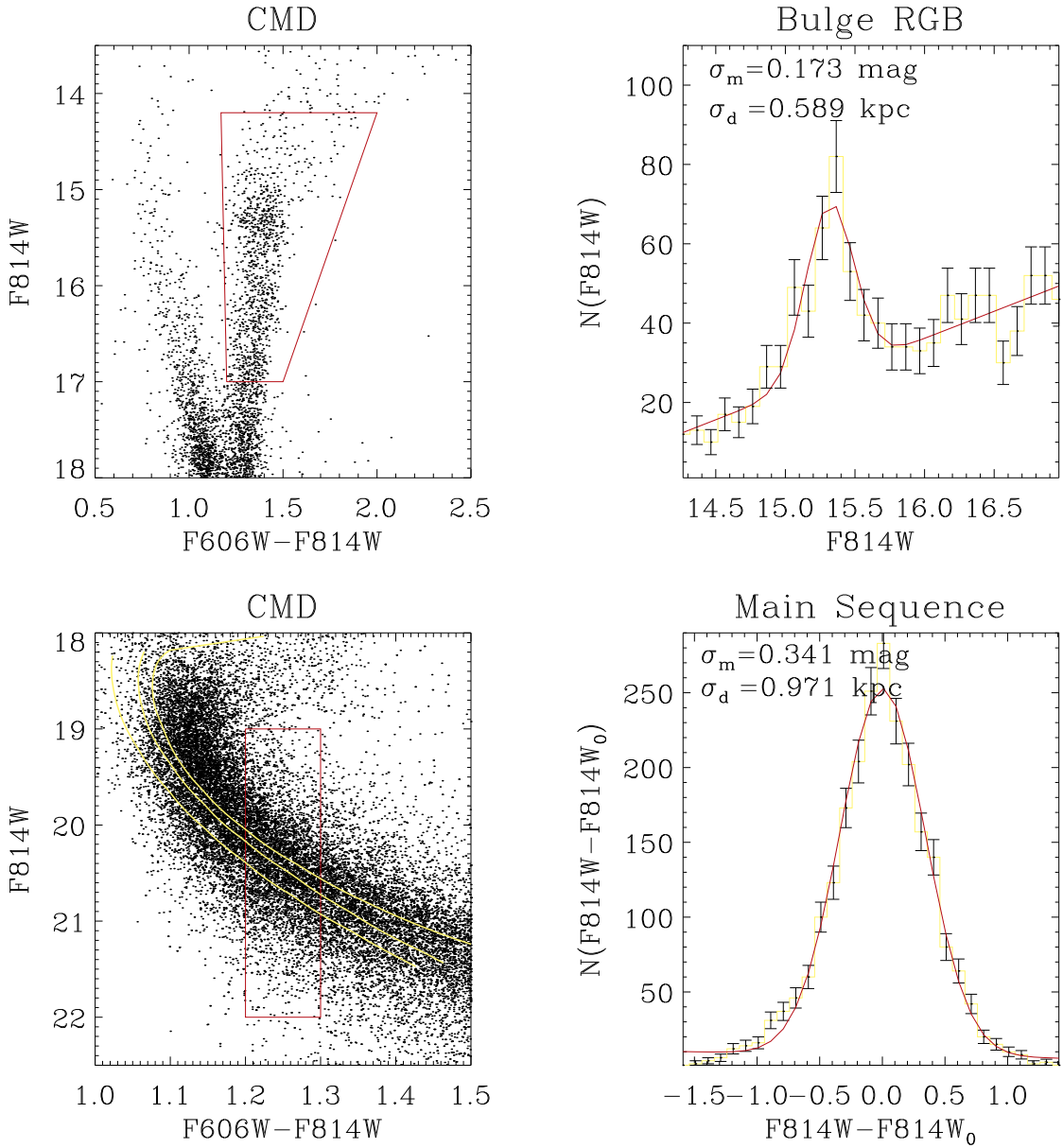}
\caption{Photometric distance estimates using two different regions of the CMD. Left panels give the selection regions, right-hand panels give the range of magnitudes extracted for the regions. {\it Top  Row:} Evolved bulge population. {\it Bottom row:} Main sequence population. an isochrone chosen to fit the bulge population {\it  within the selected region} is shown overplotted at 0,$\pm 1\sigma$~in the bulge metallicity distribution. Distance moduli ({\it Bottom Right}) were estimated relative to the zero-metallicity isochrone as interpolated to the observed color. The 1$\sigma$~spreads of distance modulus are indicated for each population.}
\label{f_distcomp}
\end{figure}

\begin{figure}
\plotone{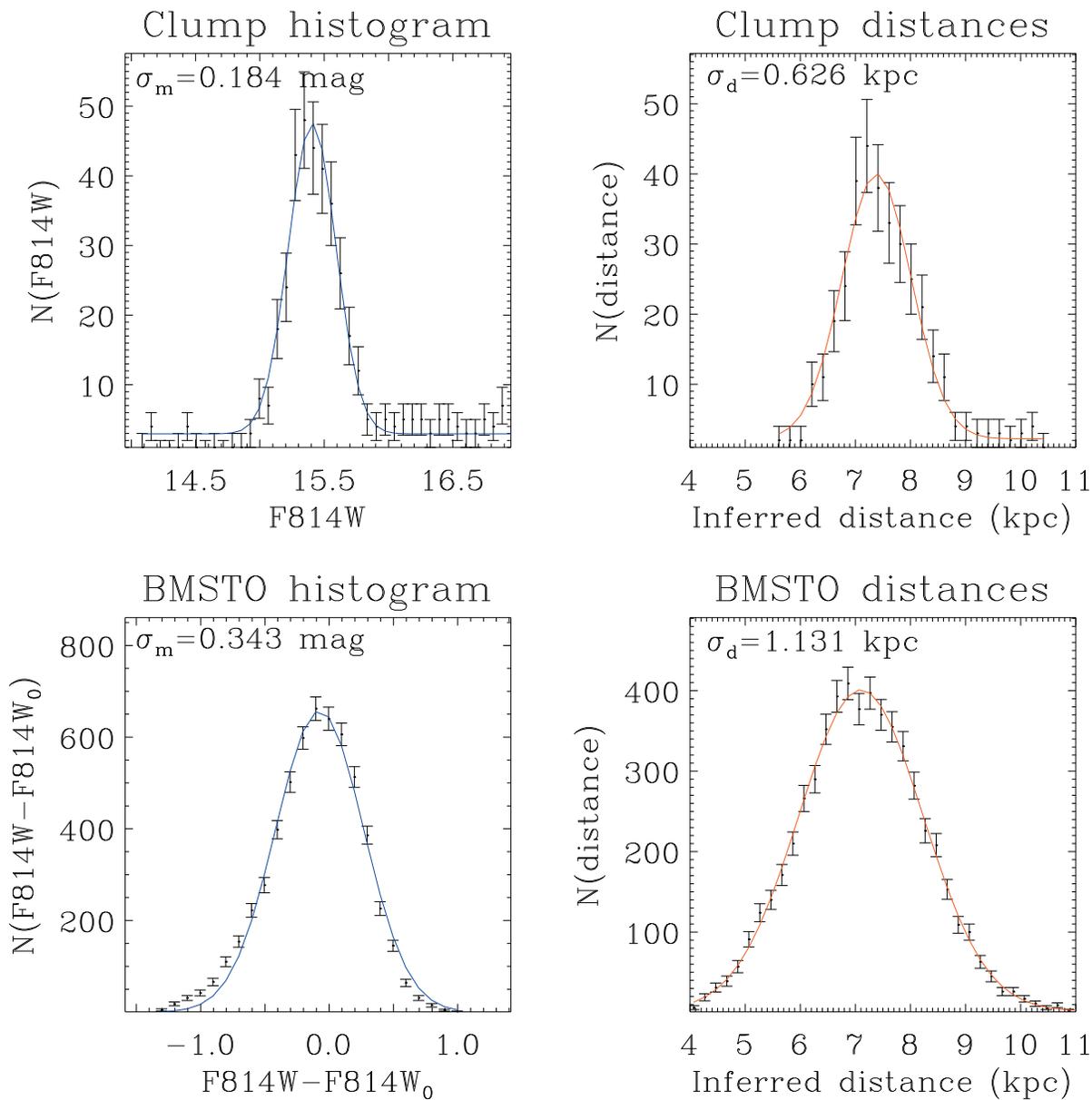}
\caption{Example comparison of inferred distance moduli from a synthetic bulge CMD, for the bulge RGB (top) and below the main sequence turn-off (BMSTO, bottom row). Regions identical to those in Figure \ref{f_distcomp} were used.}
\label{f_dtrial}
\end{figure}

\begin{figure}
\plotone{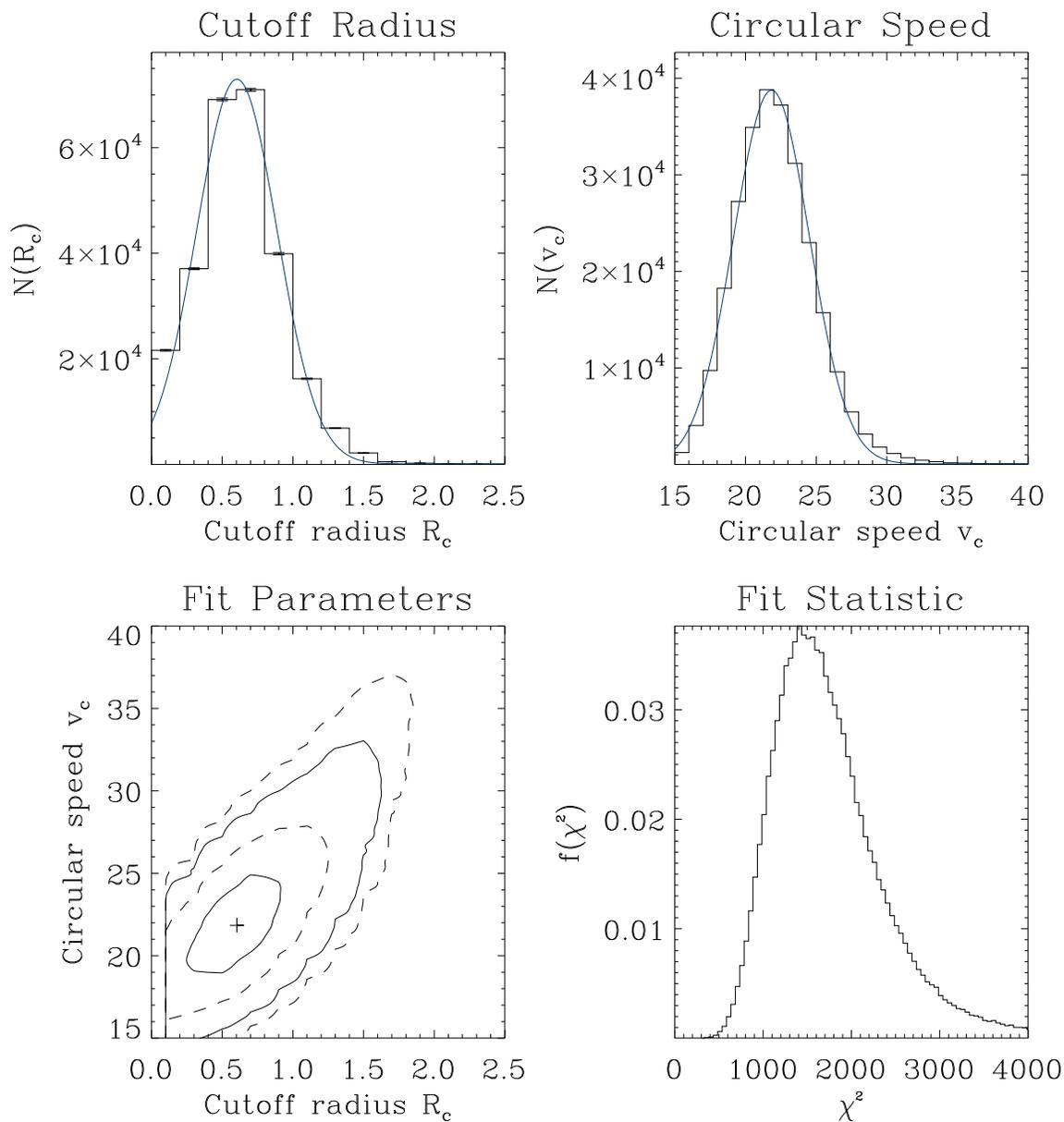}
\caption{Circular speed curve recovered from synthetic proper motion observations (Section \ref{ss_recov}). Cutoff radius $R_C = 0.35R_s$~and constant circular speed $v_C=50$~km s$^{-1}$~exterior to this cutoff were simulated. Countours represent fractions 0.5,0.1,0.01 and 0.002 of the peak in the $R_c$,$v_c$~histogram.}
\label{f_binsim}
\end{figure}

\begin{figure}
  \plotone{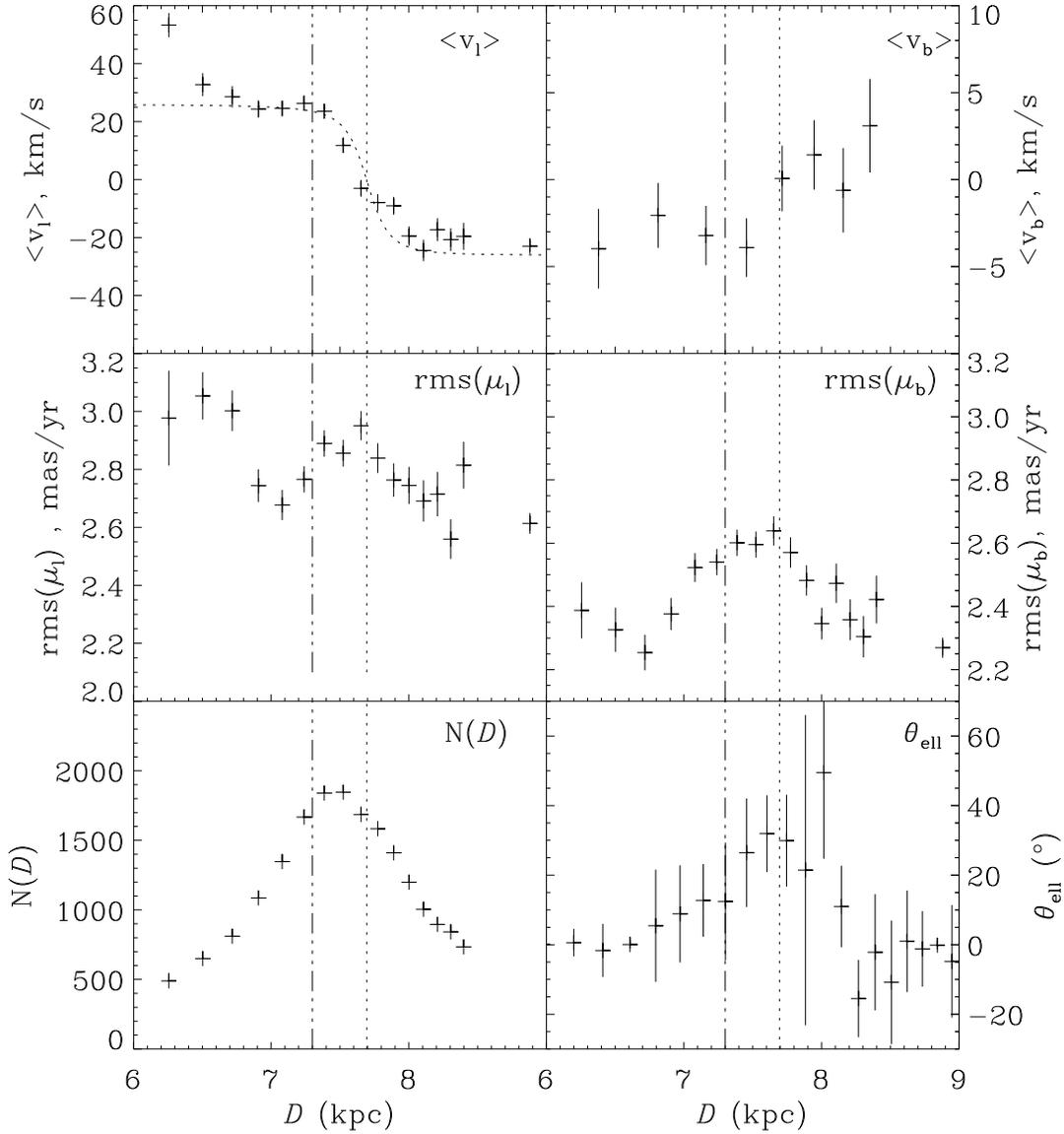}
  \caption{Observed stellar kinematics as a function of distance modulus along the line of sight. {\it Top row:} Transverse velocities in galactic longitude (left; with projection model overplotted as a dotted line) and latitude (right; binned in pairs). {\it Middle row:} Observed proper motion dispersions. {\it Bottom row:} Number of tracer stars per bin (left) and the angle of the major axis of the best-fitting velocity ellipse (right; see also Section \ref{sec_vell}). The line $\alpha=0\degr$~(vertical dotted line) is clearly displaced from the line of maximum stellar density (dot-dashed line).}
\label{fig_obs}
\end{figure}

\begin{figure}
\plotone{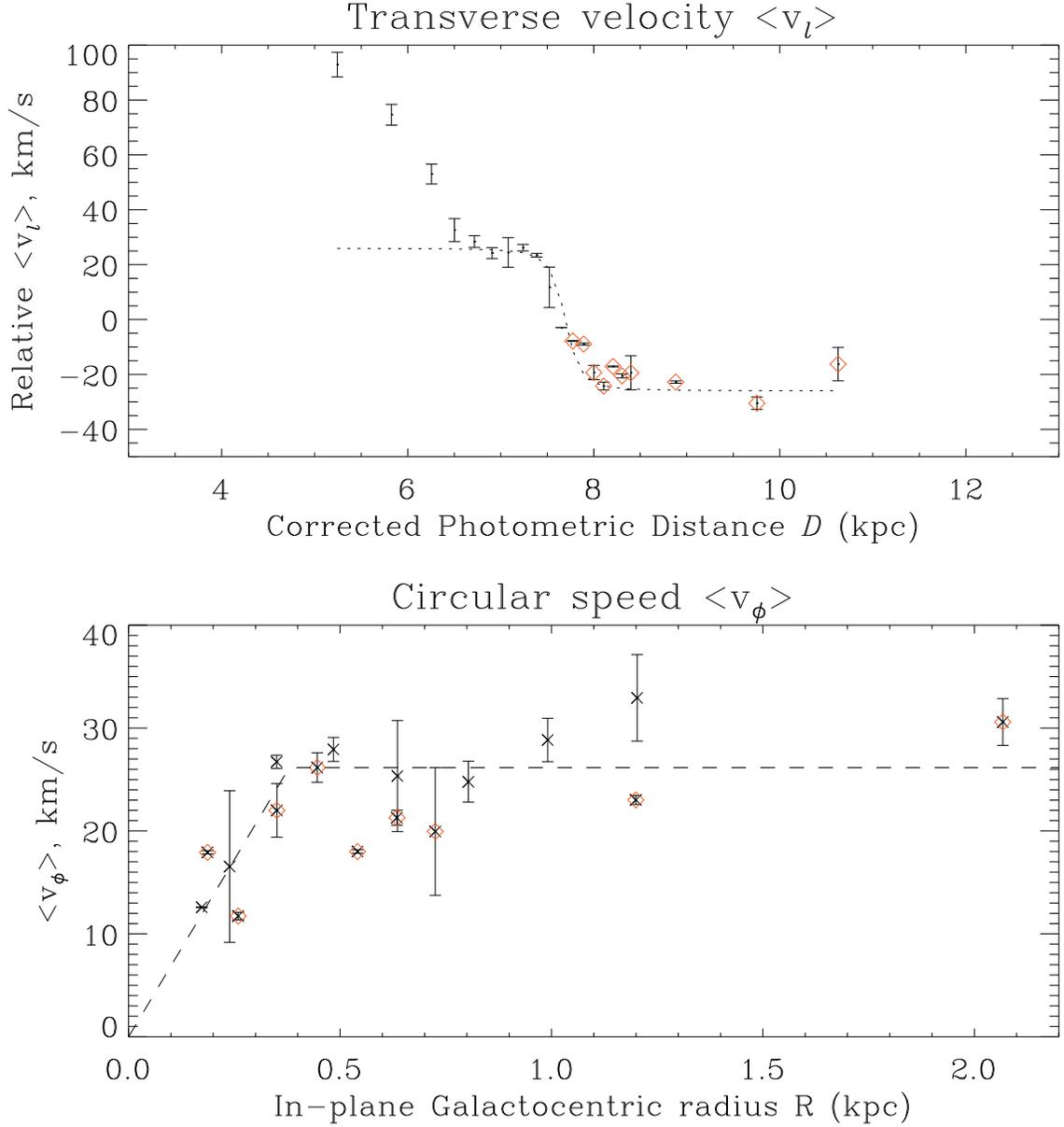}
\caption{Circular speed curve $\vmpe$~recovered from our proper motion data. {\it Top Panel:} observed transverse velocity \vml as a  function of distance along the line of sight, with an example constant-\vmp curve overplotted (dotted line). Although our distance estimate breaks down for stars in foreground and background spiral arms, the mixing of disk objects with bulge in the foreground is apparent, and we are measuring proper motions out to the far side of the bulge. {\it Bottom:} assuming \vmr and \vmz are both zero, the circular speed curve is recovered under the condition that the near-side and far-side trends (red diamonds) be symmetric. In qualitative agreement with recent radial velocity results (Rich et al. 2007), we see apparent solid body-like rotation only out to galactocentric radius $R_c$=0.3-0.4 kpc.}
\label{f_rotcurve}
\end{figure}

\begin{figure}
\plotone{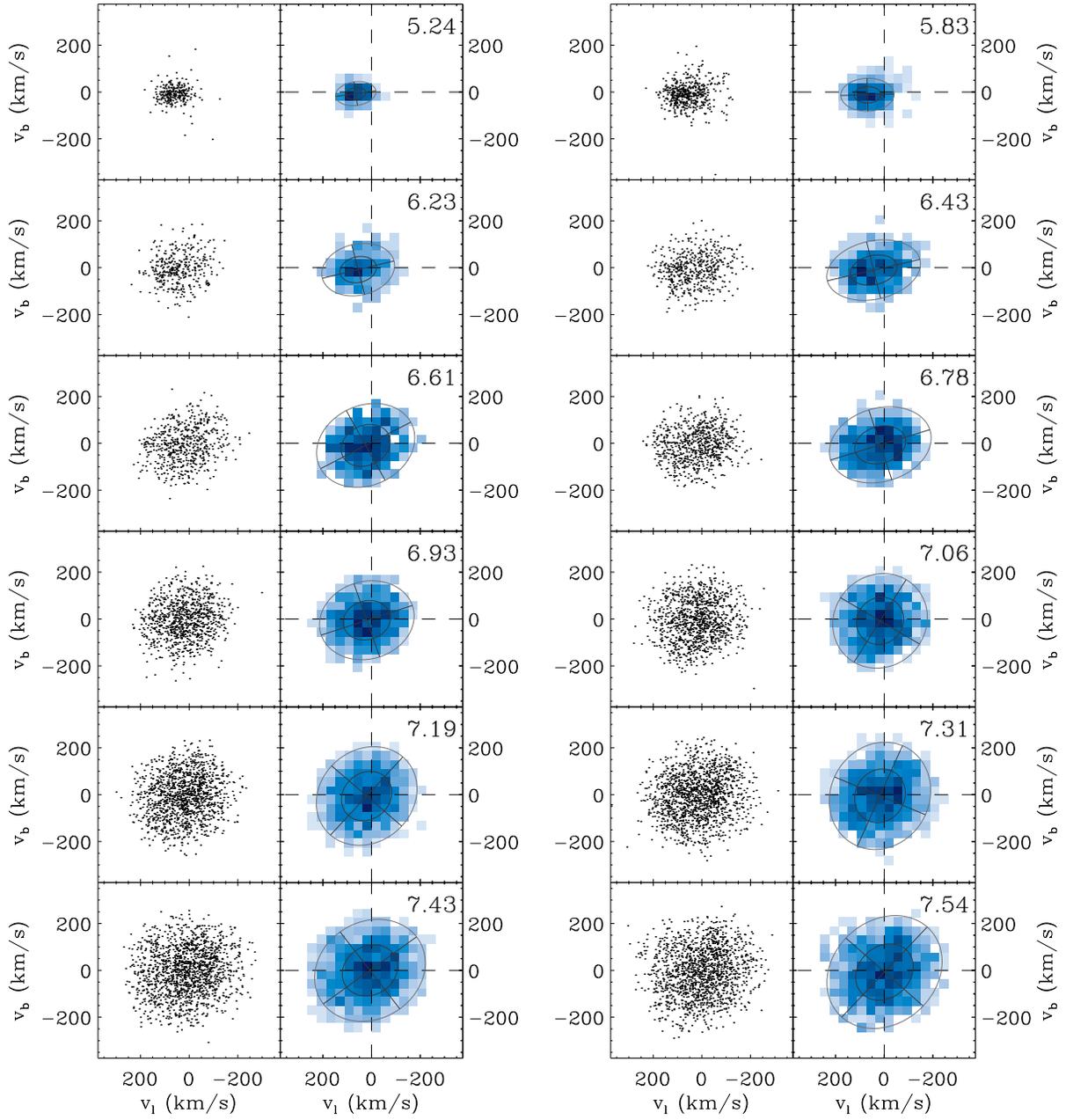}
  \caption{Observed transverse velocity distribution as a function of line of sight distance. Distance is marked in the right panel for each pair and increases reading left-right. The left panels denote the observed velocity distribution while the right panels give the 2D histograms and $1\sigma, 2\sigma$~best-fitting ellipses. This figure shows $5.2 \la d \la 7.6$kpc.}
\label{fig_ellip}
\end{figure}

\begin{figure}
\plotone{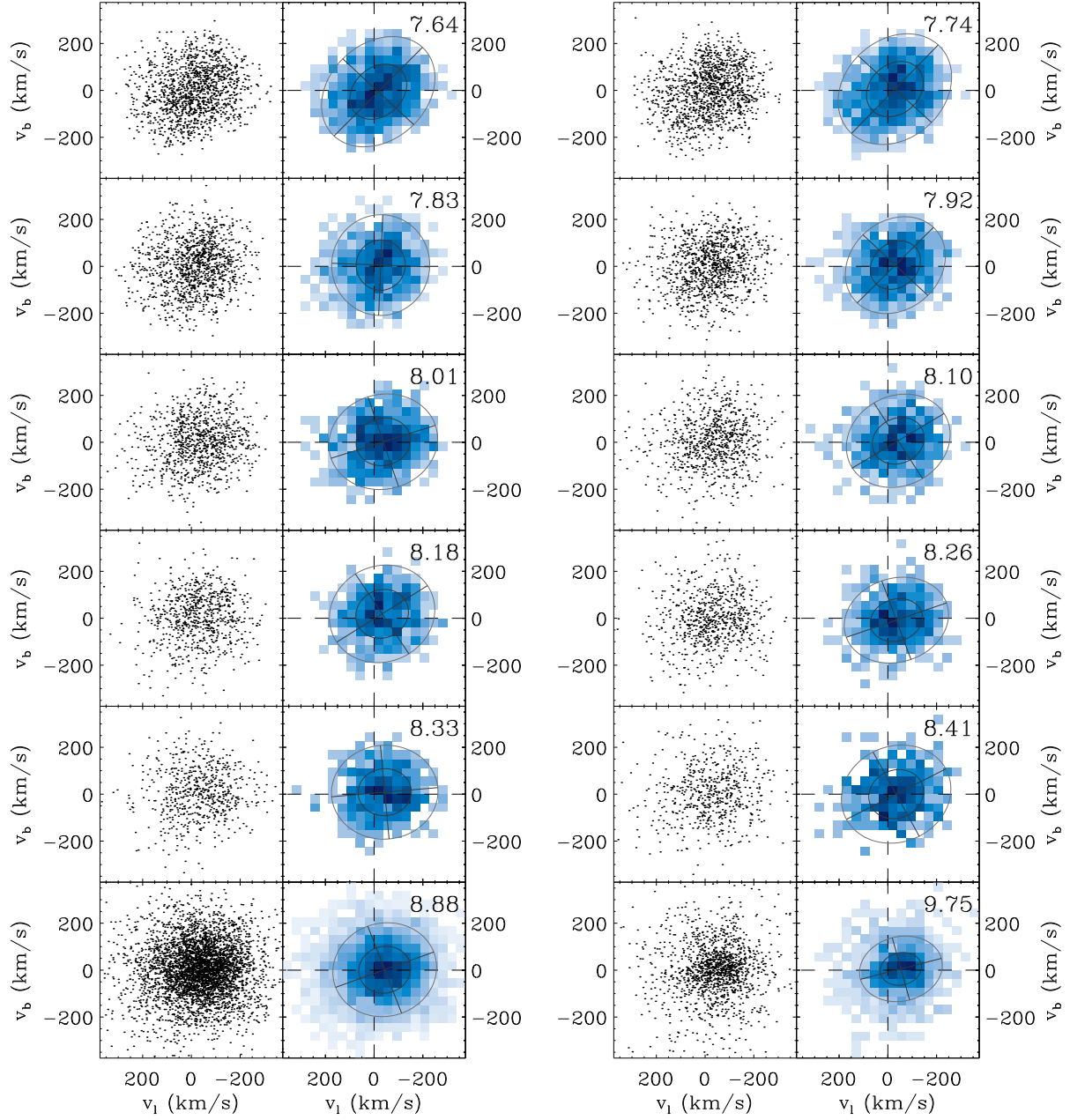}
  \caption{As for figure \ref{fig_ellip}, but for $7.6 \la d \la
    10$~kpc.}  
\label{fig_ellip2}
\end{figure}

\begin{figure}
\plottwo{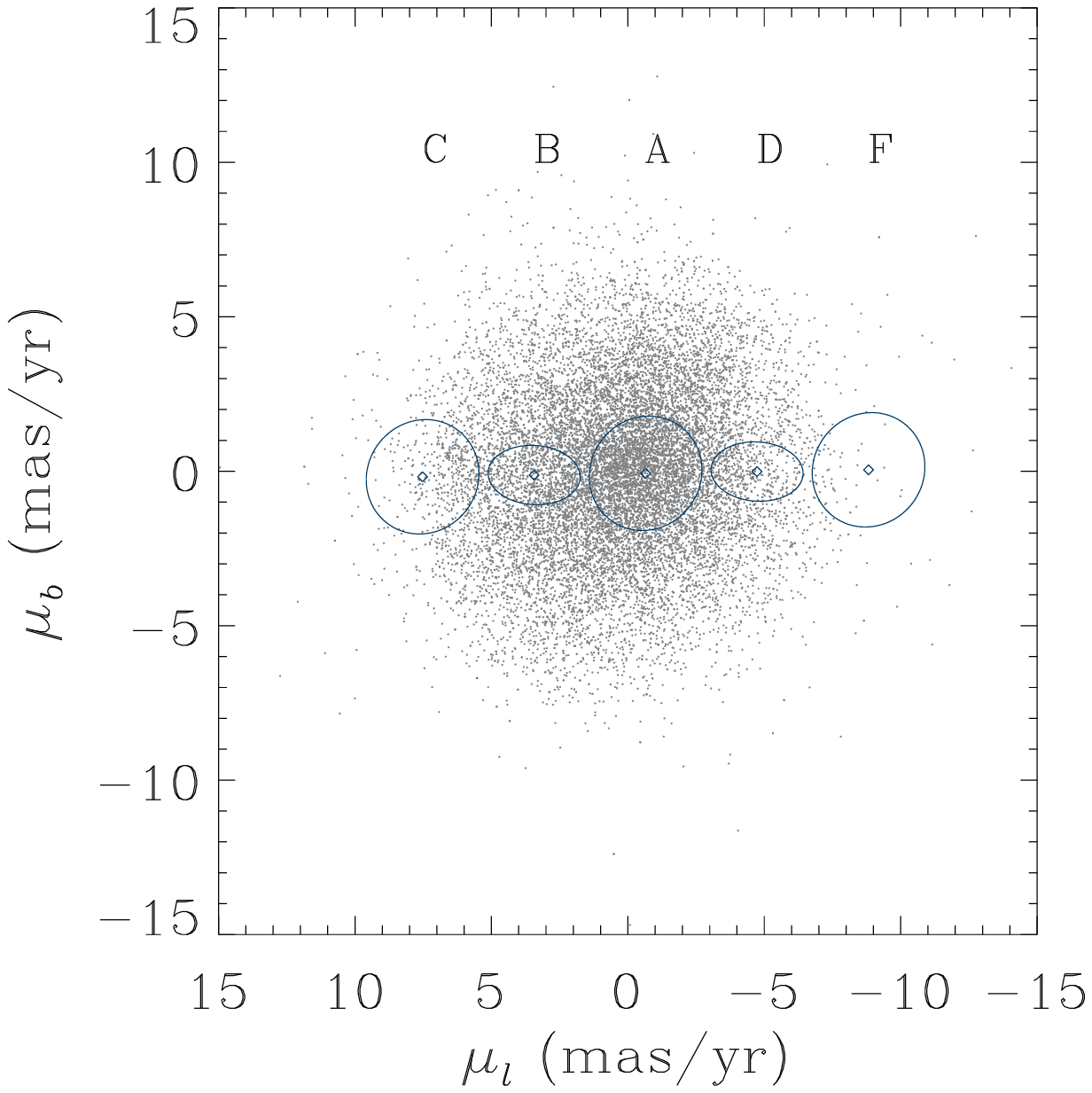}{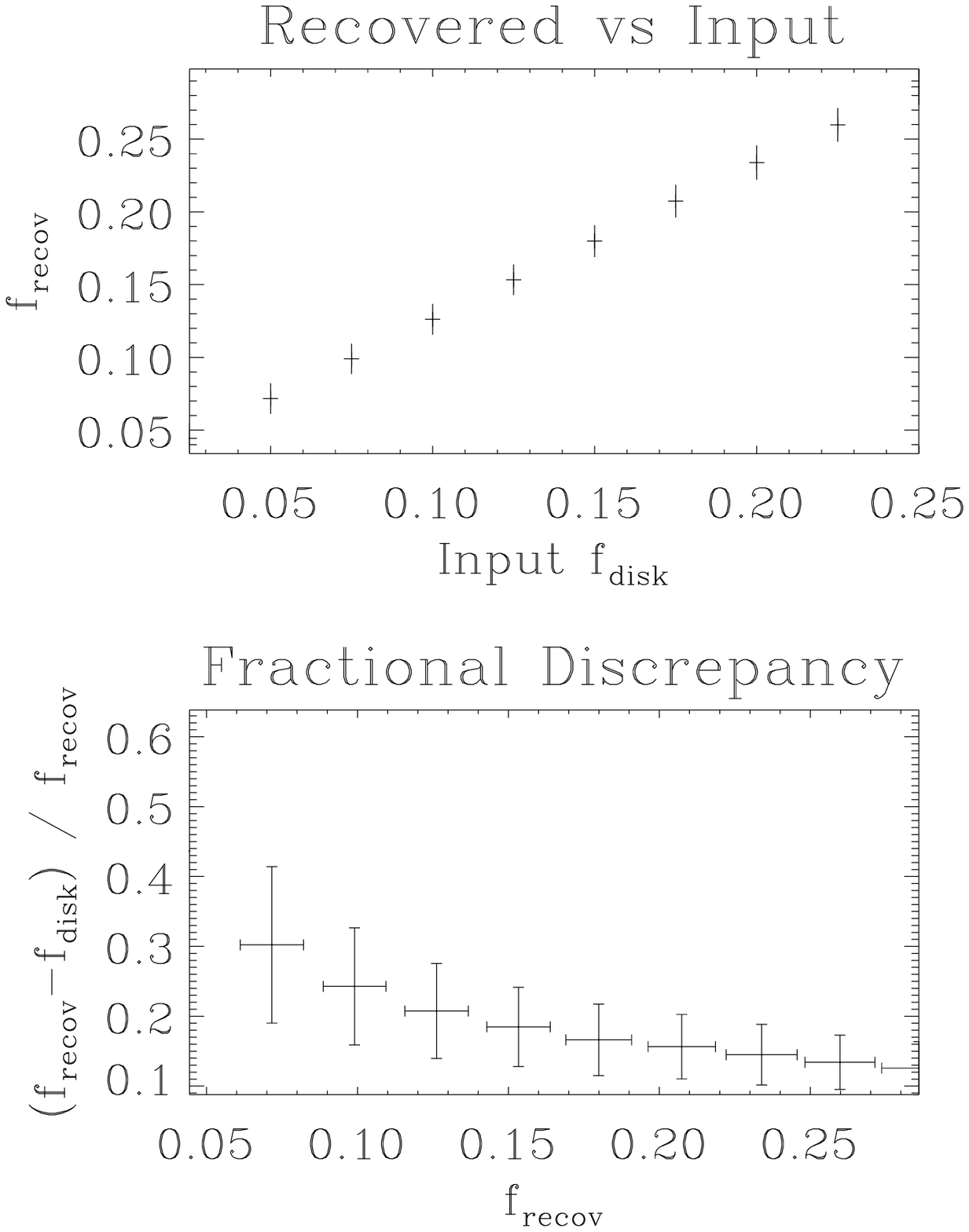}
\caption{Estimating the fractional contribution of the bulge. {\it  Left:} vector point diagram of a random subsample from the full set of stars with proper motion measurements, with 0.7$\sigma$~ellipses with axis-ratios representative of the mean-disk and mean-bulge respectively. The initial estimate for the disk population is approximated by $N_{disk} \simeq N_B - N_D$~in the regions above, that of the bulge is approximated by $N_{bulge} \simeq N_A-N_C+N_F$. {\it Right:} Difference between the disk fraction recovered in this manner, $f_{recov}$, and an input population $f_disk$~simulated using the proper motion ellipses fit previously, for $10^4$~trials.}
\label{f_fbulge}
\end{figure}

\begin{figure}
\plotone{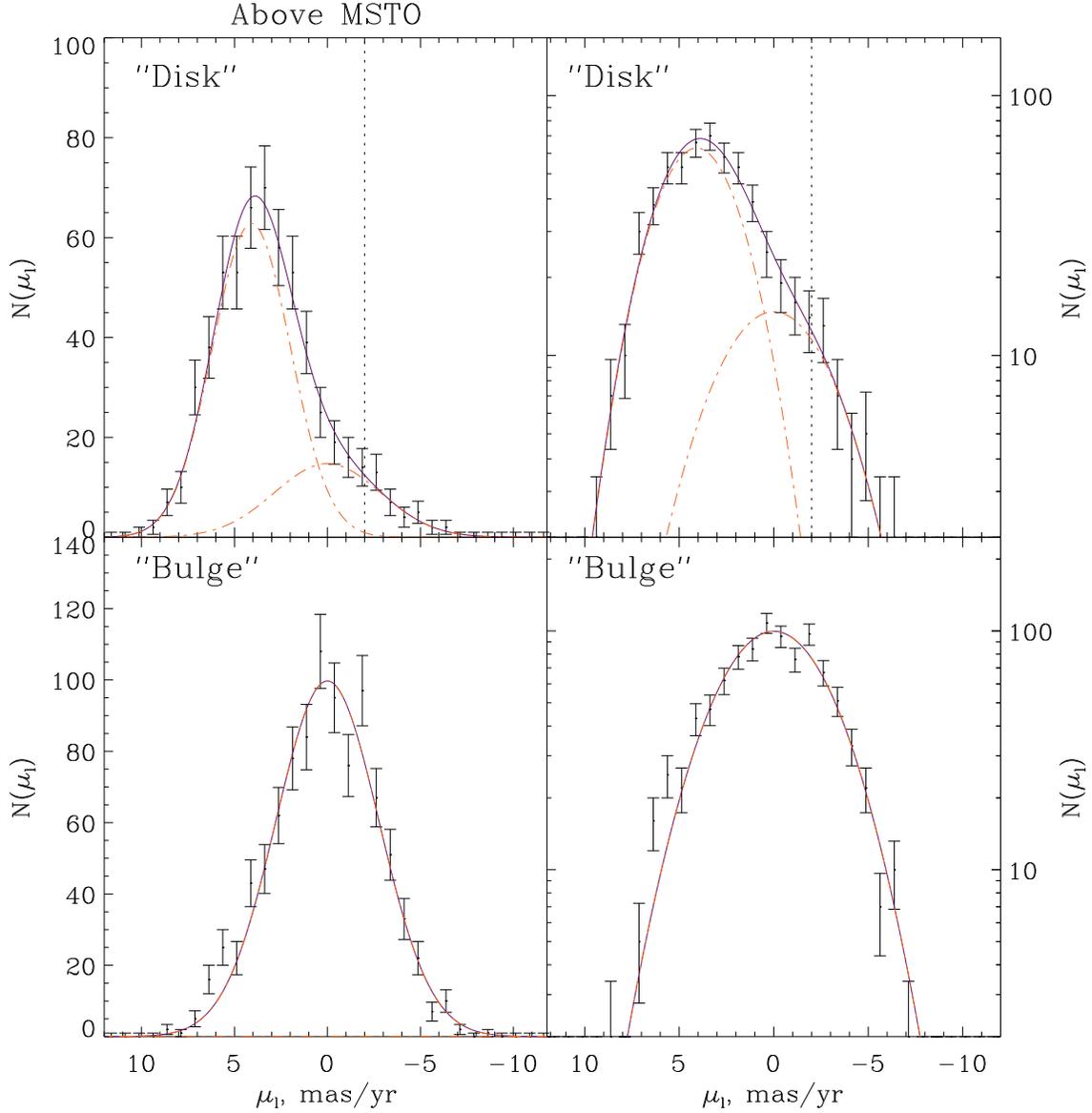}
\caption{Latitudinal proper motions $\mu_l$~for objects above the Bulge main sequence turn-off. The nominal selection regions in the CMD for the Bulge and disk are those of Figures \ref{f_short} and \ref{f_short2}. {\it Top Row:} $\mu_l$~for objects in the disk region of the CMD, plotted on linear and log scales. {\it Bottom Row:} $\mu_l$~for the Bulge region of the CMD. The vertical dashed line shows our $\mu_l$~cutoff for bulge membership. The nominal disk population contains a significant population with kinematics indistinguishable from the Bulge. These objects appear to be an {\it apparently} young population in the Bulge, and may represent Blue Stragglers or a comparatively recent epoch of star formation in the Bulge. See also Figure \ref{f_bulge}.}

\label{f_strag}
\end{figure}

\begin{figure}
\plotone{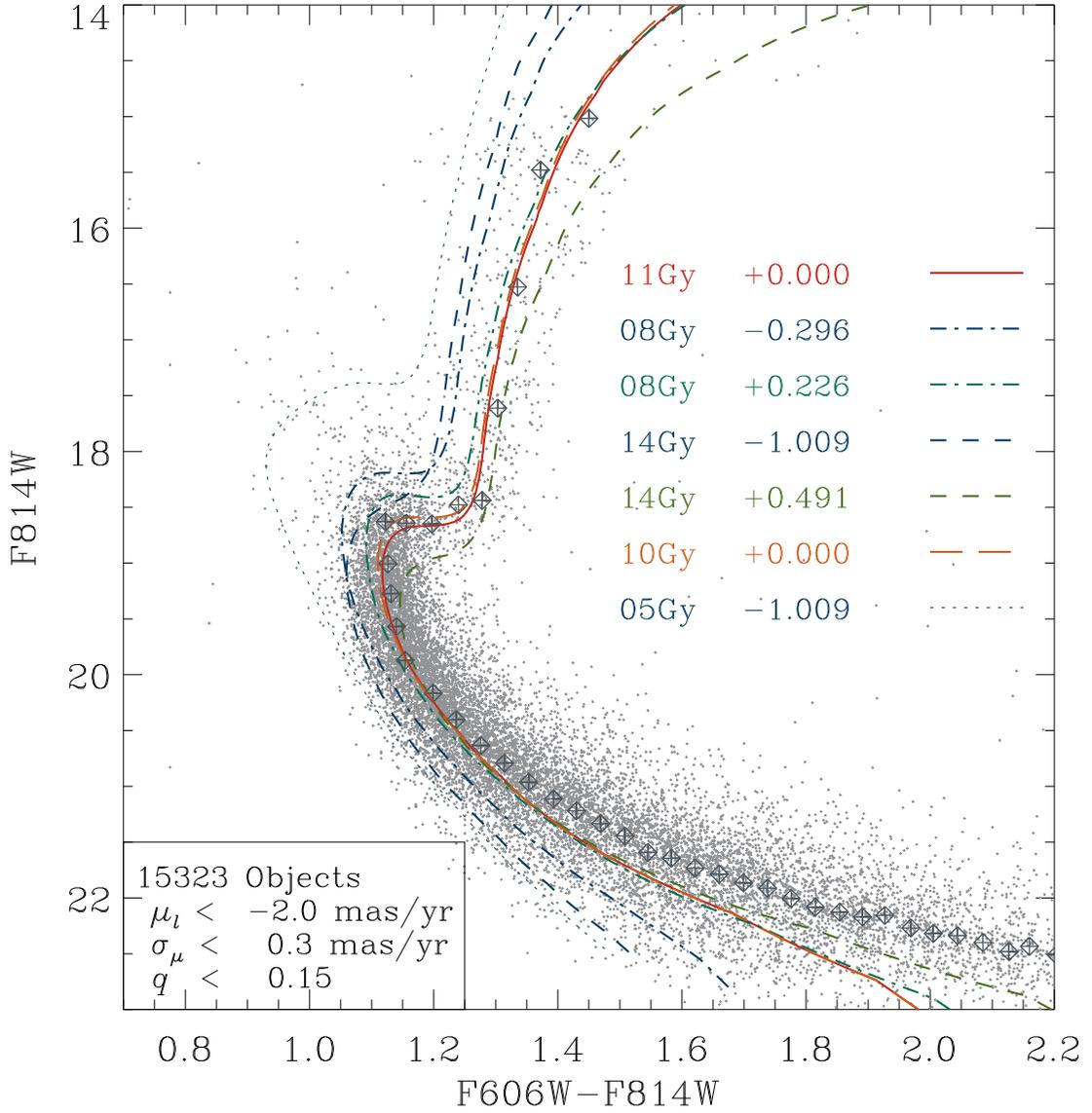}
\caption{Proper motion-selected bulge objects (Section \ref{ss_ndisk}), using similar mean proper motion criteria to Kuijken \& Rich (2002) but with a 6$\sigma$~detection requirement imposed. This CMD was divided into bins and the median computed (diamonds); below the MSTO the uncertain binary fraction causes an artificial apparent age effect, so we focus on the region above the MSTO for comparison. An alpha-enhanced, solar-metallicity isochrone at 11Gyr represents the median sequence well above the turn-off. Also shown are sequences at metallicity [Fe/H]=(-1.009,-0.226, +0.491) and ages (8, 10, 14) Gyr to bracket the Bulge population above the MSTO. Also shown is a very young, very metal-poor population (dotted line).}
\label{f_bulge}
\end{figure}

\begin{figure}
\epsscale{0.5}
\plotone{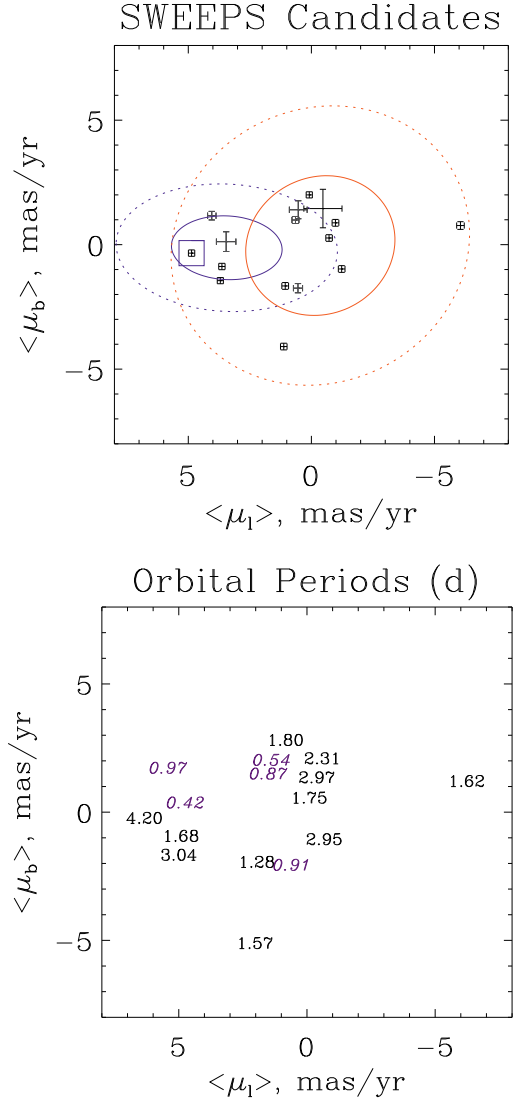}
\caption{{\it Top:} Proper motions of the sixteen SWEEPS candidates. {Left Bottom:} as above, except the candidates are marked with their orbital periods. The orbital periods of the ultrashort-period transit planet candidates are given in italics. The 1$\sigma$~and 2$\sigma$~contours of the stellar distributions of bulge (red, right) and disk (blue, left; see also Section \ref{s_cand}) are overplotted. There is an apparent clumping of objects within the 1$\sigma$~ellipse of the disk population; furthermore SWEEPS-04 (blue box; period 4.2 days), known to lie in a likely disk-dominated region of the CMD, falls close to the the mean-disk proper motion.}
\label{f_scand}
\end{figure}

\begin{figure}
\plottwo{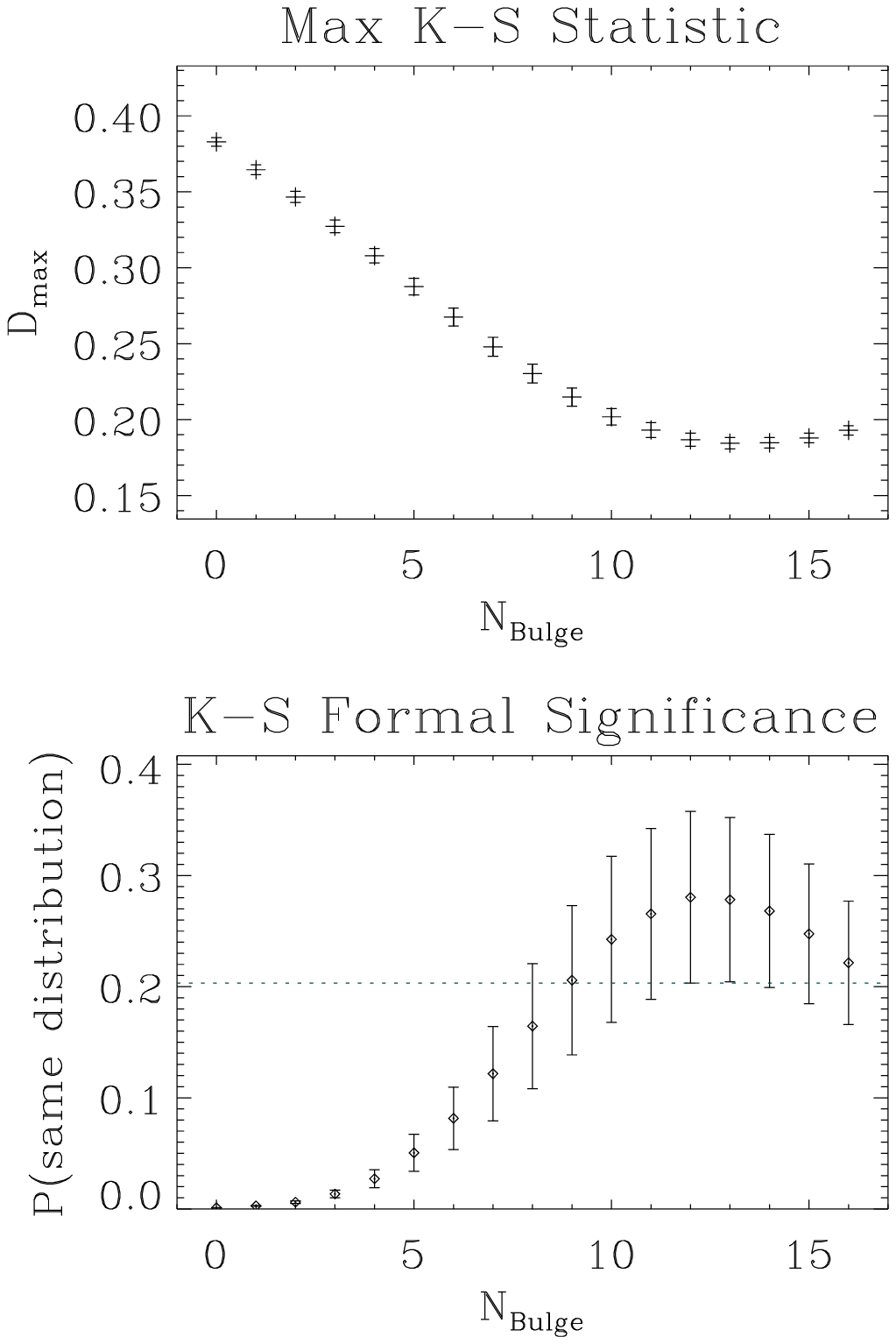}{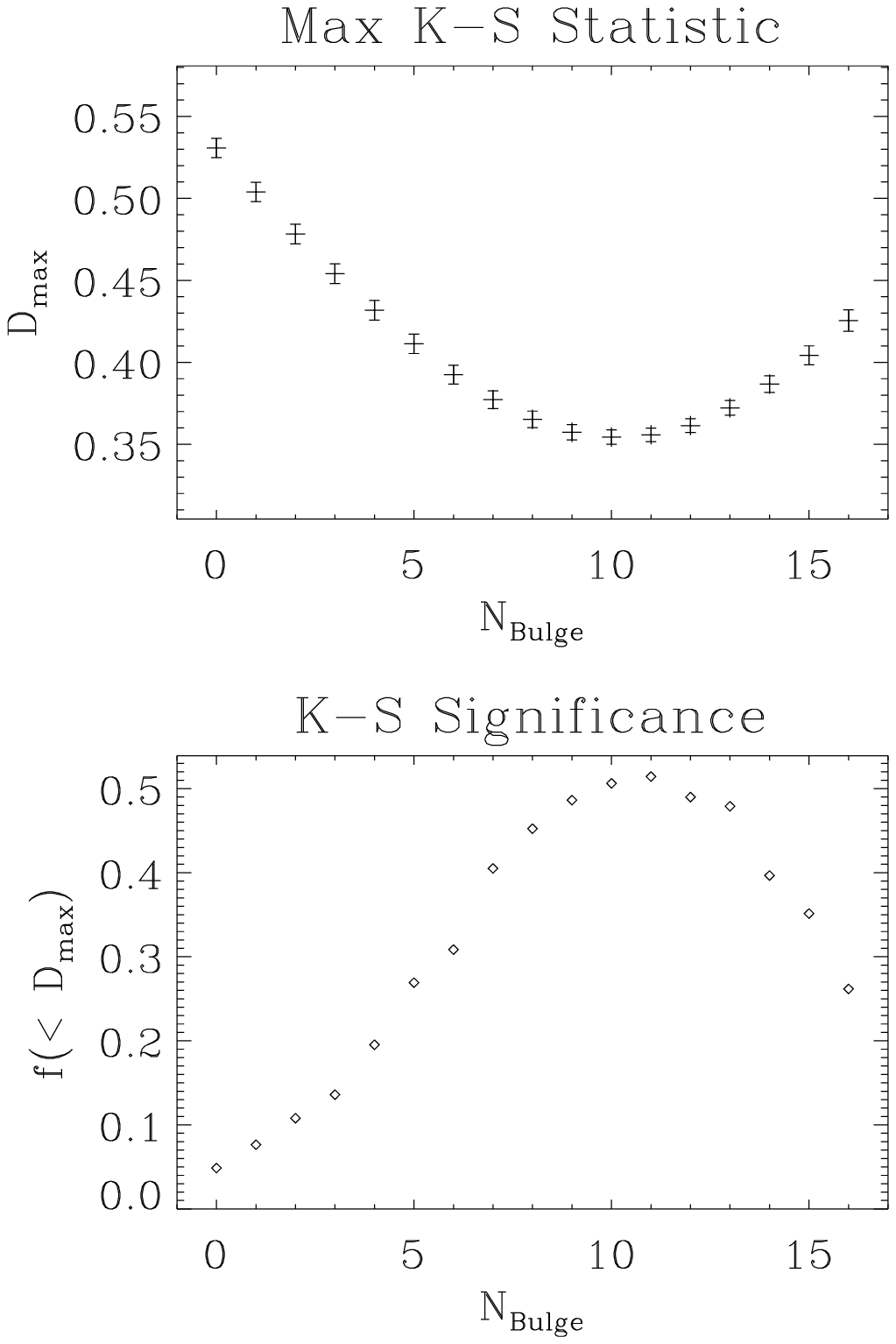}
\caption{Constraints on kinematic membership of the SWEEPS candidates. We use the angular distribution of candidates in $\{\mu_l,\mu_b\}$~space to compare the observed distribution with trials compsed of populations drawn from varying fractions of the best-fit disk and bulge distributions. {\it Left:} Maximum K-S statistic between the observed and trial dataset (top), for 10,000 trials each at a range of disk contributions to the observed population. This yields a formal probability that the observed and synthetic datasets are realisations of the same parent distribution (bottom); at the 1$\sigma$~level, a population of at least eight bulge objects is consistent with the peak. {\it Right:} as before, this time using the 2-D K-S test on all the sixteen candidates.}
\label{f_candkin}
\end{figure}

\begin{figure}
\plotone{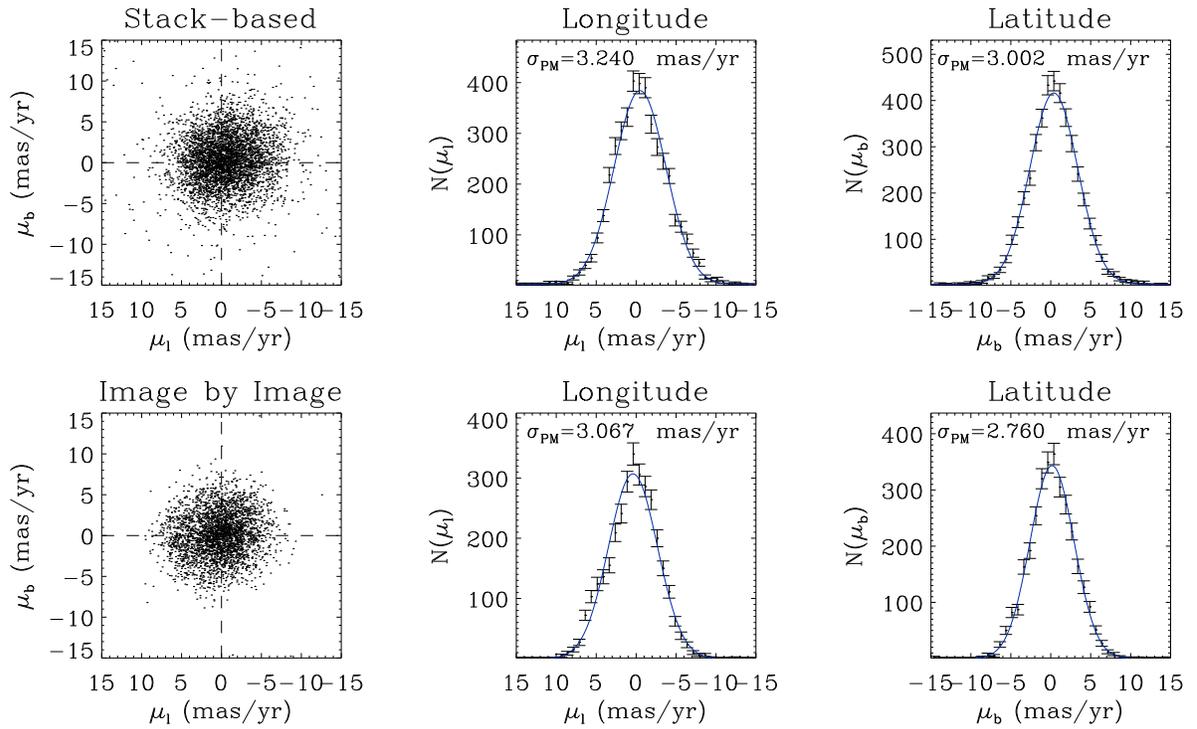}
\caption{Comparison of proper motion measurements (in mas yr$^{-1}$) when computed by comparison to positions using the optimal stack for photometry (top) and using an image-by-image approach for both epochs (bottom).}
\label{fig:compar}
\end{figure}

\begin{figure}
 \plotone{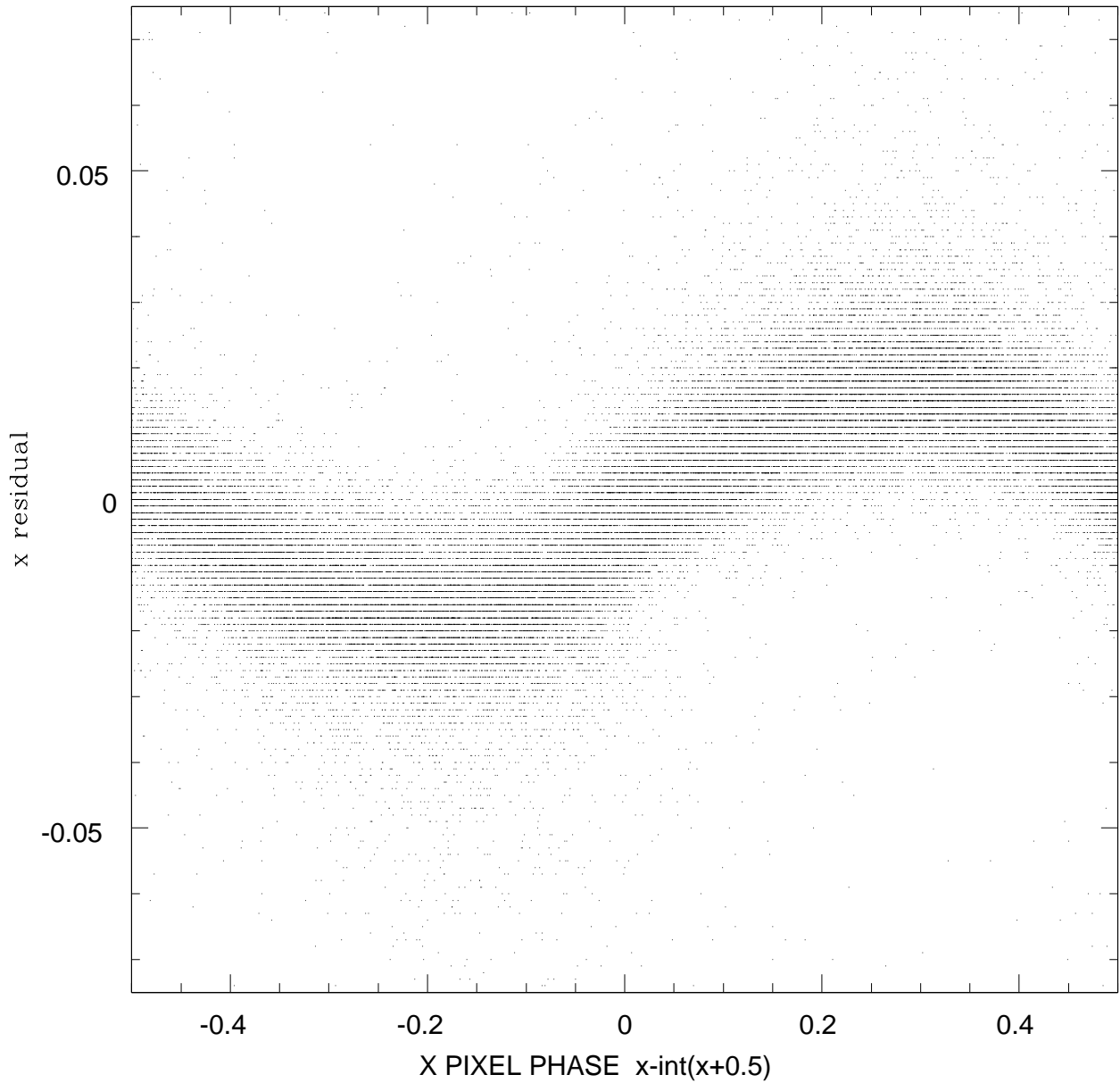}
\caption{Alteration of the ePSF in the production of the twice-oversampled superimage used for SWEEPS photometry. The image model computed in Sahu et al. (2006) provides an estimate of the full pixel-flux at each position $x,y$~in this oversampled space, thus we may de-interlace the superimage into four \flt - type images. When the Anderson \& King (2006) techniques are used to measure positions on each of the four images, a difference in position as a function of pixel phase between pairs of de-interlaced images becomes apparent. This illustrates that the combination of images into the superimage has subtly changed the ePSF of the scene. While optimal for photometry, use of the SWEEPS superimage can lead to a position systematic at the 0.02-pixel level.}
\label{f_system}
\end{figure}

\begin{figure}
  \epsscale{1.1}
  \plottwo{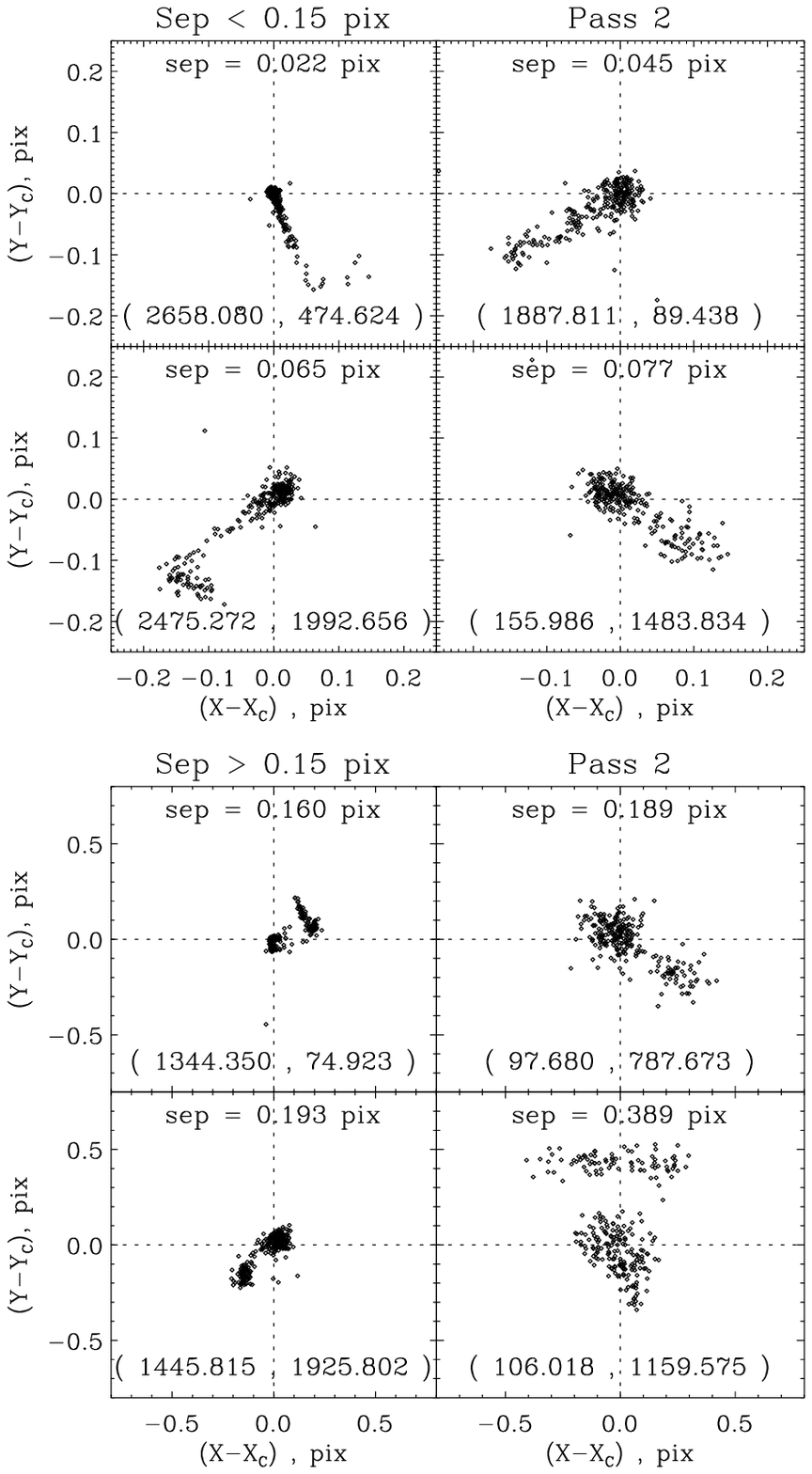}{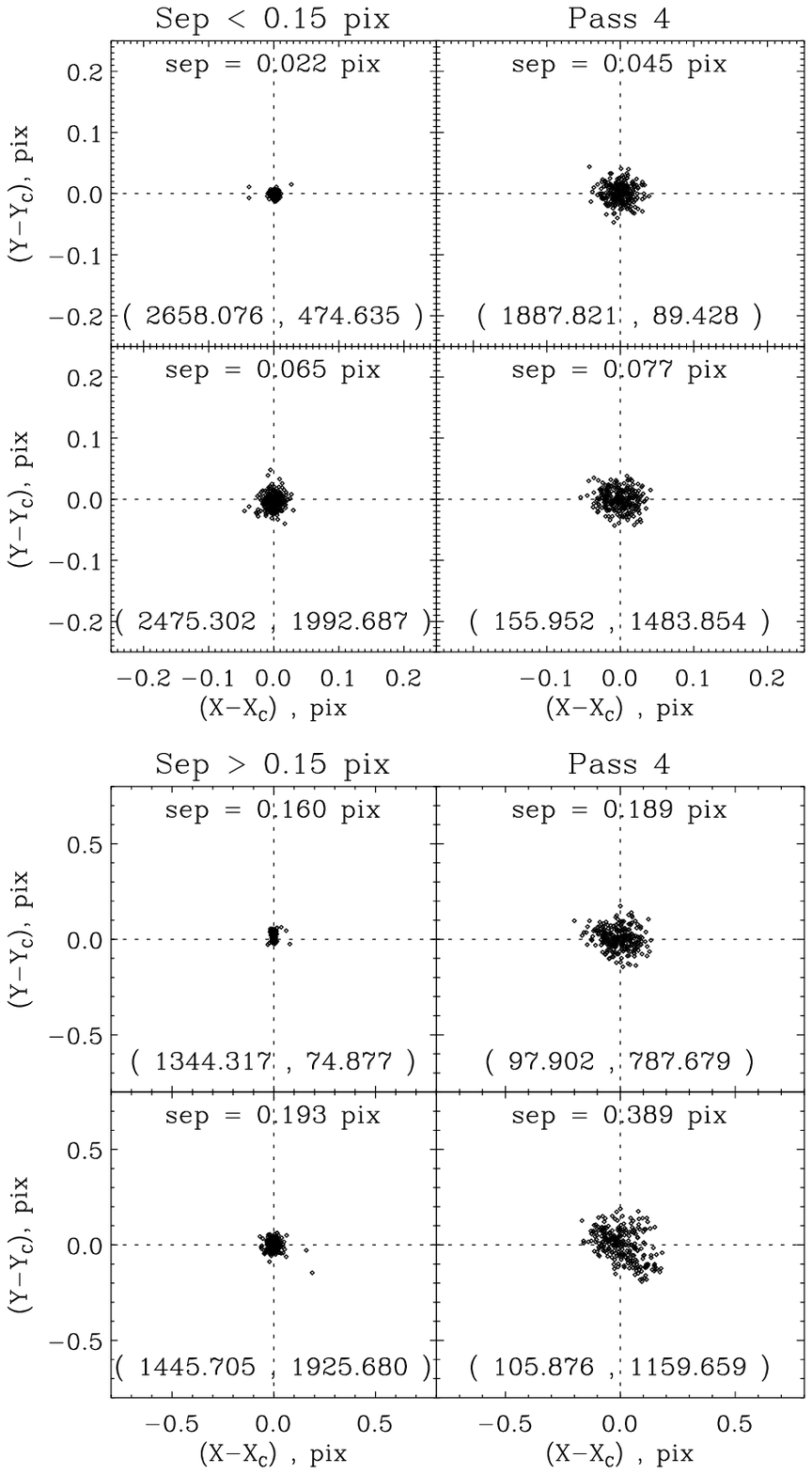}
  \caption{For $\sim6\%$~of objects, the techniques of Section \ref{s_sigma} fail to converge on a single solution for mean position ($X_c$,$Y_c$). Objects for which the fitted peaks separate by $<0.15$~pix show a main clustering and a trail or in some cases another island of points ({\it Left Top}). Objects with wider fitted separation show clearer separation between two clusters ({\it Left Bottom}). Measurements within the island with the greatest number of points are selected for further passes; this usually coincides with the island with lowest scatter ({\it Right Top \& Right Bottom}).}
 \label{f_bimod}
\end{figure}

\begin{figure}
 \plotone{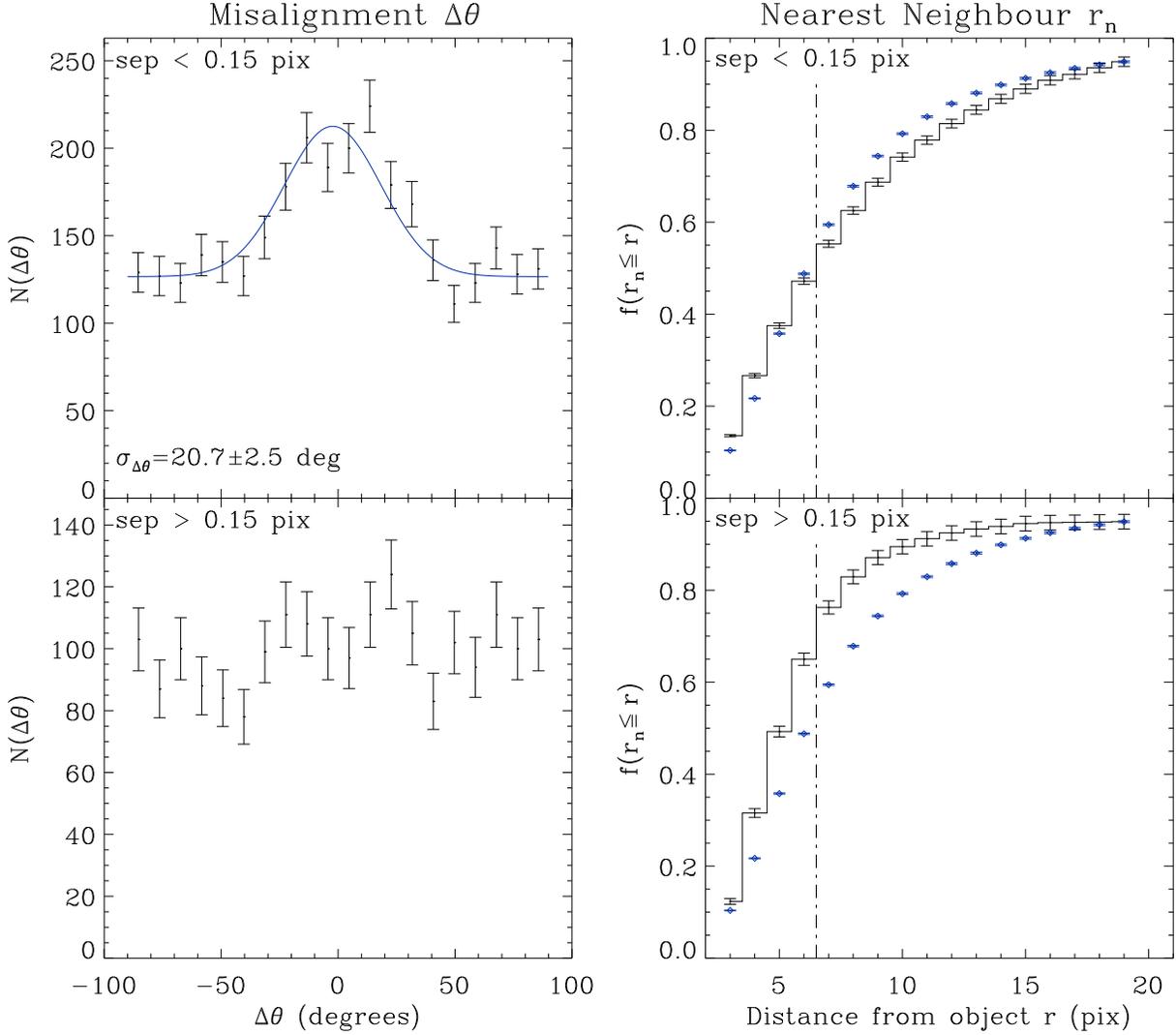}
 \caption{Relating the properties of the bimodal position-estimates to the location of the nearest bright neighbor (i.e. that is at least as bright as the object measured). {\it Left:} misalignment $\Delta\theta$~between the line joining the two measurement peaks and the line joining the object to the nearest bright neighbor, for the subset of each population where the nearest bright neighbor is within 6 pixels of the object in question. {\it Right:} fraction of objects within each population for which the nearest bright neighbor lies within distance $r$~pixels (solid lines), compared to the fraction for the entire population in the field of view (symbols).}
 \label{f_bimod2}
\end{figure}

It thus appears clear that a significant fraction of both classes of
object with bimodal position estimates, represent different
manifestations of the effects of crowding by stellar neighbors. The
former class of bimodal object is found all throughout the sample
diagram, while the high-separation class of bimodal object coincides
with the population showing anomalously high position-scatter during
the second pass of the photometry (Figure \ref{fig_rms}). Where two
clouds of points are produced for an object, we select the cloud with
the highest number of measurements; visual inspection suggests that in
most cases this clump shows the lower scatter of the two clouds and is
interpreted as the true position of the star (Figure \ref{f_bimod},
Right). This selection takes place at the same time as sigma-clipping
when the master-list is updated after each pass of the photometry
(Section \ref{s_sigma}). The resulting master position-list now shows
rather small internal variation (Figure \ref{fig_rms}). We remind the
reader that when selecting a clean-bulge sample, our selection on the
crowding-index $q$~will remove most of the bimodal objects from
consideration.


\section{Time Variation in Charge Transfer Efficiency}\label{appenC}

ACS/WFC has its horizontal shift registers along the Detector-X
direction at the top and bottom of the detector, so that during
readout charge is transferred in the positive Y-direction (WFC1) and
negative Y-direction (WFC2). It thus incorporates a Charge Transfer
Efficiency (CTE) effect; the signal measured from a star is dependent
on the detector-Y position of that star due to the number of transfers
of the signal during readout. This effect may be observed in a number
of ways; the same scene may be observed at two different integration
times or pointings and the change in magnitude or position observed
plotted against detector-Y position (e.g. Riess \& Mack 2004). Or, a
sharply-defined feature in the CMD may be measured and its variation
over the CCD tracked if there are enough stars in the frame
(e.g. Brown et al. 2005). The effect on measured magnitudes for
ACS/WFC is a ``V''-shaped pattern in instrumental magnitude versus
detector-Y; objects nearest the chip-gap are farthest from the shift
registers and thus suffer the greatest signal decrease during
readout. Recently it has been pointed out that CTE can also affect the
detected position of stars on the detector; the signature behaviour is
a trend in measured position with detector-Y, with a discontiunity at
the inter-chip gap (Kozhurina-Platais et al. 2007).

\begin{figure}
\plotone{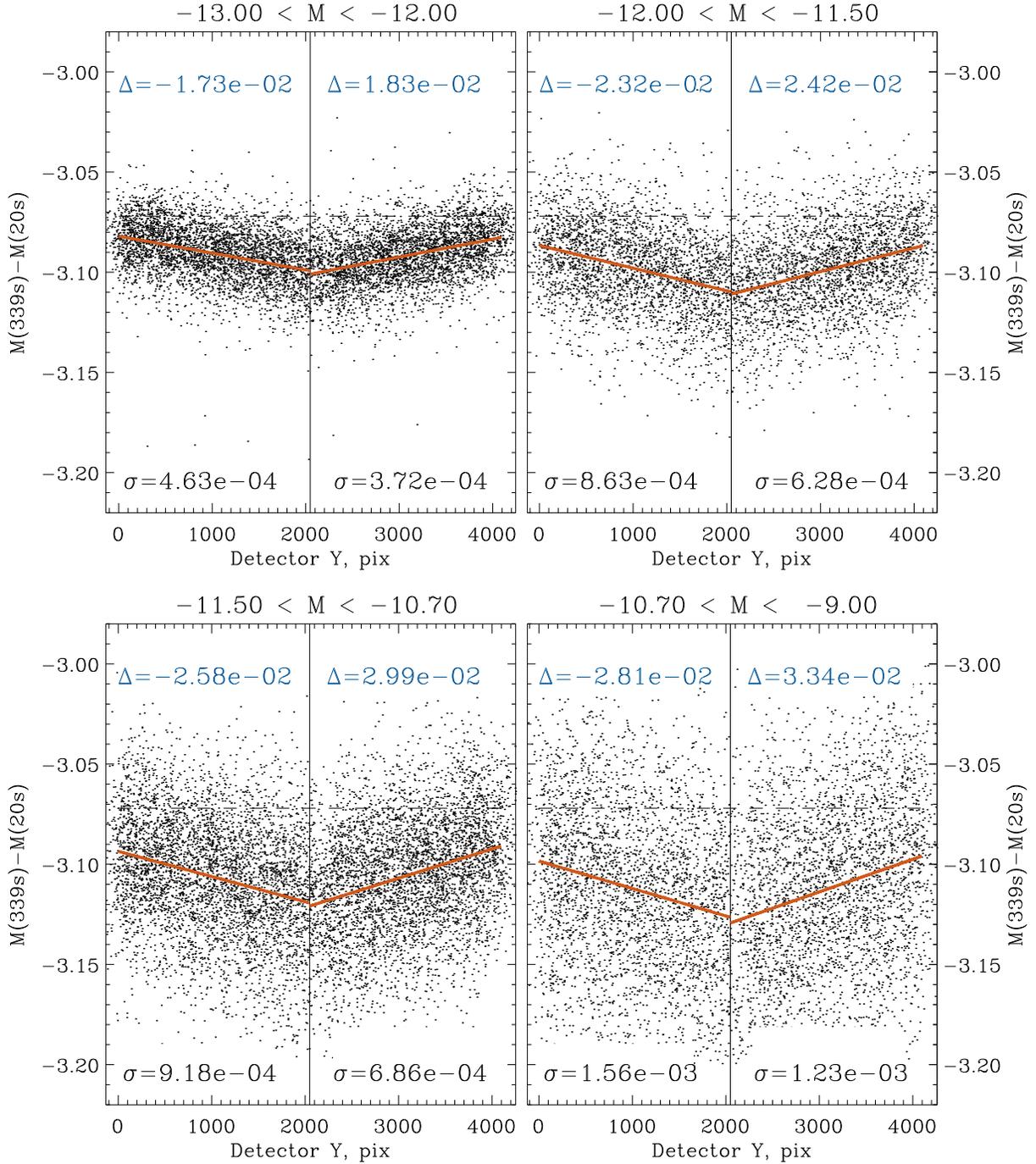}
\caption{Differential Charge Transfer Efficiency (CTE). Within the 2004 epoch, comparison of extracted instrumental magnitudes at 339s and at 20s shows position-dependence typical of CTE effects, where the apparent brightness depends on the distance over the chip the flux must travel at readout. Titles give the instrumental magnitude range in the 339s exposures, the dashed line the expected magnitude difference from the differences in exposure-time alone. Inset numbers give variation amplitude across each chip and the scatter $\sigma$~in the fit due to measurement errors.}
\label{f_cte}
\end{figure}

\begin{figure}
\plotone{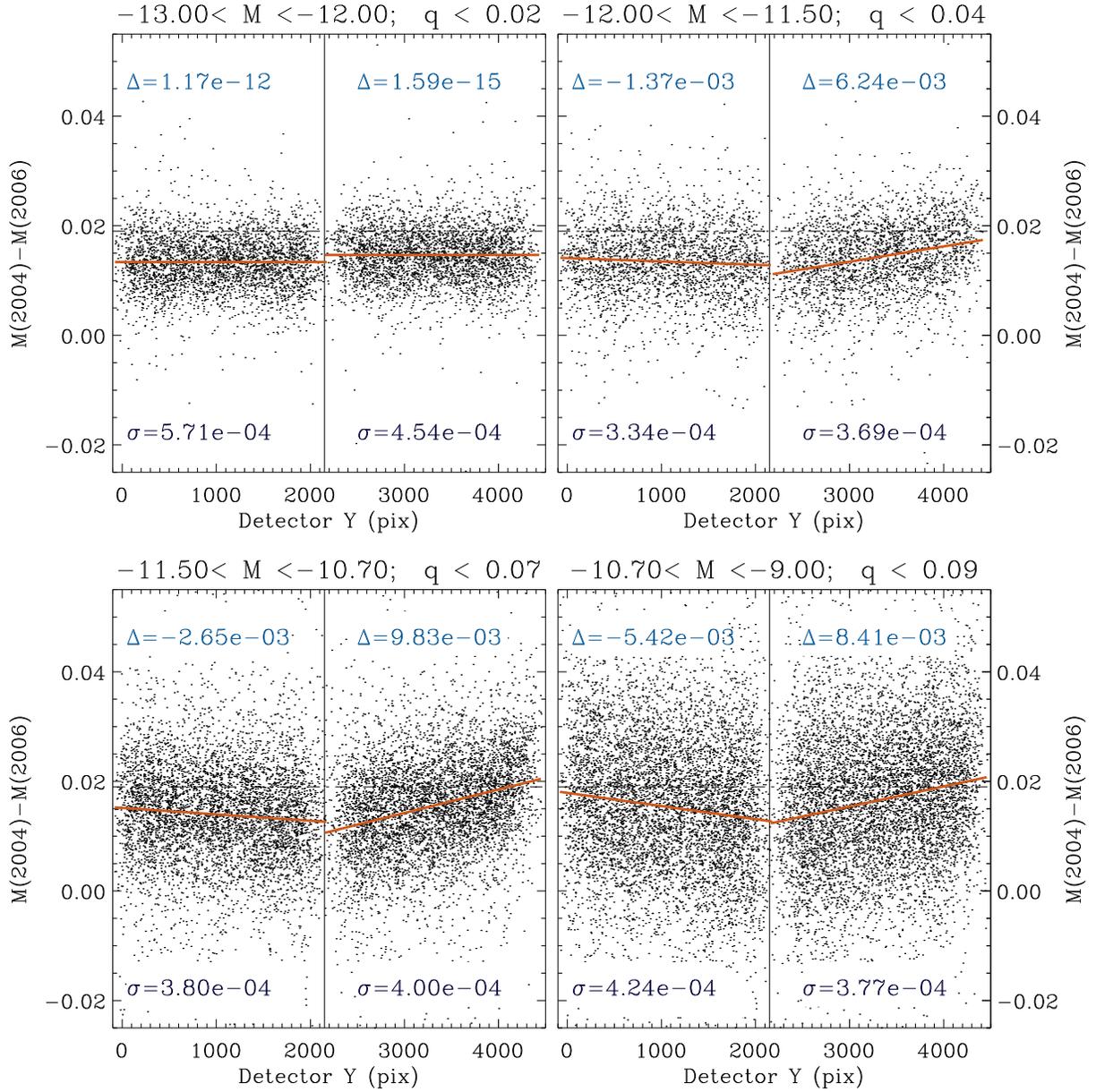}
\caption{When the magnitude difference between {\it epochs} is examined, the instrumental magnitudes recorded from the 345s integrations in 2006 are brighter than those from the 339s integrations in 2004, but by slightly less than the 0.019 mag predicted from the integration times alone; the pattern with Y-position suggests a significant component due to {\it differential} CTE, i.e. the CTE has degraded slightly between the epochs. Symbols as Figure \ref{f_cte}}
\label{f_cte2}
\end{figure}

We thus searched for evidence of any CTE-effects in our data to assess
its impact on proper motion measurements. Brown et al. (2005) used the
$\sim3$-mag difference between the horizontal branch and
subgiant branch to probe this distribution and concluded that the
tendency of the high stellar background to fill pixel wells was
reducing CTE effects to beneath the 5~mmag level. This process is
difficult to apply to our data because similar features with low
intrinsic dispersion are not available in the faint magnitude regime
at which CTE is expected to be significant; when following a similar
procedure to assess the magnitude spread as a function of position we
see largely random variation with Detector-Y which we attribute to
intrinsic variation in the scene. Instead, we establish that CTE
effects are present in the SWEEPS data by measuring the difference in
instrumental magnitude between 339s and 20s exposures within the 2004
epoch. Because the CTE effect is more pronounced for fainter objects,
it will affect measurements in the short exposures more than the long;
thus the V-shaped CTE fingerprint is expected. Indeed, it is clearly
observed (Figure \ref{f_cte}), at the 20-30mmag level and with an
amplitude $\Delta$~that increases with object faintness. Next, we
compare instrumental magnitude measurements of objects observed at
nearly identical exposure times between the two observation epochs
(339s in 2004 and 345s in 2006). The CTE signal is again clearly
visible at the 2-10mmag level, and also increases in amplitude
$\Delta$~with increasing object faintness. This is a {\it
  differential} CTE measurement, in that the CTE of the detector has
degraded somewhat in the two years between measurements (Figure
\ref{f_cte2}).

\begin{figure}
\plotone{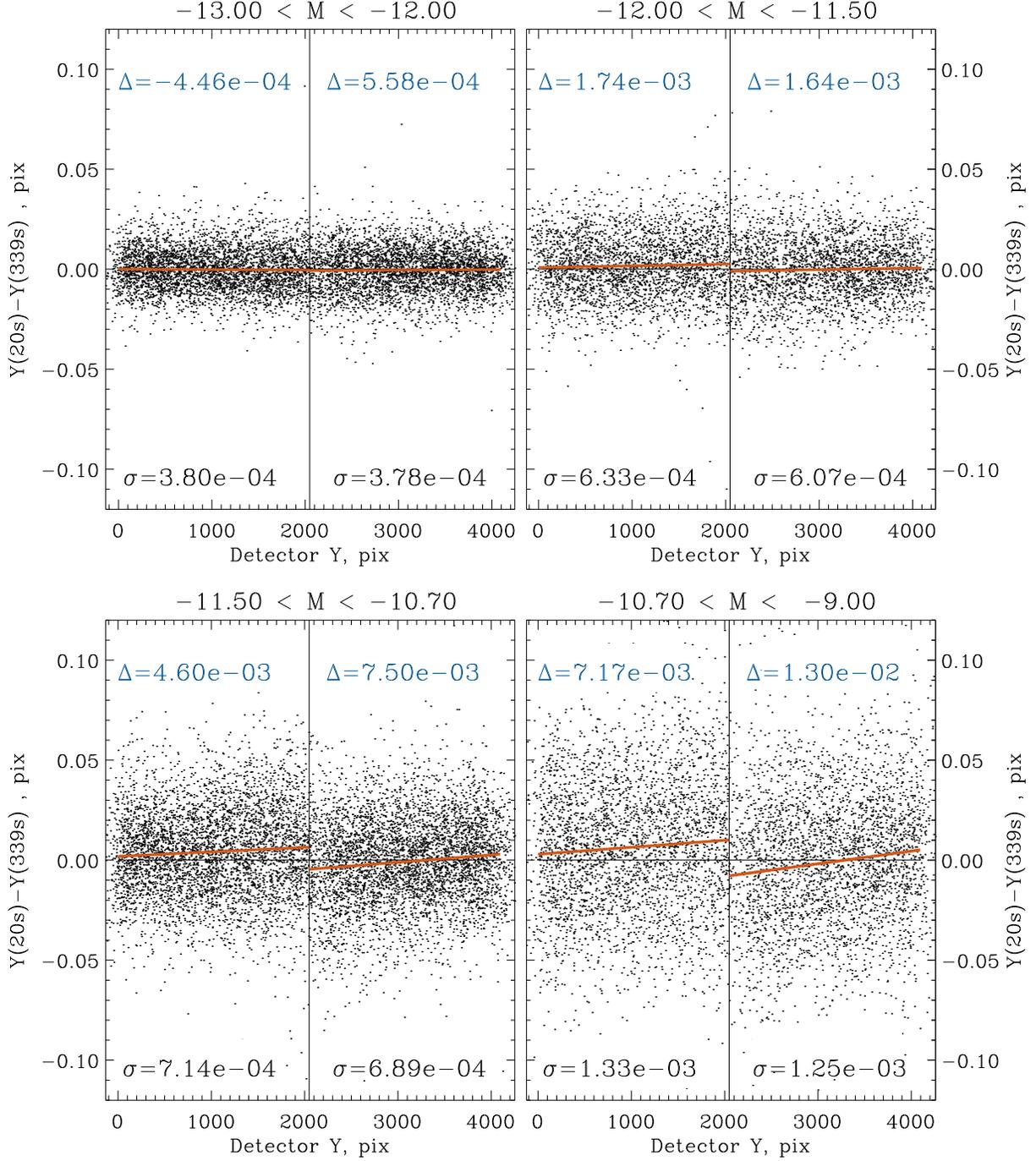}
\caption{Astrometric CTE effects. The Position-dependent CTE shift $\Delta Y$~is given as a function of position on the detector, when comparing the same stars at exposure time 339s and 20s. The titles give the instrumental magnitude range in the 339s exposures. The amplitude of variation across the chip $\Delta$~is given as well as the expected error $\sigma$~on these fits due to the measurement scatter.}
\label{f_cte_pos1}
\end{figure}
\begin{figure}
\plotone{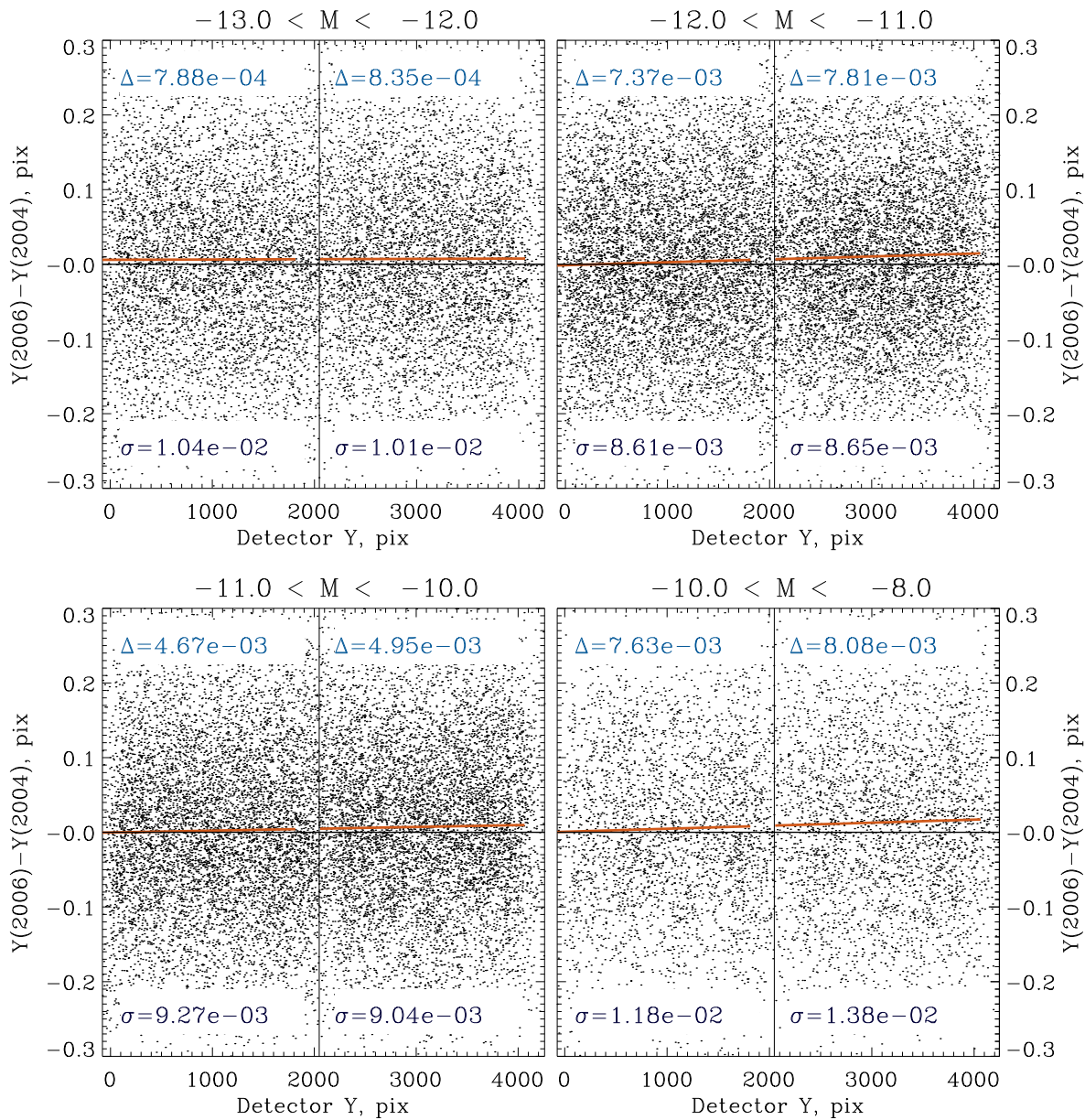}
\caption{As Figure \ref{f_cte_pos1}, this time giving the position discrepancy between the 339s exposures in 2004 with the 349s exposures in 2006. Here the scatter $\sigma$~in fitted trends is much larger due to the {\it intrinsic} motion of the stars; no differential CTE effect is detectable.}
\label{f_cte_pos}
\end{figure}

To estimate the CTE contribution to astrometry and thus proper motion
measurements, we compare position-measurements between 339s and 20s
exposures in the 2004 epoch. This is then scaled by the size of the
inter-epoch CTE magnitude effect to estimate the contribution of CTE
to the proper motion measurements. Positions from the 20s exposures
were transformed to the median frame of the 2004s exposures using
local transformations (Section \ref{sub_pm}), with objects in
likely-bulge regions of the CMD and at instrumental magnitude ($-13
\le M_{\rm{inst}} < -12$) in the 339s epoch used as tracers for the
transformation. A linear trend was fit to the position-difference
$\Delta Y$~as a function of detector-Y (polynomials were also tried
but found to give no advantage over the linear trends) and the range
of $\Delta Y$~across each chip measured. Errors on the trends were
estimated by simulating a large number of trials assuming no intrinsic
variation with detector-Y and computing the standard deviation of
recovered $\Delta Y$~ranges. The astrometric CTE effect noted by
Kozhurina-Platais et al. (2007) is clearly present in the 2004 epoch
(Figure \ref{f_cte_pos1}). The size of the trend increases with target
faintness compared to the tracer stars. No trend is detected in the
tracer-star magnitude range, while the trend reaches 10~millipix in
the ($-10.7 \le M_{\rm{inst}} < -9$) range. Assuming the astrometric
CTE signal to scale with the magnitude CTE signal, we thus expect an
astrometric CTE signal between the deep observations in the 2004 and
2006 epochs of perhaps 0.5-3~millipix at the faintest instrumental
magnitudes. However, when we search for such a signal in the
position-differences between epochs, such a signal is not detectable
above the scatter caused by {\it intrinsic} motion of the stars
themselves between the two epochs (Figure \ref{f_cte_pos}). Thus the
predicted astrometric effect due to {\it differential} CTE is less
than 0.2 mas, and thus not a significant source of error.

\pagebreak

\pagebreak

\clearpage

\clearpage

\clearpage

\clearpage

\clearpage

\clearpage

\clearpage

\clearpage

\clearpage

\clearpage

\clearpage

\clearpage

\clearpage

\clearpage

\clearpage

\clearpage

\clearpage

\clearpage

\clearpage

\epsscale{1.0}
\clearpage

\clearpage

\epsscale{1.0}
\clearpage

\clearpage

\clearpage

\end{document}